%% file: ANA-EXOT-2019-15-PAPER.tex
\documentclass[cernpreprint,atlasdraft=false,texlive=2016,UKenglish,texmf]{atlasdoc}
 
\usepackage{atlaspackage}
\usepackage{atlasbiblatex}
 
\usepackage{atlasphysics}
 
\graphicspath{{logos/}{figures/}}

\usepackage{multirow}
\DeclareOldFontCommand{\bf}{\normalfont\bfseries}{\mathbf}
 
\usepackage{ANA-EXOT-2019-15-PAPER-defs}

\AtlasTitle{Search for pair production of third-generation scalar leptoquarks decaying into a top quark and a $\tau$-lepton in $pp$ collisions at $\sqrt{s}=13~\tev$ with the ATLAS detector}
 
\AtlasVersion{2.3}
\AtlasAbstract{A search for pair production of third-generation scalar leptoquarks decaying into a top quark and a $\tau$-lepton is presented. The search is based on a dataset of $pp$ collisions at $\sqrt{s}=13~\tev$ recorded with the ATLAS detector during Run~2 of the Large Hadron Collider, corresponding to an integrated luminosity of 139~fb$^{-1}$. Events are selected if they have one light lepton (electron or muon) and at least one hadronically decaying $\tau$-lepton, or at least two light leptons. In addition, two or more jets, at least one of which must be identified as containing $b$-hadrons, are required. Six final states, defined by the multiplicity and flavour of lepton candidates, are considered in the analysis. Each of them is split into multiple event categories  to simultaneously search for the signal and constrain several leading backgrounds. The signal-rich event categories require at least one hadronically decaying $\tau$-lepton candidate and exploit the presence of energetic final-state objects, which is characteristic of signal events. No significant excess above the Standard Model expectation is observed in any of the considered event categories, and 95\% CL upper limits are set on the production cross section as a function of the leptoquark mass, for different assumptions about the branching fractions into $t\tau$ and $b\nu$. Scalar leptoquarks decaying exclusively into $t\tau$ are excluded up to masses of $1.43~\tev$ while, for a branching fraction of 50\% into $t\tau$, the lower mass limit is $1.22~\tev$.}
 
\author{The ATLAS Collaboration}
 
\AtlasRefCode{EXOT-2019-15}
\PreprintIdNumber{CERN-EP-2020-241}
 
\AtlasJournal{JHEP}
\AtlasJournalRef{JHEP 06 (2021) 179}
\AtlasDOI{10.1007/JHEP06(2021)179}

\hypersetup{pdftitle={ATLAS document},pdfauthor={The ATLAS Collaboration}}
 
\begin{document}
 
\maketitle
 
\tableofcontents

\section{Introduction}
\label{sec:intro}

The similarities between the quark and lepton sectors of the Standard Model (SM), which exhibit a similar structure, raise the possibility of an existing underlying symmetry connecting the two sectors. Consequently, many extensions of the Standard Model of particle physics contain leptoquarks (LQ)~\cite{pati_salam_colour,georgi_glashow_unification,dimop_suss,techni2, techni3, string, comp}, hypothetical particles that carry non-zero baryon and lepton quantum numbers and are charged under all SM gauge groups.
In particular, they are triplets with respect to the strong interaction, and have fractional electric charge. A LQ state can have either spin 0 (scalar LQ) or spin 1 (vector LQ), and only the former is considered in this paper.
Because of their quantum numbers, LQs couple simultaneously to both quarks and leptons, enabling direct transitions between the two.
Scalar LQs are assumed to couple to the quark--lepton pair via a Yukawa interaction, with coupling constants that can vary across fermion generations, including the possibility of mixing between different quark and lepton generations.
Consequently, scalar LQs can mediate processes that violate lepton flavour universality, and have been proposed as an explanation for measurements of $B$-meson decays that exhibit tantalising deviations from SM predictions~\cite{Hiller:2014yaa, Gripaios:2014tna, Freytsis:2015qca, Bauer:2015knc, DiLuzio:2017chi, Buttazzo:2017ixm, Cline:2017aed}.
The assumption that LQs can only interact with leptons and quarks of the same generation follows the minimal Buchm\"{u}ller--R\"{u}ckl--Wyler (BRW) model~\cite{Buchmueller:1987LQ}, which is adopted in this paper.
The quark--lepton--LQ coupling is determined by two parameters: a model parameter $\beta$ and the coupling parameter $\lambda$. Consequently, the coupling to the charged lepton is given by $\sqrt{\beta} \lambda$, while the coupling to the neutrino is given by $\sqrt{1-\beta} \lambda$.
 
In $pp$ collisions, LQs are mainly produced in pairs ($\mathrm{LQ\overline{LQ}}$) via gluon--gluon fusion and quark--antiquark annihilation, mediated by the strong interaction. There are also lepton-mediated $t$- and $u$-channel production processes that depend on the unknown strength of the Yukawa interaction. However, their contribution can usually be neglected for values of $\lambda \lesssim 1$, and particularly in the case of third-generation LQs ($\mathrm{LQ}_3$), as they would require third-generation quarks in the initial state. The $\mathrm{LQ}$ pair-production cross section can therefore, to a very good approximation, be taken to depend only on the assumed value of the LQ mass ($m_\mathrm{LQ}$) for a given LQ spin and centre-of-mass energy.
Furthermore, it is assumed that the value of $\lambda$ is such that LQs have narrow decay widths of about 0.2\% of $m_\mathrm{LQ}$, so that on-shell production dominates.
Single LQ production in association with a lepton is also possible, but the cross section depends on the strength of the Yukawa interaction and it is not considered
in this paper.
 
The most recent searches from the ATLAS and CMS experiments for pair production of LQs coupling to third-generation quarks and leptons were performed using 36.1~fb$^{-1}$ of $pp$ collisions at $\sqrt{s}=13~\tev$ at the Large Hadron Collider (LHC). The ATLAS results, many of which are reinterpretations of previously published searches for supersymmetric particles, are summarised in Ref.~\cite{EXOT-2017-30}. The different ATLAS searches are not combined statistically and the results are presented as a function of the LQ mass and the branching ratio into charged leptons ($\BR$) for two different classes of LQ signals: up-type LQs ($\LQu \to b\tau / t\nu$) and down-type LQs ($\LQ \to t\tau / b\nu$), which have different electric charges. Both types of LQs are excluded for masses below $800~\gev$ independently of $\BR$. For the limiting cases of $\BR=1$ and  $\BR=0$, masses below $1000~\gev$ and $1030~\gev$ ($970~\gev$ and $920~\gev$) are excluded for $\LQu$ ($\LQ$). Searches for LQs with off-diagonal couplings to third-generation quarks and first- or second-generation leptons have also been performed~\cite{Aad:2020iuy,Aad:2020jmj}. 
The CMS experiment has performed searches for leptoquarks~\cite{CMS-SUS-18-001, CMS-B2G-16-028, CMS-B2G-16-027, CMS-EXO-17-016, CMS-EXO-17-029}, obtaining similar mass exclusions.
 
This paper presents a dedicated search for the pair production of $\LQ$ in the $t\tau t\tau$ decay mode. This search uses the full Run~2 dataset of $pp$ collisions at $\sqrt{s}=13~\tev$ recorded with the ATLAS detector and corresponding to an integrated luminosity of 139~fb$^{-1}$.
Events are selected if they have at least one light lepton (electron or muon, denoted by $\ell$) and at least one hadronically decaying $\tau$-lepton, or at least two light leptons. In addition, two or more jets, at least one of which must be identified as containing $b$-hadrons, are required.
Six final states, defined by the multiplicity and flavour of lepton candidates, are considered in the analysis. Each of them is split into multiple event categories.
The most sensitive event categories require at least one hadronically decaying $\tau$-lepton candidate and exploit the presence of energetic final-state objects, which is characteristic of signal events.
In those event categories the final discriminating variable used is the scalar sum of the transverse momenta of all selected leptons, the selected jets and the missing transverse momentum; this variable peaks at much higher values for the signal than for the background.
The main background contributions arise from top-quark--antitop-quark ($\ttbar$) production with a jet or photon misidentified as a light lepton or with a jet misidentified as a hadronically decaying $\tau$-lepton, and from SM processes yielding multiple leptons in the final state, such as $\ttbar$ production in association with a vector boson or a Higgs boson, and diboson production. The rest of the event categories are designed to be enriched in the most relevant backgrounds. A maximum-likelihood fit is performed across event categories to search for the signal and constrain several leading backgrounds simultaneously. Given the low background yields and good signal-to-background separation provided by the final discriminating variable used in the signal-rich event categories, the search sensitivity is determined by the limited number of data events rather than by the systematic uncertainties of the background estimation.
This search is performed in the LQ mass range between $500~\gev$ and $1600~\gev$ as a function of $\BR$. By considering LQ masses down to $500~\gev$, the coverage of this search partly overlaps with that of Ref.~\cite{EXOT-2017-30}, for which masses below $800~\gev$ were excluded independently of $\BR$. At the same time, this search significantly extends the reach to higher LQ masses.

 
\section{ATLAS detector}
\label{sec:detector}
 
The ATLAS detector~\cite{PERF-2007-01} at the LHC covers almost the entire solid angle around the collision point,\footnote{ATLAS
uses a right-handed coordinate system with its origin at the nominal interaction point (IP) in the centre of the detector.
The $x$-axis points from the IP to the centre of the LHC ring, the $y$-axis points upward, and the $z$-axis coincides with the
axis of the beam pipe. Polar coordinates ($r$,$\phi$) are used in the transverse plane, $\phi$ being the azimuthal angle around
the beam pipe. The pseudorapidity is defined in terms of the polar angle $\theta$ as $\eta = - \ln \tan(\theta/2)$.
Angular distance is measured in units of $\Delta R\equiv \sqrt{(\Delta\eta)^2+(\Delta\phi)^2}$.}
and consists of an inner tracking detector surrounded by a thin superconducting solenoid producing a 2~T axial magnetic field,
electromagnetic and hadronic calorimeters, and a muon spectrometer (MS) incorporating three large toroidal magnet assemblies.
The inner detector contains a high-granularity silicon pixel detector, including the insertable B-layer~\cite{ATLAS-TDR-2010-19,PIX-2018-001},
and a silicon microstrip tracker, together providing a precise reconstruction of tracks of charged particles in the pseudorapidity range $|\eta|<2.5$.
The inner detector also includes a transition radiation tracker that provides tracking and electron identification information for $|\eta|<2.0$.
The calorimeter system covers the pseudorapidity range $|\eta| < 4.9$. Within the region $|\eta|< 3.2$, electromagnetic (EM) calorimetry is
provided by barrel and endcap high-granularity lead/liquid-argon (LAr) electromagnetic calorimeters, with an additional thin LAr presampler
covering $|\eta| < 1.8$ to correct for energy loss in material upstream of the calorimeters. Hadronic calorimetry is provided by
a steel/scintillator-tile calorimeter, segmented into three barrel structures within $|\eta| < 1.7$, and two copper/LAr hadronic endcap calorimeters.
The solid angle coverage is completed with forward copper/LAr and tungsten/LAr calorimeter modules optimised for electromagnetic and
hadronic measurements, respectively. The muon spectrometer measures the trajectories of muons with $|\eta|<2.7$ using multiple layers of
high-precision tracking chambers located in a toroidal field of approximately 0.5~T and 1~T in the central and endcap regions of ATLAS, respectively.
The muon spectrometer is also instrumented with separate trigger chambers covering $|\eta|<2.4$.
A two-level trigger system~\cite{TRIG-2016-01}, consisting of a hardware-based first-level trigger followed by
a software-based high-level trigger (HLT), is used to reduce the event rate to a maximum of around 1~kHz for offline storage.
 
\section{Data and simulated event samples}
\label{sec:datamc}
 
A dataset of $pp$ collisions at $\sqrt{s}=13~\tev$ collected by the ATLAS experiment during 2015--2018 and corresponding to an integrated luminosity of $139~\ifb$ is used. The uncertainty in the integrated luminosity is 1.7\%~\cite{ATLAS-CONF-2019-021}, obtained using the LUCID-2 detector \cite{LUCID2} for the primary luminosity measurements. The number of additional $pp$ interactions per bunch crossing (pile-up) in this dataset ranges from about 8 to 70, with an average of 34. Only events recorded under stable beam conditions and for which all detector subsystems were known to be in a good operating condition are used. The trigger requirements are discussed in Section~\ref{sec:events}.
 
Monte Carlo (MC) simulation samples were produced for the different signal and background processes using the configurations shown in Table~\ref{tab:mcconfig}, with the samples used to estimate the systematic uncertainties in parentheses.
All simulated samples, except those produced with the \textsc{Sherpa}~2.2.1~\cite{Gleisberg:2008ta} event generator, utilised \textsc{EvtGen} 1.2.0~\cite{Lange:2001uf} to model the decays of heavy-flavour hadrons.
Pile-up was modelled using events from minimum-bias interactions generated with \textsc{Pythia}~8.186~\cite{Sjostrand:2007gs} with the A3 set of tuned parameters~\cite{ATL-PHYS-PUB-2011-014} (referred to as the `tune'), and overlaid onto the simulated hard-scatter events according to the luminosity profile of the recorded data.
The generated events were processed through a simulation~\cite{Aad:2010ah} of the ATLAS detector geometry and response
using \textsc{Geant4}~\cite{Agostinelli:2002hh}, and through the same reconstruction software as the dataset of $pp$ collisions.
Corrections were applied to the simulated events so that the particle candidates' selection
efficiencies, energy scales and energy resolutions match those determined from data control samples.
The simulated samples are normalised to their cross sections, and computed to the highest order available in perturbation theory.
 
Samples used to model the $\LQ$ signal were generated at next-to-leading order (NLO) in QCD with \MGMCatNLO~2.6.0~\cite{Alwall:2014hca}, using the LQ model of Ref.~\cite{Mandal:2015lca} that adds parton showers to previous fixed-order NLO QCD calculations~\cite{Kramer:2004df,Kramer:1997hh}, and the NNPDF 3.0 NLO~\cite{Ball:2014uwa} parton distribution function (PDF) set.
The parton shower (PS) and hadronisation were modelled using \textsc{Pythia} 8.230~\cite{Sjostrand:2007gs} with the A14 tune~\cite{ATL-PHYS-PUB-2014-021}. \textsc{MadSpin}~\cite{Artoisenet:2013madspin} was used for the decay of the scalar $\LQ$. The coupling parameter $\lambda$ was set to 0.3, resulting in the $\LQ$ width of about $0.2\%$ of its mass~\cite{Buchmueller:1987LQ,Belyaev:2005LQ}. The charge of $\LQ$ is set to $1/3e$, implying that it decays into either a $t\tau$ or $b\nu$ pair.
Most signal samples were produced for a model parameter of $\beta = 0.5$, which corresponds to identical amplitudes for the $\LQ \to t\tau$ and $\LQ \to b\nu$ processes and, therefore, similar branching ratios for the two decay modes. The signal samples had a mixture of final states so that desired branching ratios $\BR$ were obtained by reweighting the samples based on generator information. These samples were produced for $\LQ$ mass values between $500~\gev$ and $800~\gev$, in steps of $100~\gev$, and between $800~\gev$ and $1.6~\tev$, in steps of $50~\gev$. Additional samples for $\beta = 1$ were generated for the same $\LQ$ mass values between $800~\gev$ and $1.5~\tev$, to gain statistical precision in high-sensitivity signal regions.
The leptoquark signal production cross sections were taken from calculations~\cite{Beenakker:2016lwe,Beenakker:1997ut,Beenakker:2010nq,Beenakker:2016gmf} of direct top-squark pair production, as both are massive, coloured, scalar particles with the same production modes. The calculations were at approximate next-to-next-to-leading order (NNLO) in QCD with resummation of next-to-next-to-leading logarithmic (NNLL) soft gluon terms, with uncertainties determined by variations of the factorisation and renormalisation scales, the strong coupling constant $\alpha_\textrm{S}$, and the PDFs. The cross sections do not include lepton $t$-channel contributions, which are neglected in Ref.~\cite{Mandal:2015lca} and may lead to corrections at the percent level~\cite{Borschensky:2020hot}.
Uncertainties affecting the modelling of the signal acceptance were estimated from the envelope of independent pairs of renormalisation and factorisation scale variations by a factor of 0.5 and 2, by propagating the PDF+$\alpha_\textrm{S}$ uncertainties following the PDF4LHC15 prescription~\cite{Butterworth:2015oua}, and by considering two alternative samples generated with settings that increase or decrease the amount of QCD radiation~\cite{ATL-PHYS-PUB-2016-004}.
 
Samples used to model the $\ttbar$ and single-top-quark background were generated with the NLO generator \textsc{Powheg-Box} v2~\cite{Frixione:2007nw,Nason:2004rx,Frixione:2007vw,Alioli:2010xd,Alioli:2009je,Re:2010bp} using the NNPDF3.0 NLO PDF set. In the $\ttbar$ sample, the \textsc{Powheg-Box} model parameter $h_{\textrm{damp}}$, which controls matrix element (ME) to PS matching and effectively regulates the high-$\pt$ radiation, was set to 1.5 times the top-quark mass. Overlaps between the $\ttbar$ and $tW$ final states were avoided by using
the diagram removal scheme~\cite{Frixione:2005vw}.
The parton shower, hadronisation, and underlying event were modelled by \textsc{Pythia}~8.210 with the NNPDF2.3 LO~\cite{Ball:2012cx} PDF set in combination with the A14 tune.
Uncertainties affecting the modelling of the acceptance and event kinematics of $\ttbar$ events due to the choice of PS and hadronisation model, the NLO ME-to-PS matching, and the effects of initial- and final-state QCD radiation~\cite{ATL-PHYS-PUB-2016-020} are estimated by comparing the nominal predictions with those obtained using the alternative simulated samples (see Table~\ref{tab:mcconfig}).
The $\ttbar$ and single-top-quark simulated samples are normalised to the cross sections calculated at NNLO in QCD including the resummation of NNLL soft gluon terms~\cite{Czakon:2011xx,Kidonakis:2011wy,Kidonakis:2010ux,Kidonakis:2010tc}.

Samples for $\ttW$ and $\ttH$ production were generated using the NLO generators  \textsc{Sherpa}~2.2.1  and \textsc{Powheg-Box} v2~\cite{Hartanto:2015uka}, respectively, with the NNPDF3.0 NLO PDF set. In the case of the $\ttW$ sample, the ME was calculated for up to one additional parton at NLO and up to two partons at LO using \textsc{Comix}~\cite{Gleisberg:2008fv} and \textsc{OpenLoops}~\cite{Cascioli:2011va} and merged with the \textsc{Sherpa} parton shower~\cite{Schumann:2007mg} using the \textsc{MePs@Nlo} prescription~\cite{Hoeche:2012yf}. The generated $\ttH$ events were interfaced to \textsc{Pythia}~8.2 and the A14 tune, and with Higgs decay branching ratios calculated using \textsc{Hdecay}~\cite{deFlorian:2016spz,Djouadi:1997yw}. The cross section used to normalise the $\ttW$ (\ttH) sample is 601 (507) fb, which is computed at NLO in QCD with NLO electroweak corrections~\cite{Beenakker:2001rj,Beenakker:2002nc,Dawson:2002tg,Dawson:2003zu,Yu:2014cka,Frixione:2014qaa,Frixione:2015zaa,Alwall:2014hca, deFlorian:2016spz}.
Uncertainties in the $\ttW$ (\ttH) cross section include $\pm 12\%$ ($^{+5.8\%}_{-9.2\%}$), estimated by varying the QCD factorisation and renormalisation scales, and $\pm 4\%$ ($\pm 3.6\%$)
from PDF+$\alpha_\textrm{S}$ variations, estimated using the PDF4LHC15 prescription. Uncertainties affecting the modelling of the acceptance and event kinematics due to the choice of parton shower and hadronisation model are estimated by comparing the nominal predictions with those obtained using the alternative simulated samples (see Table~\ref{tab:mcconfig}). In the case of the $\ttW$ sample, an additional uncertainty on the modelling of the acceptance and event kinematics is considered from renormalisation and factorisation scale variations by a factor of 0.5 and 2, relative to the nominal scales. 
 
\begin{table*}[t!]
\centering
\caption{\label{tab:mcconfig}
The configurations used for event generation of signal and background processes.
The samples used to estimate the systematic uncertainties are indicated in parentheses.
$V$ refers to production of an electroweak boson ($W$ or $Z/\gamma^*$).
The matrix element order refers to the order in the strong coupling constant of the perturbative calculation.
If only one parton distribution function is shown, the same one is used for both the ME and parton shower generators;
if two are shown, the first is used for the ME calculation and the second for the parton shower.
Tune refers to the underlying-event tune of the parton shower generator. MG5\_aMC
refers to \MGMCatNLO 2.2, 2.3, or 2.6;
\textsc{Pythia} 6 refers to version 6.427~\cite{Sjostrand:2000wi}; \textsc{Pythia} 8
refers to version 8.2; \textsc{Herwig++} refers to version 2.7~\cite{Bahr:2008pv}; \textsc{Herwig 7} refers to version 7.0.4~\cite{Bellm:2015jjp}; \textsc{MePs@Nlo} refers to the method used in \textsc{Sherpa} to match the matrix element to the parton shower.
All samples include leading-logarithm photon emission, either modelled by the parton shower generator or by \textsc{Photos}~\cite{Golonka:2005pn}. The mass of the top quark ($m_t$) and SM Higgs boson were set to $172.5~\gev$ and $125~\gev$, respectively.}
\vspace{0.25cm}
{\small
\setlength\tabcolsep{1.5pt}
\begin{tabular}{llllll}
\hline\hline
Process & Generator & ME order & Parton shower & PDF & Tune  \\
\hline\hline
$\LQLQ$ & MG5\_aMC  & NLO & \textsc{Pythia} 8\ & NNPDF3.0 NLO  & A14 \\
$\ttbar$ & \textsc{Powheg-Box} & NLO & \textsc{Pythia} 8 & NNPDF3.0 NLO/ & A14  \\
&  &  &  & NNPDF2.3 LO  & \\
& (\textsc{Powheg-Box}) & (NLO) & (\textsc{Herwig} 7) &  (NNPDF3.0 NLO/ &  (H7-UE-MMHT) \\
&  &  &  & MMHT2014 LO ) & \\
& (MG5\_aMC) & (NLO) & (\textsc{Pythia} 8) &  (NNPDF3.0 NLO/ &  (A14) \\
&  &  &  & NNPDF2.3 LO)  & \\
& (\textsc{Powheg-Box} & (NLO) & (\textsc{Pythia} 8) &  (NNPDF3.0 NLO/ &  (A14Var3CUp~\cite{ATL-PHYS-PUB-2014-021}) \\
& $h_{\textrm{damp}}=3 m_t$)  &  &  & NNPDF2.3 LO ) & \\
$\ttbar W$  & \textsc{Sherpa 2.2.1} & \textsc{MePs@Nlo} & \textsc{Sherpa} &  NNPDF3.0 NNLO & \textsc{Sherpa} default \\
& (MG5\_aMC) & (NLO) & (\textsc{Pythia} 8) & (NNPDF3.0 NLO/ & (A14)   \\
&  &  &  & NNPDF2.3 LO) & \\
$\ttbar (Z/\gamma^* \to \ell^+\ell^-)$ & MG5\_aMC & NLO & \textsc{Pythia} 8 & NNPDF3.0 NLO/ & A14  \\
&  &  &  & NNPDF2.3 LO & \\
& (\textsc{Sherpa 2.2.0}) & (LO multileg) & (\textsc{Sherpa}) &  (NNPDF3.0 NLO) & (\textsc{Sherpa} default)\\
$\ttbar \to W^+bW^-\bar{b}\ell^+\ell^-$ & MG5\_aMC & LO & \textsc{Pythia} 8 & NNPDF3.0 LO & A14   \\
\ttH & \textsc{Powheg-Box}  & NLO & \textsc{Pythia} 8\ & NNPDF3.0 NLO / & A14 \\
&  &  &  & NNPDF2.3 LO & \\
& (\textsc{Powheg-Box}) & (NLO) & (\textsc{Herwig} 7) &  (NNPDF3.0 NLO/ &  (H7-UE-MMHT) \\
&  &  &  & MMHT2014 LO \cite{Harland-Lang:2014zoa}) & \\
Single top & \textsc{Powheg-Box} & NLO & \textsc{Pythia} 8 & NNPDF3.0 NLO/ & A14   \\
($t$-, $Wt$-, $s$-channel) &  &  &  & NNPDF2.3 LO &  \\
$t (Z/\gamma^*)$ & MG5\_aMC & LO & \textsc{Pythia} 6  & CTEQ6L1 & Perugia2012~\cite{Skands:2010ak}  \\
$t W (Z/\gamma^*)$ & MG5\_aMC & NLO & \textsc{Pythia} 8 & NNPDF2.3 LO  & A14  \\
$t\bar t t$, $t\bar t t\bar t$ & MG5\_aMC & LO & \textsc{Pythia} 8 & NNPDF2.3 LO & A14 \\
$t\bar t W^+ W^-$ & MG5\_aMC & LO & \textsc{Pythia} 8 & NNPDF2.3 LO & A14  \\
$VV$, $qqVV$, $VVV$ & \textsc{Sherpa} 2.2.2  & \textsc{MePs@Nlo} & \textsc{Sherpa} & NNPDF3.0 NNLO & \textsc{Sherpa} default  \\
$VH$  & \textsc{Pythia} 8 & LO  & \textsc{Pythia} 8 & NNPDF2.3 LO & A14 \\
$W$+jets & \textsc{Sherpa} 2.2.1 & \textsc{MePs@Nlo} & \textsc{Sherpa} & NNPDF3.0 NLO & \textsc{Sherpa} default \\
$Z$+jets & \textsc{Sherpa} 2.2.1 & \textsc{MePs@Nlo} & \textsc{Sherpa} & NNPDF3.0 NLO & \textsc{Sherpa} default \\
\hline\hline
\end{tabular}
}
\end{table*}
 
The samples for $\ttdy$ and diboson ($VV$) production follow Ref.~\cite{ATL-PHYS-PUB-2016-002,ATL-PHYS-PUB-2016-004}. For $\ttdy$, the inclusive $\ttll$ ME is computed, including off-shell $Z$ and $\gamma^*$ contributions with $m(\ell^+\ell^-) > 1~\gev$. A dedicated $\ttbar$ sample, including rare $t \to Wb \gamma^*(\to \ell^+\ell^-)$ radiative decays and requiring $m(\ell^+\ell^-) > 1~\gev$, referred to as the $\ttbar \to W^+bW^-\bar{b}\ell^+\ell^-$ sample, was added to the $\ttdy$ sample and together these form the $\ttdy$ (high mass) sample. The contribution from internal photon conversions ($\gamma^* \to \ell^+\ell^-$) with $m(\ell^+\ell^-) < 1~\gev$ is modelled by QED multiphoton radiation in the inclusive $\ttbar$ sample and is referred to as $\ttbar\gamma^*$ (low mass).
Care was taken to avoid both double-counting of contributions and uncovered regions of phase space when combining the different simulated samples. The cross section for $\ttbar (Z/\gamma^* \to \ell^+\ell^-)$ production is 167~fb, computed at NLO in QCD and electroweak couplings~\cite{Alwall:2014hca,Frixione:2015zaa}. The uncertainties from QCD scale and PDF+$\alpha_\textrm{S}$ variations are $\pm 12\%$ and $\pm 4\%$ respectively. The LO cross section from the $\ttbar \to W^+bW^-\bar{b}\ell^+\ell^-$ sample is scaled by a factor of 1.54, based on comparisons between the NNLO+NLL and LO cross sections for $\ttbar$ production~\cite{Cacciari:2011hy, Baernreuther:2012ws, Czakon:2012zr, Czakon:2012pz, Czakon:2013goa}, and assigned a 50\% normalisation uncertainty, to cover possible residual effects in the predicted yield due to the simplified normalisation procedure used and/or the fact that the event kinematics were modelled using a LO simulation.
Uncertainties affecting the modelling of the acceptance and event kinematics for the $\ttdy$ sample include the same QCD scale and tune variations as considered for the \ttH sample, PDF variations using the PDF4LHC15 prescription, and a comparison with an alternative LO multileg sample (see Table~\ref{tab:mcconfig}).
Diboson backgrounds are normalised using the cross sections computed by \textsc{Sherpa} 2.2.2. To cover possible mismodellings in the associated heavy-flavour production predicted by the parton shower, a 50\% normalisation uncertainty is assigned and treated as correlated between the $WZ$+$\geq$1$c$ and $WZ$+$\geq$1$b$ subprocesses.
The remaining rare background contributions listed in Table~\ref{tab:mcconfig} are normalised using their NLO theoretical cross sections, except for the $t\bar t t$ process, for which a LO cross section is used. To account for the fact that many of these processes are predicted using a LO simulation, and to cover possible mismodellings in the extreme kinematic regime probed by this search, a 50\% normalisation uncertainty is assigned to all of them.
 
 
\section{Event reconstruction}
\label{sec:objects}
 
Interaction vertices from the $pp$ collisions are reconstructed from at least two tracks with transverse momentum ($\pt$) larger than $500~\mev$ that are consistent with originating from the
beam collision region in the $x$--$y$ plane. If more than one primary vertex candidate is found, the candidate for which the associated tracks form the largest sum of squared $\pt$~\cite{ATL-PHYS-PUB-2015-026} is selected as the hard-scatter primary vertex.
 
Electron candidates are reconstructed from energy clusters in the electromagnetic calorimeter that are associated with inner-detector tracks~\cite{PERF-2017-01}.
They are required to satisfy $\pt>10$~GeV and $|\eta_{\mathrm{cluster}}|<2.47$, excluding the transition region between the endcap and barrel calorimeters ($1.37<|\eta_{\mathrm{cluster}}|<1.52$).
Loose and tight electron identification working points are used~\cite{EGAM-2018-01}, based on a
likelihood discriminant employing calorimeter, tracking and combined variables that provide separation between electrons and jets.
The associated track of an electron candidate is required to have at least two hits in the pixel detector and seven hits total in the pixel and silicon-strip detectors combined. For the tight identification working point, one of these pixel hits must be in the innermost layer (or the next-to-innermost layer if the module traversed in the innermost layer is non-operational), and there must be no association with a vertex from a reconstructed photon conversion~\cite{PERF-2017-02} in the detector material (termed a `material conversion' in this paper).
 
Muon candidates are reconstructed by matching tracks connecting track segments in different layers of the muon spectrometer to tracks found in the inner detector.
The resulting muon candidates are re-fitted using the complete track information from both detector systems~\cite{PERF-2015-10}.
They are required to satisfy $\pt > 10~\gev$ and $|\eta| < 2.5$.
Loose and medium muon identification working points are used~\cite{PERF-2015-10}.  Medium muon candidates with $\pt>800~\gev$ are in addition required to have hits in at least three MS stations (referred to as the `high-$\pt$ working point'), in order to maximise the momentum resolution for the muon track
and thus suppress backgrounds with high-$\pt$ muons arising from momentum mismeasurements.
 
Electron (muon) candidates are matched to the primary vertex by requiring that the significance of their transverse impact parameter, $d_0$,
satisfies $|d_0/\sigma(d_0)|<5\,(3)$, where $\sigma(d_0)$ is the measured uncertainty in $d_0$,
and by requiring that their longitudinal impact parameter, $z_0$, satisfies $|z_0 \sin\theta|<0.5$~mm, where $\theta$ is the track's polar angle.
To further suppress leptons from heavy-flavour hadron decays, misidentified jets, or photon conversions (collectively referred to as `non-prompt leptons'),
lepton candidates are also required to be isolated in the tracker and in the calorimeter. A track-based lepton isolation criterion is defined by calculating the quantity $I_R = \sum \pt^{\textrm{trk}}$, where
the scalar sum includes all tracks (excluding the lepton candidate itself) within the cone defined by $\Delta R<R_{\textrm{cut}}$ around the 
direction of the lepton.  The value of $R_{\textrm{cut}}$ is the smaller of $r_{\textrm{min}}$ and $10~\gev/\pt^\ell$, where
$r_{\textrm{min}}$ is set to 0.2 (0.3) for electron (muon) candidates and where $\pt^\ell$ is the lepton $\pt$.
All lepton candidates must satisfy $I_R/\pt^\ell < 0.15$.
Additionally, electrons (muons) are required to satisfy a calorimeter-based isolation criterion: the sum of the transverse energy within a cone of size
$\Delta R=0.2$ around the lepton, after subtracting the contributions from pile-up and the energy deposit of the lepton itself, is required to be less than 20\% (30\%) of $\pt^\ell$.
Muons are required to be separated by $\Delta R > 0.2$ from any selected jets (defined below). If two electrons are closer than $\Delta R=0.1$, only the one with the higher \pt\ is considered. An electron lying within $\Delta R=0.1$ of a selected muon is rejected.
 
Light leptons of different qualities are used in the analysis, as summarised in Table~\ref{tbl:tightleps}. `Loose' light leptons simply satisfy the corresponding identification criteria, as well as the isolation and impact parameter requirements discussed above. They are used in the event preselection, and to define non-overlapping analysis channels (see Section~\ref{sec:event_selection}).  `Tight' and/or `Very Tight' light leptons are then required, depending on the analysis channel, to improve the rejection of particular reducible backgrounds (see Section~\ref{sec:eventcat}). They are discussed further in the following. Uncertainties in light-lepton reconstruction, identification, isolation, and trigger efficiencies are taken into account, but have a negligible impact in the analysis.
 
Despite the fact that leptons in decays of hadrons that contain bottom- and charm-quarks are highly suppressed by the selection criteria described above, several analysis channels considered in this search (see Section~\ref{sec:events}) require additional suppression of backgrounds containing non-prompt leptons, and other processes where the electron charge is incorrectly assigned.
Non-prompt leptons are further rejected using a boosted decision tree (BDT) discriminant based on isolation and variables that are used in the calculation of the multivariate $b$-tagging discriminant (see description below)
referred to as the non-prompt lepton BDT~\cite{HIGG-2017-02}.
The efficiency at the chosen working point for muons (electrons) that satisfy the calorimeter- and track-based isolation criteria
is about 80\% (65\%) for $\pt \sim 20~\gev$ and reaches a plateau of 95\% (90\%) at $\pt \sim 45~\gev$.
The corresponding rejection factor\footnote{The rejection factor is defined as the reciprocal of the efficiency.} against leptons from the decay of $b$-hadrons is about 3.5 (10), after resolving ambiguities between overlapping reconstructed objects.
Very Tight muon candidates are Tight muons that pass the non-prompt lepton BDT requirement (referred to as the `non-prompt-lepton veto').
To further suppress material conversions, additional requirements on the associated track $\pt$ and on the ratio of the electron's calorimeter energy to
its track's momentum are applied to tight electrons.
Tight electrons with incorrect charge assignment are rejected using a BDT discriminant based on calorimeter and tracking quantities~\cite{PERF-2017-01}. An efficiency of 88\% for isolated electrons with correct charge assignment is obtained, with a rejection factor of $\sim$3.3 for isolated electrons with incorrect charge assignment.
The resulting electron candidates are further split into three classes: `Material Conversion', `Internal Conversion', and `Very Tight'.
Material conversion candidates have a reconstructed displaced vertex with radius $r > 20$~mm that includes the track associated with the electron.\footnote{The beampipe and insertable B-layer inner radii are 23.5~mm and 33~mm, respectively.} The invariant mass of the associated track and the closest (in $\Delta\eta$) opposite-charge track reconstructed in the silicon detector, calculated at the conversion vertex, is required to be $<100~\mev$. Internal conversion candidates, which correspond to the internal photon conversions (see Section~\ref{sec:datamc}), are required to fail the requirements for material conversions, and the di-track invariant mass, this time calculated at the primary vertex, is also required to be $<100~\mev$.
Therefore, Very Tight electron candidates are Tight electrons that satisfy the non-prompt-lepton veto, the charge-misassignment veto, the internal-conversion veto, and the material-conversion veto requirements,
and have $|\eta| < 2$. The last requirement rejects a small fraction of electrons with a large charge misassignment rate because of the limited number of hits used in the track reconstruction.
 
\begin{table*}
\centering
\caption{\label{tbl:tightleps}
Summary of requirements applied to define Loose (L), Tight (T), and Very Tight (T*) light leptons.
The quality cuts for tight muon identification depend on the transverse momentum of the muon candidates.
}
\vspace{0.25cm}
\small
\begin{tabular}{l|ccc|ccc}
\hline\hline
& \multicolumn{3}{c|}{$e$} & \multicolumn{3}{c}{$\mu$} \\
& L & T & T* & L & T & T* \\
\hline\hline
Identification  & loose & tight & tight & loose & medium & medium  \\
& & & & & or high-$p_{\mathrm T}$ & or high-$p_{\mathrm T}$ \\
Isolation & \multicolumn{3}{c|}{Yes} & \multicolumn{3}{c}{Yes}  \\
Non-prompt-lepton veto &  No & No & Yes & No & No & Yes \\
Electron charge-misassignment veto & No & No  & Yes & \multicolumn{3}{c}{---} \\
Electron material-conversion veto & No & No & Yes & \multicolumn{3}{c}{---} \\
Electron internal-conversion veto & No & No & Yes & \multicolumn{3}{c}{---} \\
$|d_0|/\sigma_{d_0}$  &  \multicolumn{3}{c|}{$<5$~~~~} & \multicolumn{3}{c}{$<3$~~~~} \\
$|z_0 \sin \theta|$ [mm] & \multicolumn{3}{c|}{$< 0.5$} & \multicolumn{3}{c}{$< 0.5$} \\
\hline\hline
\end{tabular}
\end{table*}

Hadronically decaying $\tau$-lepton candidates ($\tauh$) are reconstructed from energy clusters in the calorimeters and associated inner-detector tracks~\cite{PERF-2013-06, ATL-PHYS-PUB-2015-045}. They are required to have either one or three associated tracks (referred to as `one-prong' and `three-prong' $\tauh$ candidates, respectively), with a total charge of $\pm 1e$.
The candidates are required to satisfy $\pt>25~\gev$ and $|\eta|<2.5$, excluding the EM calorimeter's transition region, and to originate from the primary vertex.
A recurrent neural network discriminant using calorimeter- and tracking-based variables is used to identify real $\tauh$ candidates and reject jet backgrounds (referred to as `fake $\tauh$ candidates') ~\cite{ATL-PHYS-PUB-2019-033}.
Loose and medium identification working points are used, and the selected $\tauh$ candidates are referred to as `Loose' and `Medium', respectively. The loose working point has a target efficiency of 85\% (75\%) for one-prong (three-prong) $\tauh$ candidates, with an expected rejection factor against light-jets of 21 (90). The corresponding efficiencies and rejections for the medium working point are 75\% (60\%) and 35 (240) for one-prong (three-prong) $\tauh$ candidates, respectively.
Electrons that are reconstructed as one-prong $\tauh$ candidates are removed using a BDT with an efficiency (rejection factor) of 95\% (30--100) for real (fake) $\tauh$ candidates depending on the $\pt$. Additionally, $\tauh$ candidates are required to be separated by $\Delta R > 0.2$ from any selected electron or muon candidates.
The $\tauh$ reconstruction and identification efficiencies and the $\tauh$ energy scale in the simulation are calibrated to those measured in a data control sample of $Z \to \tau^+\tau^-$ events~\cite{ATLAS-CONF-2017-029}, and the associated uncertainties are considered in the analysis.
The uncertainty in the $\tauh$ identification efficiency is split into eight uncorrelated components, corresponding to different $\tauh$ $\pt$ ranges and separately for one-prong and three-prong candidates. It is approximately 2.5\% (3.0\%) for one-prong (three-prong) $\tauh$ candidates with $\pt<300~\gev$, and 3.5\% (6.5\%) for $\pt \geq 300~\gev$.
The uncertainty in the $\tauh$ energy scale is about 1.2\% (3.0\%) for one-prong (three-prong) $\tauh$ candidates~\cite{ATLAS-CONF-2017-029}, and is split into eight independent components.
An additional correction and associated uncertainties are estimated for the probability of misidentification of electrons as $\tauh$ candidates using a data control sample of $Z \to e^+e^-$ events.

The inputs for jet reconstruction are built by combining measurements from both the tracker and the calorimeter using the particle flow (PFlow) algorithm~\cite{PERF-2015-09,1805279}. Jet candidates are reconstructed from such PFlow objects using the anti-$k_t$ algorithm with a radius parameter $R=0.4$~\cite{Cacciari:2008gp,Cacciari:2011ma}.
After subtracting the expected energy contribution from pile-up following the jet area method~\cite{Cacciari:2008gn}, the jet energy scale (JES) and resolution (JER) are corrected to particle level using MC simulation, and then calibrated in situ using $Z$+jets, $\gamma$+jets and multijet events~\cite{1805279}.
Jets are required to satisfy $\pt > 25~\gev$ and $|\eta| < 2.5$. A jet-vertex tagger (JVT) is used to remove jets associated with pile-up vertices and having $\pt < 60~\gev$ and $|\eta| < 2.4$~\cite{PERF-2014-03}.  Any jets within $\Delta R = 0.2$ of a selected electron or a $\tauh$ candidate are rejected.
Uncertainties associated with jets arise from the JES and JER, and the efficiency to pass the JVT requirement.
The largest contribution results from the JES, whose uncertainty dependence on jet $\pt$ and $\eta$, jet flavour, and pile-up treatment is split into 27 uncorrelated components
that are treated independently in the analysis~\cite{1805279}.  The total JES uncertainty varies from 1\% to 4\% depending on the jet \pT. A total of seven uncorrelated uncertainty components affecting the JER are also considered.

Jets containing $b$-hadrons are identified ($b$-tagged) via an algorithm~\cite{FTAG-2018-01,ATL-PHYS-PUB-2017-013} that uses multivariate techniques to combine information about the impact parameters of displaced tracks and the topological properties of secondary and tertiary decay vertices reconstructed within the jet. For each jet, a value for the multivariate $b$-tagging discriminant is calculated. A jet is considered $b$-tagged if this value is above the threshold corresponding to an average 77\% efficiency to tag a $b$-quark jet, with a light-jet\footnote{`Light jet' refers to a jet originating from the hadronisation of a light quark ($u$, $d$, $s$) or a gluon.} rejection factor of about 140, a charm-jet ($c$-jet) rejection factor of about 4, and a $\tauh$-jet rejection factor of about 17, as determined for jets with
$\pt >20~\gev$ and $|\eta|<2.5$ in simulated $\ttbar$ events.
Correction factors derived from dedicated calibration samples enriched in $b$-jets, $c$-jets, or light jets, are applied to the simulated samples~\cite{FTAG-2018-01,ATLAS-CONF-2018-001,ATLAS-CONF-2018-006}. In the case of $\tauh$-jets, for which no dedicated calibration sample exists, the correction factors derived for $c$-jets are used.
Uncertainties in these corrections include a total of nine independent sources affecting $b$-jets and five independent sources affecting $c$-jets.
Six sources of uncertainty affecting light jets are also considered.
An additional uncertainty is included for the extrapolation of these corrections to jets
with $\pt$ beyond the kinematic reach of the data calibration samples used ($\pt>300~\gev$ for $b$- and $c$-jets,
and $\pt>750~\gev$ for light jets); it is taken to be correlated among the three jet flavours.
Finally, an uncertainty related to the application of $c$-jet scale factors to $\tauh$-jets is considered.
The approximate relative size of the $b$-tagging efficiency uncertainty is 2\% for $b$-jets, 10\% for $c$-jets and $\tauh$-jets, and 30\% for light jets.
 
The missing transverse momentum $\vec{p}_{\mathrm{T}}^{\mathrm{miss}}$ (with magnitude $\met$) is defined as the negative vector sum of the
$\pt$ of all selected and calibrated objects in the event, including a term to account for momentum from soft particles
in the event that are not associated with any of the selected objects~\cite{PERF-2016-07}.
This soft term is calculated from inner-detector tracks matched to the selected primary vertex, which makes it more resilient to
contamination from pile-up interactions. Uncertainties associated with energy scales and resolutions of leptons and jets
are propagated to $\vec{p}_{\mathrm{T}}^{\mathrm{miss}}$. Additional uncertainties originating from the modelling
of the underlying event, in particular its impact on the $\pt$ scale and resolution
of unclustered energy, are negligible.

 
\section{Search strategy}
\label{sec:events}
 
The search discussed in this paper targets $\LQ$ pair production in the $t\tau t\tau$ final state, thus being particularly sensitive to high values of $\BR$. In this decay mode, there is a high probability that the final state contains at least one light lepton from a semileptonic top-quark decay or a leptonic $\tau$-lepton decay, which is used to trigger the event and to help suppress multijet backgrounds. The presence of additional $\tauh$ candidates and/or additional light leptons is exploited to further reduce SM backgrounds and improve the search sensitivity. The final state of interest also contains two energetic $b$-jets, and may contain additional light jets from initial- or final-state radiation and/or from a hadronically decaying $W$ boson in one of the top-quark decays. The multiple sources of leptons in the event motivate the definition of different analysis channels depending on the multiplicity of light leptons, the multiplicity of $\tauh$ candidates, and the electric charges of light leptons (see Section~\ref{sec:event_selection}). The analysis channels are subdivided into different event categories (see Section~\ref{sec:eventcat}) so that a maximum-likelihood fit is performed across event categories to search for the signal and constrain several leading backgrounds simultaneously. The requirement of multiple leptons in the event implies the presence of multiple neutrinos, which makes the kinematic reconstruction of the top quarks and consequently of the LQ invariant mass difficult. Nevertheless, the decay of a pair of massive LQs results in energetic final-state objects, which is exploited in the most sensitive analysis channels, both in optimising the event selection in the different categories considered and in defining a powerful event variable used in the statistical analysis to discriminate the signal from the background. Further details of the search strategy are provided in the following sections.
 
\subsection{Event selection}
\label{sec:event_selection}
The events used in the analysis are selected with high efficiency using single-lepton and dilepton triggers~\cite{TRIG-2016-01}, which use electron and muon signatures.
Single-lepton triggers with low $\pt$ threshold and lepton isolation requirements are combined in a logical OR with higher-threshold triggers without isolation requirements to give maximum efficiency.
Single-electron triggers with a $\pt$ threshold of 24 (26)~$\gev$ in the 2015 (2016, 2017 and 2018) data-taking period(s) and isolation requirements are used along with triggers with a 60~$\gev$ threshold and no isolation requirement, and with a 120 (140)~$\gev$ threshold with looser identification criteria.
For single-muon triggers, the lowest $\pt$ threshold is 20 (26)~$\gev$ in 2015 (2016--2018), while the higher $\pt$ threshold is 50~$\gev$ for all periods.
The dielectron triggers require two electrons that satisfy loose identification criteria with different $\pt$ thresholds:
12~$\gev$ in 2015, 17~$\gev$ in 2016, and 24~$\gev$ in 2017--2018 .
Dimuon triggers utilise asymmetric $\pt$ thresholds for leading (subleading) muons: 18 (8)~$\gev$ in 2015 and  22 (8)~$\gev$ in 2016--2018.
An electron+muon trigger requires events to have an electron candidate satisfying loose identification with a 17~$\gev$ threshold and a muon candidate with a 14~$\gev$ threshold for all periods.
 
Events selected by the trigger are required to satisfy basic preselection requirements. They must have at least one primary vertex candidate. Events are required to contain either one light lepton and at least one $\tauh$ candidate, or at least two light leptons. At this stage, the light leptons and $\tauh$ candidates satisfy the Loose selection criteria (see Section~\ref{sec:objects}) and have $\pt > 10~\gev$ and $\pt>25~\gev$, respectively. Furthermore, the leading light lepton in the event is required to have $\pt > 25~\gev$. Events with one light lepton must have been selected by a single-lepton trigger, whereas events with at least two light leptons are required to be selected by a logical OR of the single-lepton and dilepton triggers. The selected light leptons are required to match, with $\Delta R < 0.15$, the corresponding leptons reconstructed by the trigger and to have a $\pt$ exceeding the trigger $\pt$ threshold by 1~$\gev$ or 2~$\gev$ (depending on the lepton trigger, lepton multiplicity criteria, and data-taking conditions), besides the $25~\gev$ requirement for the leading light leptons.
These requirements are used to ensure operating in the trigger efficiency plateau, and to apply any corrections to the simulation in order to reproduce the per-lepton trigger efficiencies measured in data~\cite{Aad:2019wsl,Aad:2020uyd}.
In addition, two or more jets, at least one of which is $b$-tagged, are required.
The trigger requirement has an efficiency of about 85\% (98\%) for signal events with one light lepton (at least two light leptons) satisfying the preselection requirements.
 
Six final states, termed `channels', are analysed, defined by the multiplicity and flavour of Loose lepton candidates with the $\pt$ requirements indicated above:
\begin{itemize}
\item \ltau:  one light lepton and at least one $\tauh$ candidate;
\item \llOStau:  two opposite-charge (denoted by OS, standing for opposite-sign) light leptons and at least one $\tauh$ candidate;
\item \llSSorllltau:  two same-charge (denoted by SS, standing for same-sign) light leptons or three light leptons, and at least one $\tauh$ candidate;
\item \llOSnotau: two OS light leptons and no $\tauh$ candidates;
\item \llSSnotau:  two SS light leptons and no $\tauh$ candidates;
\item \lllnotau:  three light leptons and no $\tauh$ candidates.
\end{itemize}
The selection criteria are orthogonal to those of the other channels so that each event only contributes to a single analysis channel.
Finally, in all analysis channels the minimum $\pt$ requirement on light leptons is raised to $25~\gev$.
The analysis channels with no $\tauh$ candidates are used for the determination of particular backgrounds, while those with at least one $\tauh$ candidate are in addition used to search for the signal.

\subsection{Event categorisation}
\label{sec:eventcat}
 
The channels are subdivided into different event categories optimised either to search for the signal (referred to as `signal regions', or SR), to obtain improved background estimates (referred to as `control regions', or CR), or to validate the estimated backgrounds (referred to as `validation regions', or VR).
In the optimisation of the SRs, different features of the LQ signal are exploited, such as the multiplicity of $\tauh$ candidates, the charge configuration of reconstructed leptons and, especially, the difference in kinematics of final-state objects between signal and background.
In particular, the effective mass ($\Meff$), defined as the scalar sum of the transverse momenta of all selected leptons, the selected jets and the missing transverse momentum, is a powerful discriminating variable between signal and background. Additional kinematic variables exploited in the optimisation of the SRs include the $\pt$ of $\tauh$ candidates, and different invariant mass variables based on dilepton pair combinations (e.g.\ the invariant mass of the two leading $\tauh$ candidates, $\Mtt$).
The CRs are defined by inverting particular selections in order to provide background-rich samples that do not overlap with the SRs.
The VRs are defined to be kinematically closer to the SRs, and they do not overlap with the other CRs and SRs.
A total of 7 SRs, 18 CRs, and 6 VRs are considered, with their definitions given below.
For a $\LQ$ signal with $\BR=1$, the acceptance times efficiency within the seven SRs is found to be about 10\%, varying only slightly with the $\LQ$ mass, with higher mass values resulting in higher acceptance times efficiency to pass the kinematic requirements.
 
In the \ltau channel, events are required to have one Tight light lepton and, either one Medium $\tauh$ candidate and no additional Loose $\tauh$ candidates, or at least two Loose $\tauh$ candidates.
A total of nine event categories are defined, which are summarised in Table~\ref{tab:1leptau}. They consist of two subcategories based on the multiplicity of $\tauh$ candidates (1 or $\geq$2), with the former subcategory further split according to the charge configuration of the selected light lepton and $\tauh$ candidate (OS or SS). The splitting between OS and SS events improves the sensitivity, since their background compositions and signal-to-background ratios are very different.
For each of these subcategories, a CR, a VR, and a SR, are defined. All SRs require one or two high-$\pt$ $\tauh$ candidates, as appropriate, a requirement that provides significant background suppression, as illustrated in Figure~\ref{fig:shape_taupt}. Further requirements are placed on additional kinematic variables, such as the invariant mass of the light lepton and the $\tauh$ candidate ($m_{\ell\tau}$) (see Figure~\ref{fig:shape_mltau}), used in the \lonetauOS and \lonetauSS SRs, or $\Mtt$ (see  Figure~\ref{fig:shape_mtautau}), used in the \ltwotau SR.

In the \llOStau channel, events are required to have two OS light leptons satisfying the Tight selection criteria, and at least one Loose or Medium $\tauh$ candidate.
A total of six event categories are defined, which are summarised in Table~\ref{tab:2leptau}. Separate SRs and VRs are defined for events with one Medium $\tauh$ candidate (and no additional Loose $\tauh$ candidates) and at least two Loose $\tauh$ candidates. Backgrounds with resonant $\ell^+\ell^-$ pairs from quarkonia or $Z$-boson decays are suppressed by requiring that the dilepton invariant mass ($m_{\ell\ell}$) satisfies $m_{\ell\ell}>12~\gev$ and $|m_{\ell\ell}-m_Z|>10~\gev$, respectively, where $m_Z$ represents the mass of the $Z$ boson. The latter requirement is referred to as the `$Z$-veto'. The event selections are further optimised based on the $\pt$ of the leading $\tauh$ candidate ($p_{{\mathrm T},1}^{\tau}$) and the minimum invariant mass of a light lepton and a $\tauh$ candidate ($m_{\ell\tau}^{\mathrm{min}}$) (see Figure~\ref{fig:shape_mltaumin}). In addition, two dedicated CRs are defined for events with one Loose or Medium $\tauh$ candidate in order to estimate correction factors to apply to the jet misidentification (also referred to as `fake') rate in the simulation for both sets of $\tauh$ identification criteria. These CRs are enriched in $Z$+jets and dileptonic $\ttbar$ events, respectively, and do not take part of the final likelihood fit. Further details of the fake-$\tauh$ background estimation can be found in Section~\ref{sec:bkg_reducible_tau}.

\begin{table*}[t!]
\centering
\caption{\label{tab:1leptau}
Summary of event categories in the \ltau channel.
All events are required to satisfy the preselection requirements.
``T'' denotes the Tight light-lepton selection criteria (see Table~\ref{tbl:tightleps}).
The $\pt$ of the leading and subleading $\tauh$ candidates are denoted by $p_{{\mathrm T},1}^{\tau}$ and $p_{{\mathrm T},2}^{\tau}$, respectively.
The transverse mass of the system formed by the selected light lepton and the missing transverse momentum is denoted by $m_{\mathrm T}(\ell, \met)$.}
\vspace*{0.25cm}
\small
\begin{tabular}{l|ccc|ccc|ccc}
\hline\hline
& \multicolumn{3}{c|}{\lonetauOS} & \multicolumn{3}{c|}{\lonetauSS} & \multicolumn{3}{c}{\ltwotau} \\
&
CR & VR & SR &
CR & VR & SR &
CR & VR & SR \\
\hline\hline
${e/\mu}$ selection & \multicolumn{3}{c|}{T} & \multicolumn{3}{c|}{T} & \multicolumn{3}{c}{T} \\
$N_{\tauh}$ & \multicolumn{3}{c|}{1} & \multicolumn{3}{c|}{1} & \multicolumn{3}{c}{$\geq2$} \\
$N_{\mathrm{jets}}$ & \multicolumn{3}{c|}{$\geq4$} & \multicolumn{3}{c|}{$\geq3$} & \multicolumn{3}{c}{$\geq2$} \\
$\tauh$ ID & \multicolumn{3}{c|}{Medium} & \multicolumn{3}{c|}{Medium} & \multicolumn{3}{c}{Loose} \\
$\ell\tauh$ charge & \multicolumn{3}{c|}{OS} & \multicolumn{3}{c|}{SS} & \multicolumn{3}{c}{---} \\
$p_{{\mathrm T},1}^{\tau}$ [\gev]& ~~$\geq 50$ & 50--150 & $\geq 150$ & ~~$\geq 50$ & 50--150 & $\geq 150$ & ~~$\geq 50$ & 50--100 & $\geq 100$ \\
$p_{{\mathrm T},2}^{\tau}$ [\gev] & \multicolumn{3}{c|}{---} & \multicolumn{3}{c|}{---} & ~~$\geq 25$ & 25--50~~ & ~~$\geq 50 $ \\
$N_{b\textrm{-jets}}$ & \multicolumn{1}{c}{~~~~$\geq2$} & \multicolumn{2}{c|}{~~~~$\geq1$} & \multicolumn{1}{c}{~~~~$\geq2$} & \multicolumn{2}{c|}{~~~~$\geq1$} & \multicolumn{1}{c}{~~~~$\geq2$} & \multicolumn{2}{c}{~~~~$\geq1$} \\
$m_{\ell\tau}$ [\gev] & --- & \multicolumn{2}{c|}{ $\geq200$ } & --- & --- & $\geq200$  &  \multicolumn{3}{c}{---} \\
$m_{\mathrm T}(\ell, \met)$ [\gev] & \multicolumn{3}{c|}{---} & \multicolumn{1}{c}{---} & --- & $\geq100$ & \multicolumn{3}{c}{---} \\
$\Mtt$ [\gev] & \multicolumn{3}{c|}{---} & \multicolumn{3}{c|}{---} & \multicolumn{1}{c}{---} & \multicolumn{2}{c}{$\geq100$} \\
$\met$ [\gev] & --- & \multicolumn{2}{c|}{ ~~$\geq80$ } & --- & --- & ~~$\geq50$ & \multicolumn{3}{c}{---} \\
$\Meff$ [\gev] & $<800$ & \multicolumn{2}{c|}{ $\geq800$ } &  $<800$ & \multicolumn{2}{c|}{ $\geq800$ } & $<800$ & \multicolumn{2}{c}{ $\geq800$ } \\
\hline \hline
\end{tabular}
\end{table*}
 
\begin{table*}[h!]
\centering
\caption{\label{tab:2leptau}
Summary of event categories in the \llOStau channel.
All events are required to satisfy the preselection requirements.
``T'' denotes the Tight light-lepton selection criteria (see Table~\ref{tbl:tightleps}).
}
\vspace*{0.25cm}
\small
\begin{tabular}{l|cccc|cc}
\hline\hline
& \multicolumn{4}{c|}{\llOSonetau} & \multicolumn{2}{c}{\llOStwotau} \\
&
CR$^{Z}$ & CR$^{t\bar{t}}$ & VR & SR & VR & SR \\
\hline\hline
${e/\mu}$ selection & \multicolumn{4}{c|}{T} & \multicolumn{2}{c}{T} \\
${e/\mu}$ combinations & $ee/\mu\mu$ & $e\mu$ & $ee/\mu\mu$ & $ee/\mu\mu/e\mu$ & \multicolumn{2}{c}{$ee/\mu\mu/e\mu$} \\
$Z$ veto & Inverted & Yes & Yes & Yes & \multicolumn{2}{c}{Yes} \\
$m_{\ell\ell}$ [\gev] & \multicolumn{4}{c|}{$>12$~~~~} & \multicolumn{2}{c}{$>12$~~~} \\
$N_{\tauh}$ & \multicolumn{4}{c|}{~~1} & \multicolumn{2}{c}{$\geq2$} \\
$\tauh$ ID & \multicolumn{2}{c}{Loose/Medium} & \multicolumn{2}{c|}{Medium} & \multicolumn{2}{c}{Loose} \\
$p_{{\mathrm T},1}^{\tau}$ [\gev] & $\geq 25$ & $\geq  25$ & 25--150 & $\geq 150$ & $\geq 25$ & ~~$\geq 75$ \\
$m_{\ell\tau}^{\mathrm{min}}$ [\gev] & --- & --- & ~$< 100$  & $\geq 100$ & --- & ~~$\geq 50$ \\
$\Mtt$ [\gev] & \multicolumn{4}{c|}{---} & <100 & $\geq 100$ \\
$\Meff$ [\gev] & --- & $<1000$~~~~ & --- & --- & --- & --- \\
\hline \hline
\end{tabular}
\end{table*}
 
\begin{table*}[h!]
\centering
\caption{\label{tab:multileptau}
Summary of event categories in the \llSSorllltau channel.
All events are required to satisfy the preselection requirements.
``T'' and ``T*'' denote the Tight and Very Tight light-lepton selection criteria (see Table~\ref{tbl:tightleps}).}
\vspace*{0.25cm}
\small
\begin{tabular}{l|ccc}
\hline\hline
& \multicolumn{3}{c}{\llSSorllltau} \\
& VR & SR-L & SR-H\\ \hline\hline
\multirow{2}{*}{${e/\mu}$ selection} &\multicolumn{3}{c}{T* (2$\ell$SS)}\\ & \multicolumn{3}{c}{T*/T (3$\ell$)} \\
$Z$ veto & \multicolumn{3}{c}{Yes} \\
$m_{\ell\ell}$ [\gev] & \multicolumn{3}{c}{$>12$} \\
$N_{\tauh}$ & \multicolumn{3}{c}{~~$\geq 1$} \\ 
$\tauh$ ID & \multicolumn{3}{c}{Loose} \\ 
$p_{{\mathrm T},1}^{\tau}$ [\gev] & 25--125 & 125--225 & $\geq 225$ \\ 
\hline\hline
\end{tabular}
\end{table*}
 
\begin{figure*}[t!]
\centering
\subfloat[]{\includegraphics[width=0.45\textwidth]{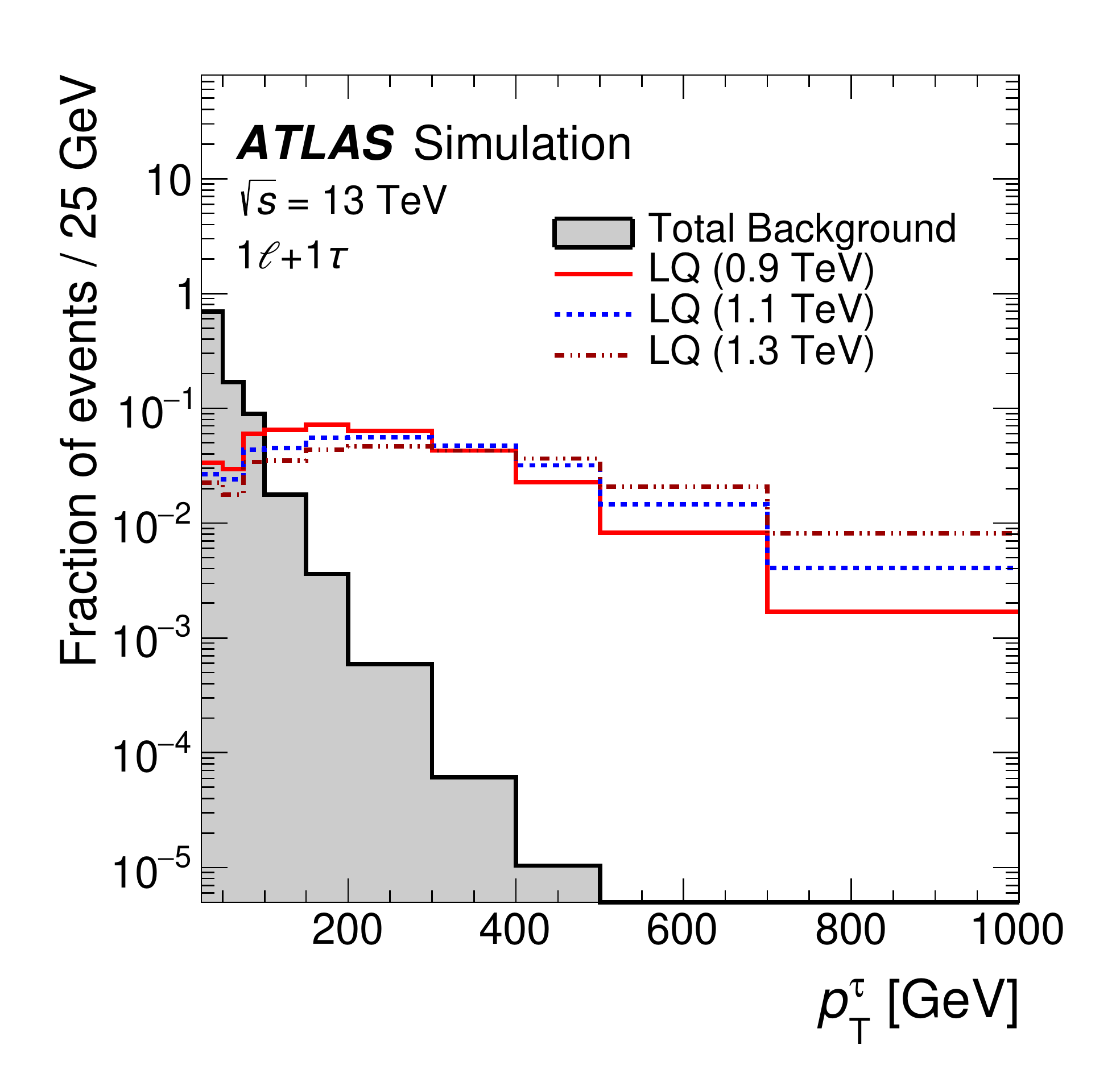}\label{fig:shape_taupt}}
\subfloat[]{\includegraphics[width=0.45\textwidth]{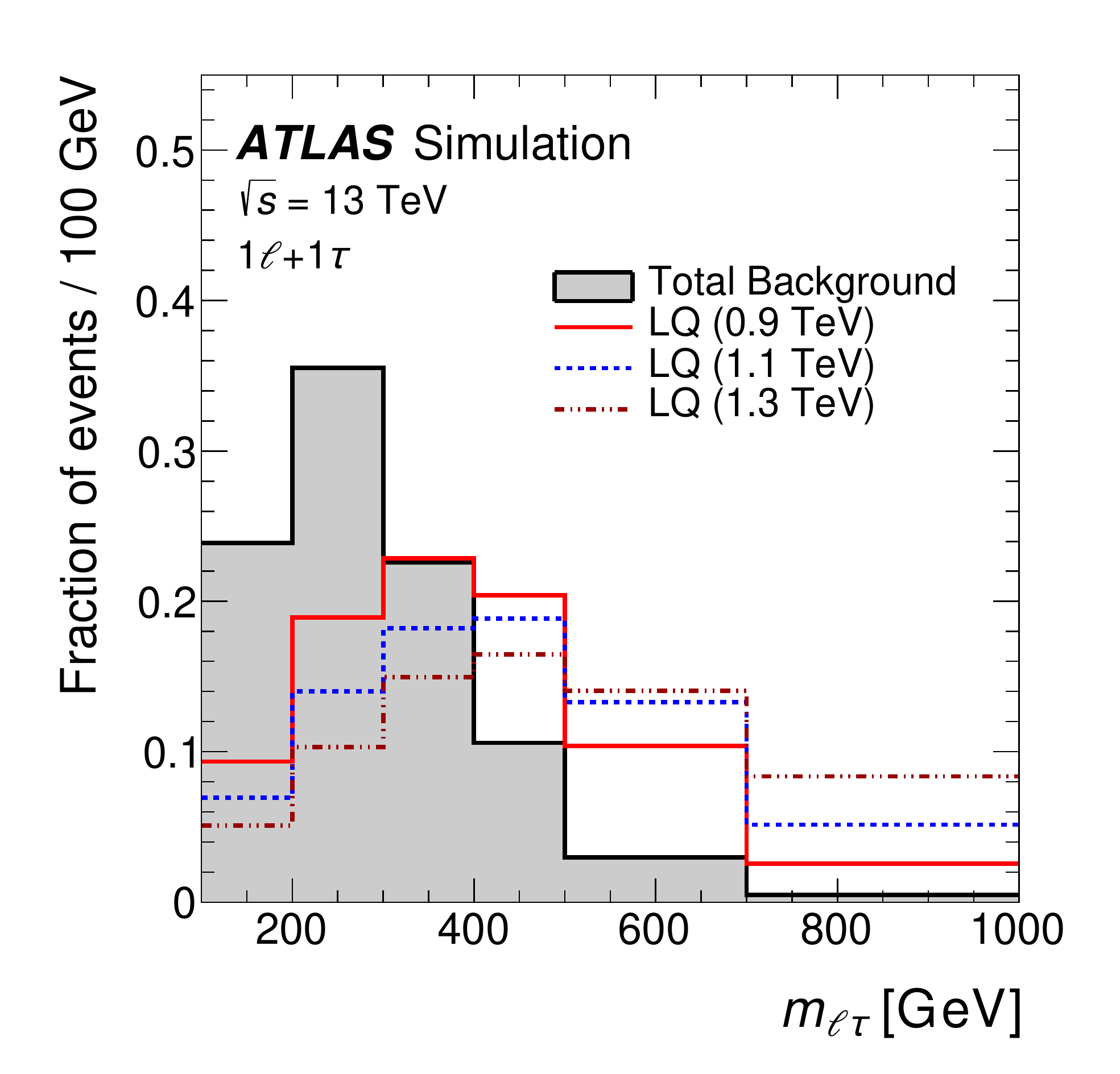}\label{fig:shape_mltau}}
\caption{Comparison of the distribution of (a) the $\pt$ of the $\tauh$ candidate ($p_{\mathrm T}^{\tau}$), and (b) the invariant mass of the light lepton and the $\tauh$ candidate ($m_{\ell\tau}$), between the total background (shaded histogram) and the LQ signal for different mass values.
The selection used corresponds to events in the \lonetau event category (a) after the preselection requirements,
and (b) after applying the additional requirement of $p_{\mathrm T}^{\tau}>150~\gev$.
The last bin in each distribution contains the overflow.
}
\label{fig:shape1}
\end{figure*}
\begin{figure*}[h!]
\centering
\subfloat[]{\includegraphics[width=0.45\textwidth]{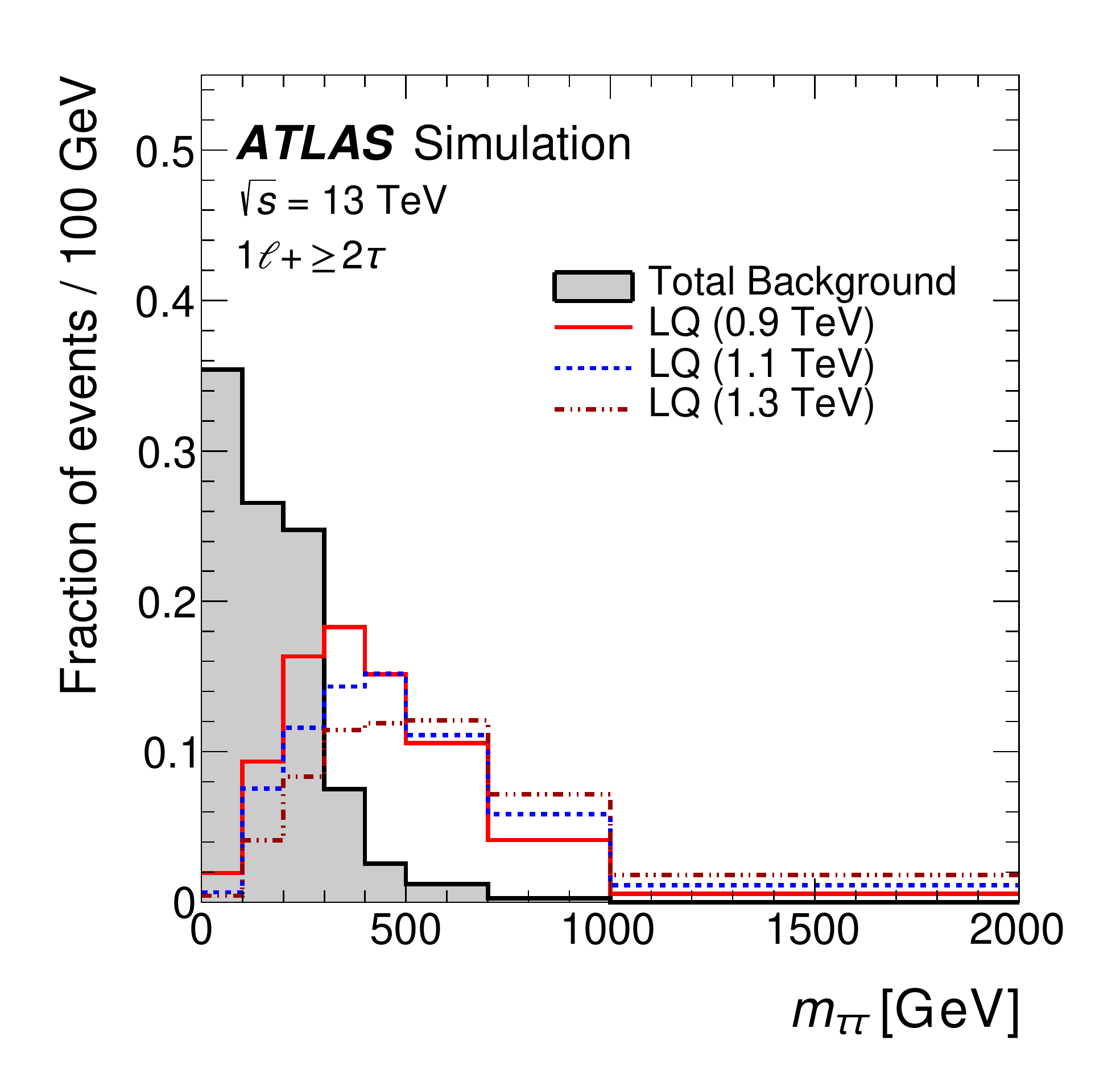}\label{fig:shape_mtautau}}
\subfloat[]{\includegraphics[width=0.45\textwidth]{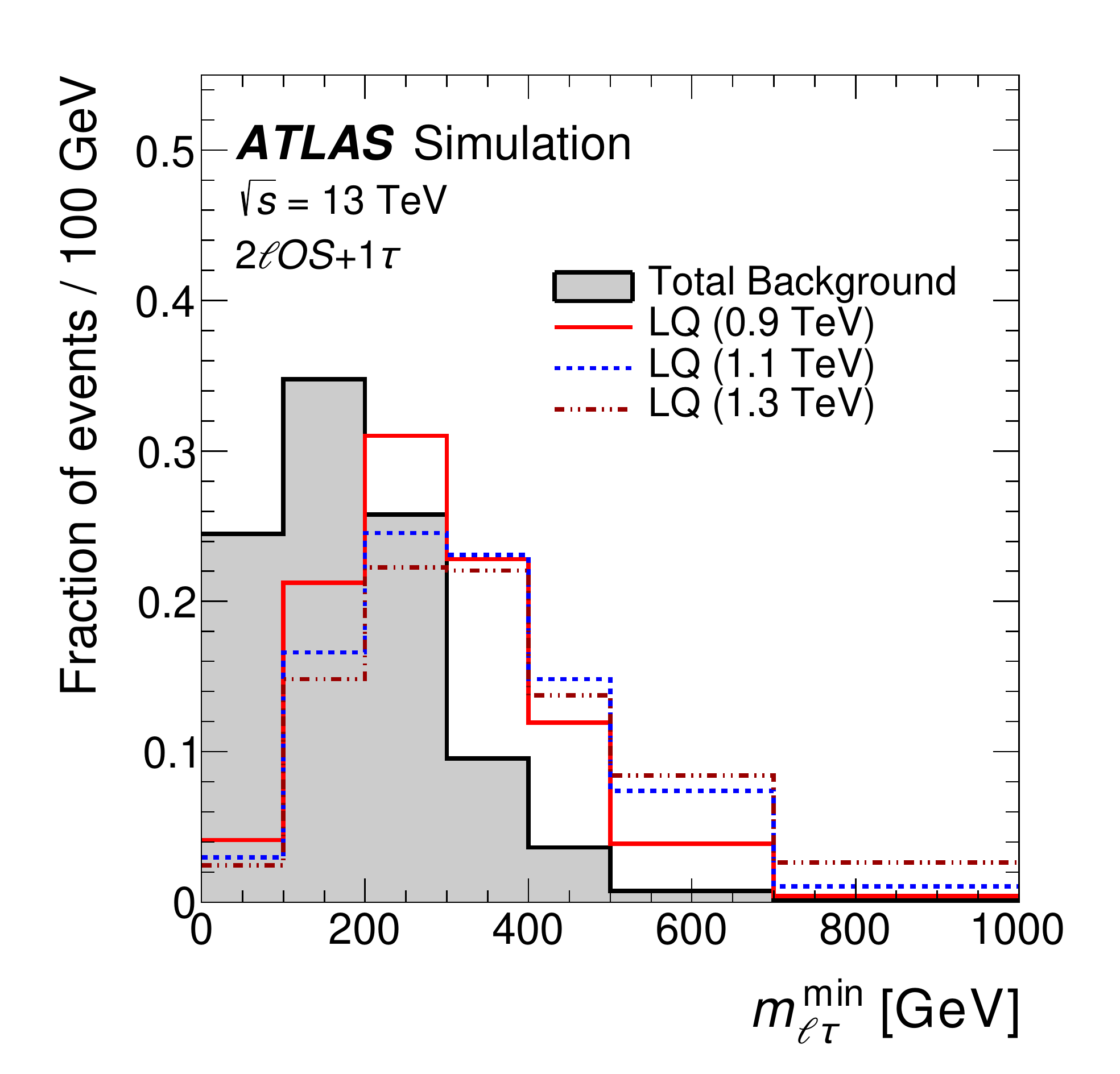}\label{fig:shape_mltaumin}}
\caption{Comparison of the distribution of (a) the invariant mass of the two leading $\tauh$ candidates ($\Mtt$), and (b) the minimum invariant mass of a light lepton and a $\tauh$ candidate ($m_{\ell\tau}^{\mathrm{min}}$), between the total background (shaded histogram) and the LQ signal for different mass values.
The selection used in (a) corresponds to events in the \ltwotau category after the requirements of $p_{{\mathrm T},1}^{\tau}>100~\gev$ and $p_{{\mathrm T},2}^{\tau}>50~\gev$,
whereas the selection used in (b) corresponds to events in the \llOSonetau category after the requirement of $p_{\mathrm T}^{\tau}>150~\gev$.
The last bin in each distribution contains the overflow.
}
\label{fig:shape2}
\end{figure*}
 
In the \llSSorllltau channel, events are required to have either two light leptons with the same charge ($2\ell$SS) or three light leptons ($3\ell$) with their charges adding up to $\pm 1$. In addition, at least one Loose $\tauh$ candidate is required. Since two SS light leptons can originate from backgrounds with non-prompt leptons, photon conversions, and electron charge misassignment (QMisID), the two SS light leptons in the event are required to satisfy the Very Tight selection criteria. In the case of $3\ell$ events, the light lepton that has opposite charge to the SS lepton pair is required to satisfy the Tight selection criteria. In addition, it is required that any $e^\pm e^\pm$, $e^+e^-$ or $\mu^+\mu^-$ pair in the event satisfies $m_{\ell\ell}>12~\gev$ and $|m_{\ell\ell}-m_Z|>10~\gev$. Similarly, $3\ell$ events are required to satisfy $|m_{3\ell}-m_Z|>10~\gev$ to eliminate potential backgrounds with $Z \rightarrow 2\ell \gamma^{*} \rightarrow 4\ell$ where one lepton has very low momentum and is not reconstructed.
Selected events fall into one of three event categories, two SRs and one VR, simply defined using $p_{{\mathrm T},1}^{\tau}$ (see Table~\ref{tab:multileptau}). Events with $p_{{\mathrm T},1}^{\tau}>225~\gev$ are assigned to the main signal region, SR-H (with the symbol ``H'' representing ``High''), which is optimal for high LQ masses, while events with $125 \leq p_{{\mathrm T},1}^{\tau} < 225~\gev$ fall into SR-L (with the symbol ``L'' standing for ``Low'') and extend the sensitivity to lower LQ masses. The VR contains the events with $25 \leq p_{{\mathrm T},1}^{\tau} < 125~\gev$.

Finally, the \llOSnotau, \llSSnotau, and \lllnotau channels require there be no $\tauh$ candidates and are primarily used to improve the background modelling, as discussed in Section~\ref{sec:background}.
Events in the \llOSnotau channel are selected by requiring an OS $e\mu$ pair with both light leptons satisfying the Tight selection criteria and no additional Loose light leptons, at least two jets, at least one $b$-tagged jet, and no Loose $\tauh$ candidates. This selection provides a $\ttbar$-rich control sample (denoted \ttbarnotau CR) that does not take part of the final likelihood fit, but that is used to derive corrections to improve the $\ttbar$ background modelling (see Section~\ref{sec:bkg_irreducible_ttbar}).
Events in the \llSSnotau channel are selected by requiring two SS light leptons satisfying the Very Tight selection criteria, except for some event categories where the internal conversion (IntC) or material conversion (MatC or Mat Conv) vetoes are inverted.
A total of eight event categories, all of which are CRs, are defined so as to be enriched in different backgrounds: $t\bar{t}$ with non-prompt electrons or muons, $t\bar{t}W$, internal conversions, and material conversions, (denoted by \lltte or \llttmu, \llttW, \llintc, and \llmatc, respectively), which are summarised in Table~\ref{tab:CRs2l0tau}. The last two CRs select events with two SS light leptons containing at least one electron that satisfies the corresponding inverted conversion veto requirement. The \lltte and \llttmu CRs select events with a SS $ee$/$\mu e$ pair and a SS $\mu\mu$/$e\mu$ pair, respectively, where the first (second) lepton denotes the leading (subleading) lepton in $\pt$. The definition of these CRs exploits the fact that in SS dilepton events from $t\bar{t}$ production the subleading lepton in $\pt$ is typically a non-prompt lepton.
In addition, the events are restricted to have two or three jets in order to suppress the contribution from $t\bar{t}W$ production.
In the case of the \llttW CR, no restriction is imposed on the light-lepton flavours, and the events are required to have at least four jets. The \lltte, \llttmu, and \llttW CRs are further split according to the charge of the light leptons ($++$ or $--$) in order to improve the discrimination between charge asymmetric and charge symmetric backgrounds (dominated by $t\bar{t}W$ and $t\bar{t}$, respectively).
Events in the \lllnotau channel are selected by requiring three light leptons satisfying the Tight or Very Tight selection criteria, with their charges adding up to $\pm 1$.
A total of four CRs are defined, which are summarised in Table~\ref{tab:CRs3l0tau}. Two CRs select events compatible with having a $Z$-boson candidate, but differing in their jet multiplicity requirements, in order to provide samples enriched in diboson (denoted by \lllVV) and $t\bar{t}Z$ backgrounds (denoted by \lllttZ), respectively. Similarly to the \llSSnotau channel, two additional CRs are defined so as to be enriched in internal- and material-conversion backgrounds, respectively, by inverting the corresponding conversion veto requirement on one of the electrons belonging to the SS lepton pair.

\begin{table*}[t!]
\centering
\caption{\label{tab:CRs2l0tau}
Summary of event categories in the \llSSnotau channel. All events are required to satisfy the preselection requirements.
``T*'' denotes the Very Tight light-lepton selection criteria (see Table~\ref{tbl:tightleps}).
Events that belong to the tt(e), tt($\mu$), and ttW categories are further split into two CRs for $++$ and $--$ charge events.
IntC and MatC stand for internal and material conversions, respectively.
The first (second) light lepton quoted in a pair denotes the leading (subleading) lepton in $\pt$.
Backgrounds with resonant $e^+e^-$ pairs from quarkonia or $Z$-boson decays due to electron charge misassignment are suppressed by requirements on the dielectron invariant mass ($m_{ee}$).}
\vspace*{0.25cm}
\small
\begin{tabular}{l|ccccc}
\hline\hline
& \multicolumn{5}{c}{\llSSnotau} \\
& \llttepm & \llttmupm & \llttWpm & \llintc & \llmatc \\
\hline\hline
${e/\mu}$ selection & \multicolumn{5}{c}{T*} \\
${e/\mu}$ combination & $ee$/$\mu e$ & $\mu\mu$/$e\mu$ & $ee$/$\mu\mu$/$e\mu$/$\mu e$  & $ee$/$e\mu$/$\mu e$ & $ee$/$e\mu$/$\mu e$ \\
Electron internal conversion veto & Yes & Yes & Yes & Inverted & Yes \\
Electron material conversion veto & Yes & Yes & Yes & Yes & Inverted \\
$N_{\mathrm{jets}}$ &2--3 & 2--3 & $\geq4$ &  $\geq2$ & $\geq2$  \\
$Z$ veto & \multicolumn{5}{c}{Yes~~~~} \\
$m_{ee}$ [\gev] & \multicolumn{5}{c}{$\geq12$~~~~~~} \\
\hline\hline
\end{tabular}
\end{table*}
\begin{table*}[t!]
\centering
\caption{\label{tab:CRs3l0tau}
Summary of four CR categories in the \lllnotau channel. All events are required to satisfy the preselection requirements.
``T'' and ``T*'' denote the Tight and Very Tight light-lepton selection criteria (see Table~\ref{tbl:tightleps}).
IntC and MatC stand for internal and material conversions, respectively.
Same-charge (opposite-charge) lepton pairs are also referred to as same-sign (opposite-sign) with abbreviation SS (OS).
The OS lepton (relative to the SS pair) is denoted $\ell_{0}$, but is not necessarily the one with highest $\pt$; the remaining SS leptons are denoted $\ell_{1}$ (closest in $\Delta R$ to $\ell_{0}$) and $\ell_{2}$ (the remaining one). }
\vspace*{0.25cm}
\small
\begin{tabular}{l|cccc}
\hline\hline
& \multicolumn{4}{c}{\lllnotau} \\
& \lllVV & \lllttZ & \lllintc & \lllmatc \\ \hline\hline
${e/\mu}$ selection & T & T & T($\ell_{0}$), T*($\ell_{1}$ and $\ell_{2}$) & T($\ell_{0}$), T*($\ell_{1}$ and $\ell_{2}$) \\
Electron internal conversion veto & Yes & Yes & Inverted($\ell_{1}$ or $\ell_{2}$) & Yes($\ell_{1}$ and $\ell_{2}$) \\
Electron material conversion veto & Yes & Yes & Yes($\ell_{1}$ and $\ell_{2}$) & Inverted($\ell_{1}$ or $\ell_{2}$) \\
$N_{\mathrm{jets}}$ & 2--3 & $\geq4$ & $\geq2$ & $\geq2$ \\
$Z$ veto & Inverted & Inverted & Yes & Yes  \\
$m_{\ell\ell}$ [\gev] & \multicolumn{4}{c}{$\geq 12$~~~~} \\
\hline\hline
\end{tabular}
\end{table*}
 
The $\Meff$ distribution is used as the final discriminating variable in all SRs. It peaks at approximately $2 m_\mathrm{LQ}$ for signal events, and at lower values for the backgrounds, as illustrated in Figure~\ref{fig:shape3}. The overall rate and composition of the background varies across the different SRs, as illustrated in Figure~\ref{fig:pieSR}. The dominant background in the \lonetauOS SR is $\ttbar$ production with both the light lepton and $\tauh$ candidate originating from the $W$ boson decays. In contrast, the main background in the \lonetauSS SR is also $\ttbar$ production, but with one jet misidentified as a $\tauh$ candidate (fake $\tauh$), one non-prompt light lepton, or an electron with misassigned charge, followed by $t\bar{t}W$ and $VV$ production. In the \ltwotau, \llOSonetau, and \llOStwotau SRs, about half of the background is also $\ttbar$ with one fake $\tauh$ candidate, while the remaining contributions arise from $t\bar{t}W$, $t\bar{t}Z/\gamma^*$, and $t\bar{t}H$ production, with varying fractions across the SRs. Finally, the
\llSSorllltau SRs are dominated by backgrounds with real leptons, with comparable contributions from $t\bar{t}W$, $t\bar{t}Z/\gamma^*$, $t\bar{t}H$, and $VV$ production. Despite their limited purity, the CRs defined above are useful for checking and correcting the background prediction (see Section~\ref{sec:background}) and constraining the related systematic uncertainties through the likelihood fit to data that also includes the SRs. The VRs are meant to provide an independent validation of the background prediction, and thus are not included in the fit.
 
\begin{figure*}[t!]
\centering
\subfloat[]{\includegraphics[width=0.45\textwidth]{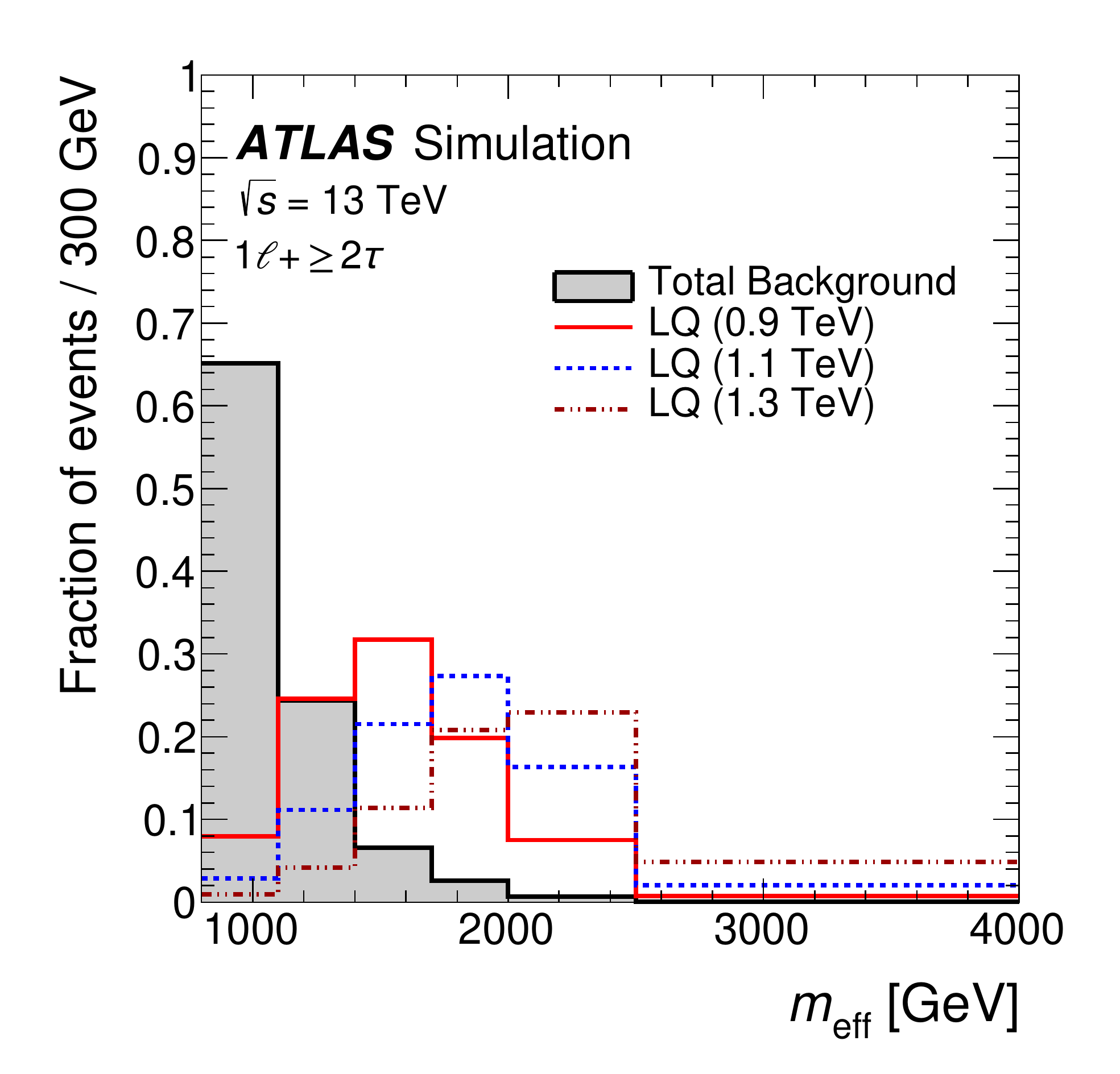}}
\subfloat[]{\includegraphics[width=0.45\textwidth]{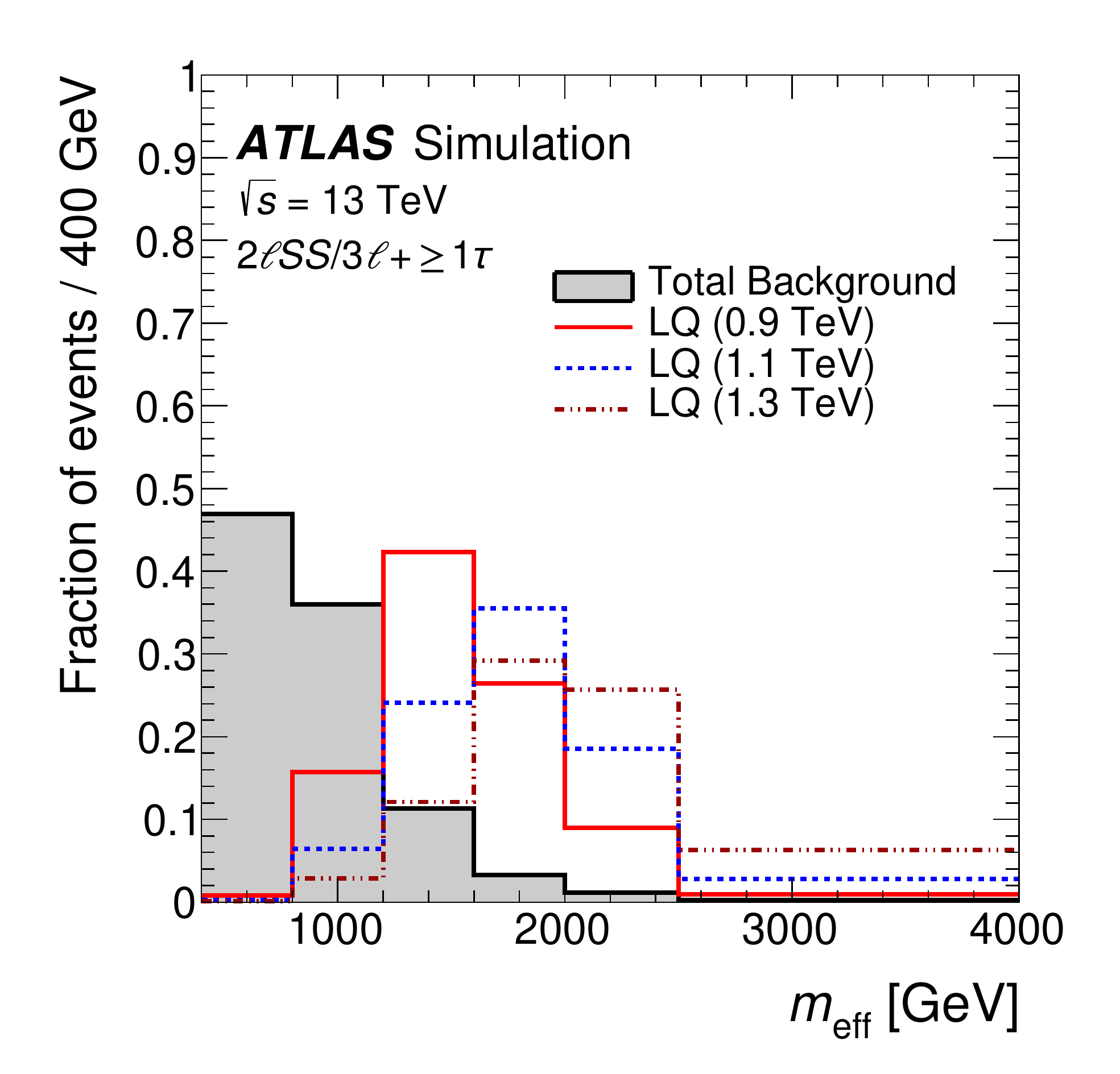}}
\caption{Comparison of the $\Meff$ distribution in (a) the \ltwotau SR, and (b) the \llSSorllltau SR-H,
between the total background (shaded histogram) and the LQ signal for different mass values.
The last bin in each distribution contains the overflow.
}
\label{fig:shape3}
\end{figure*}
 
\begin{figure*}[htbp]
\centering
\includegraphics[width=0.75\textwidth]{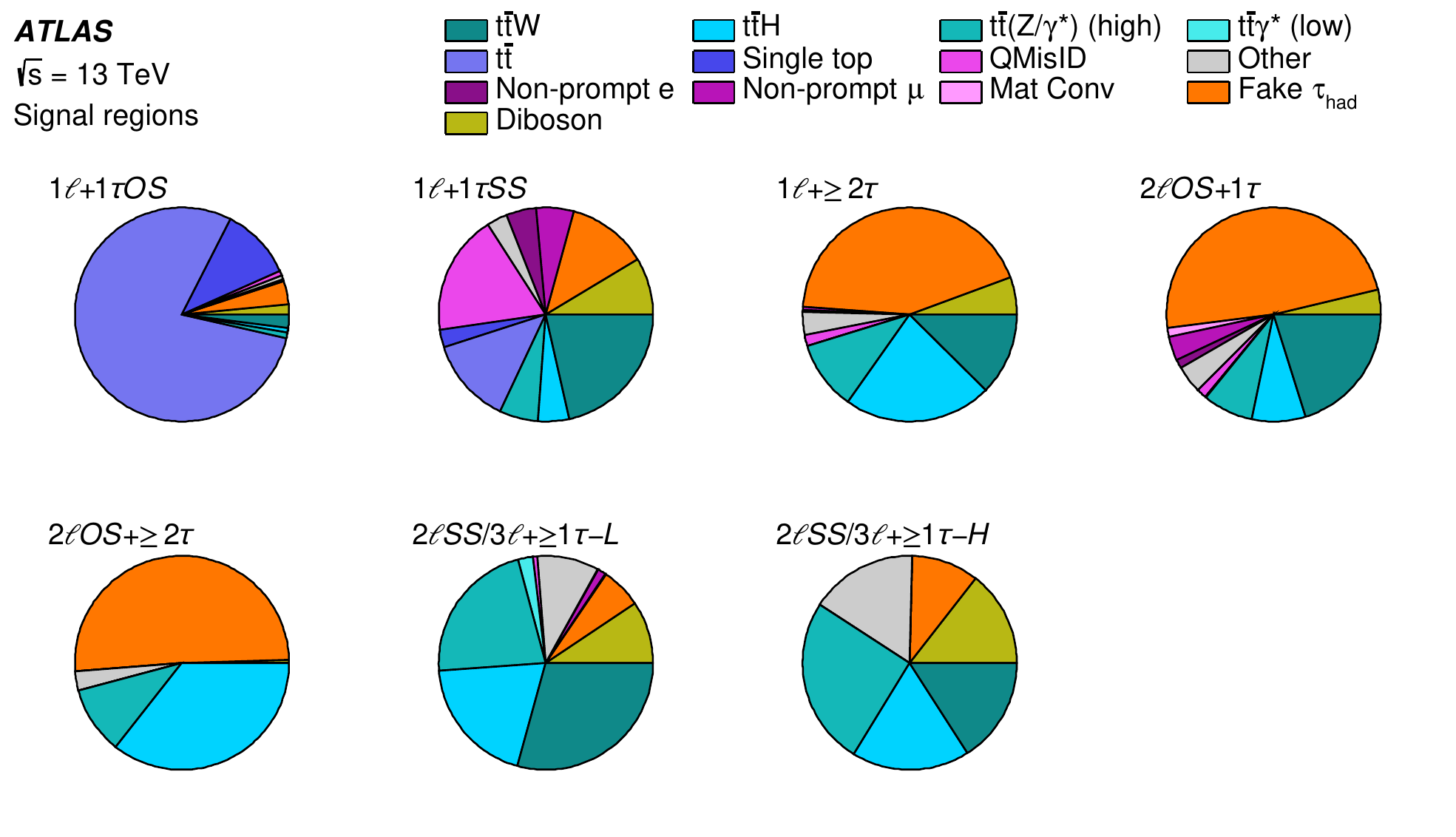}
\caption{The fractional contributions of the various backgrounds to the total predicted background in each of the seven signal region categories (see Section~\ref{sec:events}). The background estimation methods are described in Section~\ref{sec:background}. The background contributions after the likelihood fit to data under the background-only hypothesis are shown (see Section~\ref{sec:results}). }
\label{fig:pieSR}
\end{figure*}

 
\section{Background estimation}
\label{sec:background}
Backgrounds are categorised into irreducible and reducible backgrounds.
Irreducible backgrounds  (Section~\ref{sec:bkg_irreducible}) have only prompt selected leptons, i.e.\ produced in $W/Z$ boson decays, in leptonic $\tau$-lepton decays, or internal conversions. Reducible backgrounds  (Section~\ref{sec:bkg_reducible})  have prompt leptons with misassigned charge, at least one non-prompt light lepton, or fake $\tauh$ candidates.
All backgrounds are estimated using the simulated samples described in Section~\ref{sec:datamc}, which also discusses the systematic uncertainties in the modelling of these processes, so this is not repeated below. In some cases, the simulation is improved using additional corrections derived in data control samples. In particular, the event kinematics of the simulated $\ttbar$ background, or the $\tauh$ fake rate predicted by the simulation, require dedicated corrections to better describe the data. In addition, the yields of some simulated backgrounds, in particular \ttW and non-prompt-lepton backgrounds, are adjusted via normalisation factors that are determined by performing a likelihood fit to data across all event categories as discussed in Section~\ref{sec:results}.
 
\subsection{Irreducible backgrounds}
\label{sec:bkg_irreducible}
 
Background contributions with prompt leptons originate from a wide range of physics processes with their relative
importance varying by channel. In the \lonetauOS category the main irreducible background is $\ttbar$ production, followed by $\tW$ production, whereas in the rest of analysis channels the main irreducible backgrounds originate from $\ttW$ and $\ttdy$ production, followed by $\VV$ (in particular $WZ$) production. Smaller contributions originate from the following rare processes: $\tZ$, $\tWZ$, $\ttWW$, $\VVV$, $\ttt$, and $\tttt$ production.
 
\subsubsection{$\ttbar$ background}
\label{sec:bkg_irreducible_ttbar}
Detailed measurements of differential cross sections have shown that the $\ttbar$ simulation does not model the top-quark $\pT$ spectrum with sufficient accuracy, overestimating it in the high-$\pT$ tail~\cite{Aad:2019hzw,Aad:2019ntk}. In addition, the simulation underestimates the production of $\ttbar$ events with high jet multiplicity~\cite{Aad:2019ntk}. This leads to discrepancies between data and simulation in the distributions of several kinematic quantities of interest in this search, in particular the $\Meff$ variable. In order to improve the description, dedicated corrections as a function of jet multiplicity and $\Meff$ (referred to as `kinematic reweighting') are derived in the \ttbarnotau CR.
The corrections are derived by comparing the data, after subtracting small background contributions estimated from the simulation, with the predicted sum of $\ttbar$ and $\tW$ processes.\footnote{The sum of $\ttbar$ and $tW$ backgrounds is considered since they interfere at NLO.} The correction factors as a function of jet multiplicity vary from 1.05 for exactly two jets, to 1.1 for at least six jets. After correcting the jet multiplicity spectrum, a further correction as a function of $\Meff$ is derived for each jet multiplicity, and parameterised as a first-degree polynomial. For example, for exactly four jets, the resulting correction factor varies from $\sim$1.1 for $\Meff=200~\gev$ to $\sim$0.4 for $\Meff=3~\tev$, as shown in Figure~\ref{fig:meffrw_derivation}. This kinematic reweighting is applied to all (nominal and alternative) $\ttbar$ and $\tW$ simulated events, and prior to the derivation of any further corrections to improve the modelling of fake $\tauh$ candidates or non-prompt leptons (see Sections~\ref{sec:bkg_reducible_tau} and~\ref{sec:bkg_reducible_light}).
The comparison between the data and the background prediction after the kinematic reweighting, and before the likelihood fit (denoted ``pre-fit''), is illustrated in Figure~\ref{fig:meffrw_validation} for the $\Meff$ distribution in the \lonetauOS VR, which is dominated by $\ttbar$ background with a real $\tauh$ candidate. Good agreement is observed within the estimated pre-fit uncertainties. The agreement in normalisation is further improved after the likelihood fit (denoted ``post-fit"), as shown in Figure~\ref{fig:summaryVR}.
The modelling of several other kinematic quantifies such as the lepton $\pt$, $\met$, and the scalar sum of jet $\pt$, is also improved by the kinematic reweighting.
Although this kinematic reweighting is derived using $\ttbar$ dileptonic events, it is also applied to $\ttbar$ semileptonic events selected in the \lonetau channel. A systematic uncertainty from the slight difference between the slope of the nominal $\Meff$ correction factor and that derived in the \lonetauOS CR is also considered, with negligible impact on the final result.
 
\begin{figure}[t]
\centering
\subfloat[]{\includegraphics[width=0.40\textwidth]{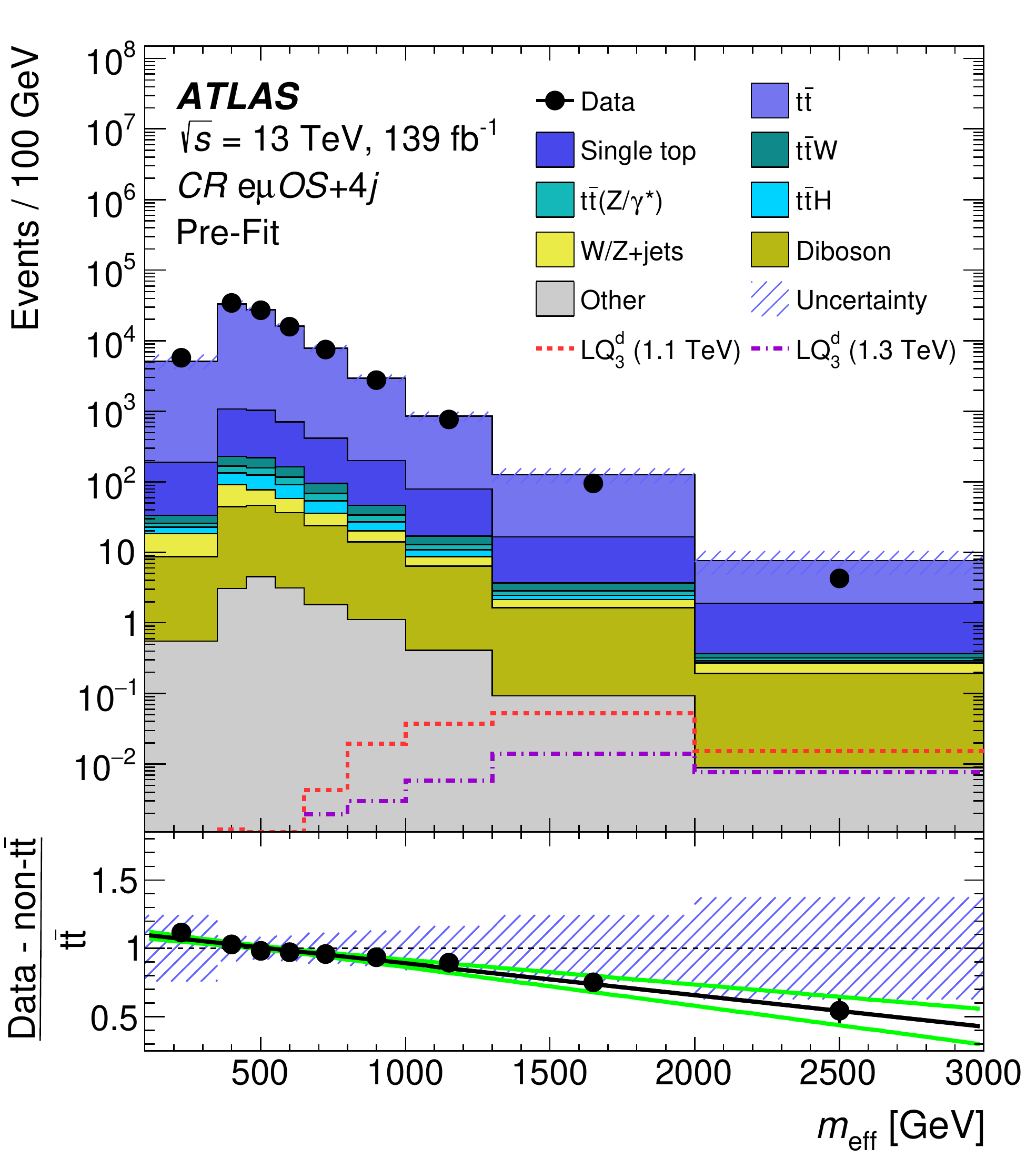}\label{fig:meffrw_derivation}}
\subfloat[]{\includegraphics[width=0.40\textwidth]{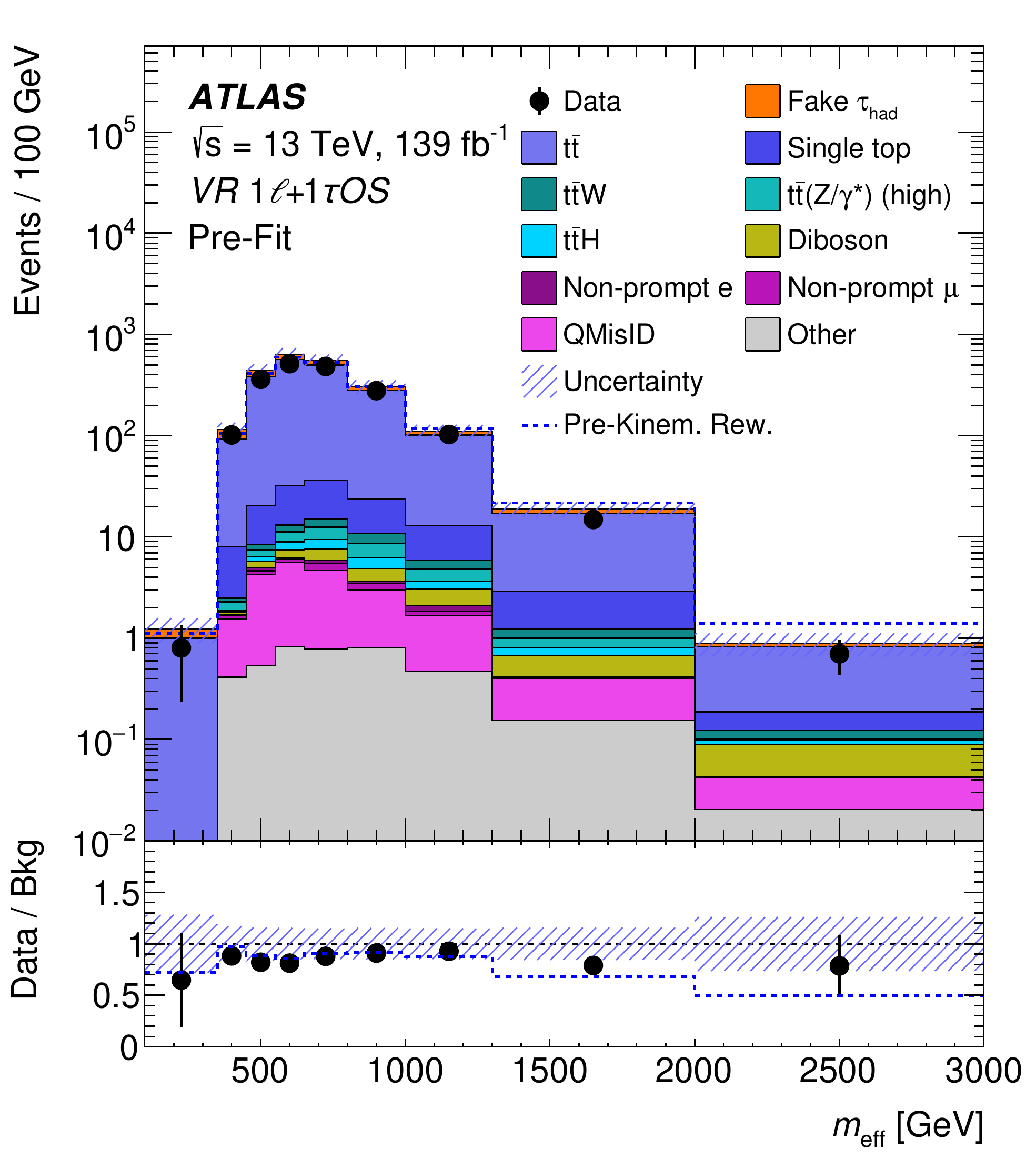}\label{fig:meffrw_validation}}
\caption{
(a) Comparison between data and the background prediction for the $\Meff$ distribution in events selected by requiring an opposite-charge (OS) $e\mu$ pair, exactly four jets, and at least one $b$-tagged jet. The background contributions shown are before the likelihood fit to data (``Pre-Fit'').
The lower panel displays the ratio of the data, after subtracting the small background contributions estimated from the simulation, to the predicted sum of $\ttbar$ and $\tW$ processes, along with the corresponding fit using a first-degree polynomial (black solid line). The associated green lines represent the estimated uncertainty in the reweighting function.
(b) Comparison of the $\Meff$ distribution between data and the pre-fit background prediction after the kinematic reweighting in the \lonetauOS VR.
The total background prediction before the kinematic reweighting (``Pre-Kinem.\ Rew.'') is shown as a dashed blue histogram.
The ratio of the data to the total pre-fit background prediction (``Bkg'') is shown in the lower panel.
The size of the combined statistical and systematic uncertainty in the background prediction is indicated by the blue hatched band.
The ratios of the data to the total pre-fit predictions before and after kinematic reweighting are shown in the lower panel.
The last bin in each figure contains the overflow.
}
\end{figure}

\subsubsection{\ttW background}
The \ttW background represents a non-negligible background in several event categories. Despite the use of state-of-the-art simulations, accurate modelling of additional QCD radiation in \ttW production remains challenging. Event categories sensitive to the \ttW background were defined in the analysis in order to study and constrain this background. These event categories are split by the sign of the sum of lepton charges (referred to as `total charge') to better discriminate the \ttW process, which has a large charge asymmetry, from other SM backgrounds that are charge symmetric. To illustrate this point, the distribution of the scalar sum of the lepton $\pt$ (denoted by $H_\textrm{T, lep}$) in the \llSSnotau channel, obtained by subtracting the distributions for events with positive total charge and with negative total charge, is shown in Figure~\ref{fig:ttW_htlep}.
In this subtraction, only the charge asymmetric processes remain visible, allowing a better assessment of the modelling of the \ttW process by the simulation. Disagreement between the data and the prefit prediction from the simulation is observed, corresponding to an overall normalisation factor that is assigned to the \ttW background, and which is determined during the likelihood fit. The measured normalisation factor is $\hat{\lambda}_{\ttW} = 1.78 \pm 0.15$, which is compatible with that determined in the SM $t\bar{t}t\bar{t}$ analysis~\cite{Aad:2020klt}, and with a previous measurement of the \ttW production cross section~\cite{Aaboud:2019njj}.
Agreement is improved after the application of the background corrections resulting from the likelihood fit, in particular the above \ttW normalisation factor, as shown in Figure~\ref{fig:ttW_meff} for the $\Meff$ distribution.

\begin{figure}[t]
\centering
\subfloat[]{\includegraphics[width=0.40\textwidth]{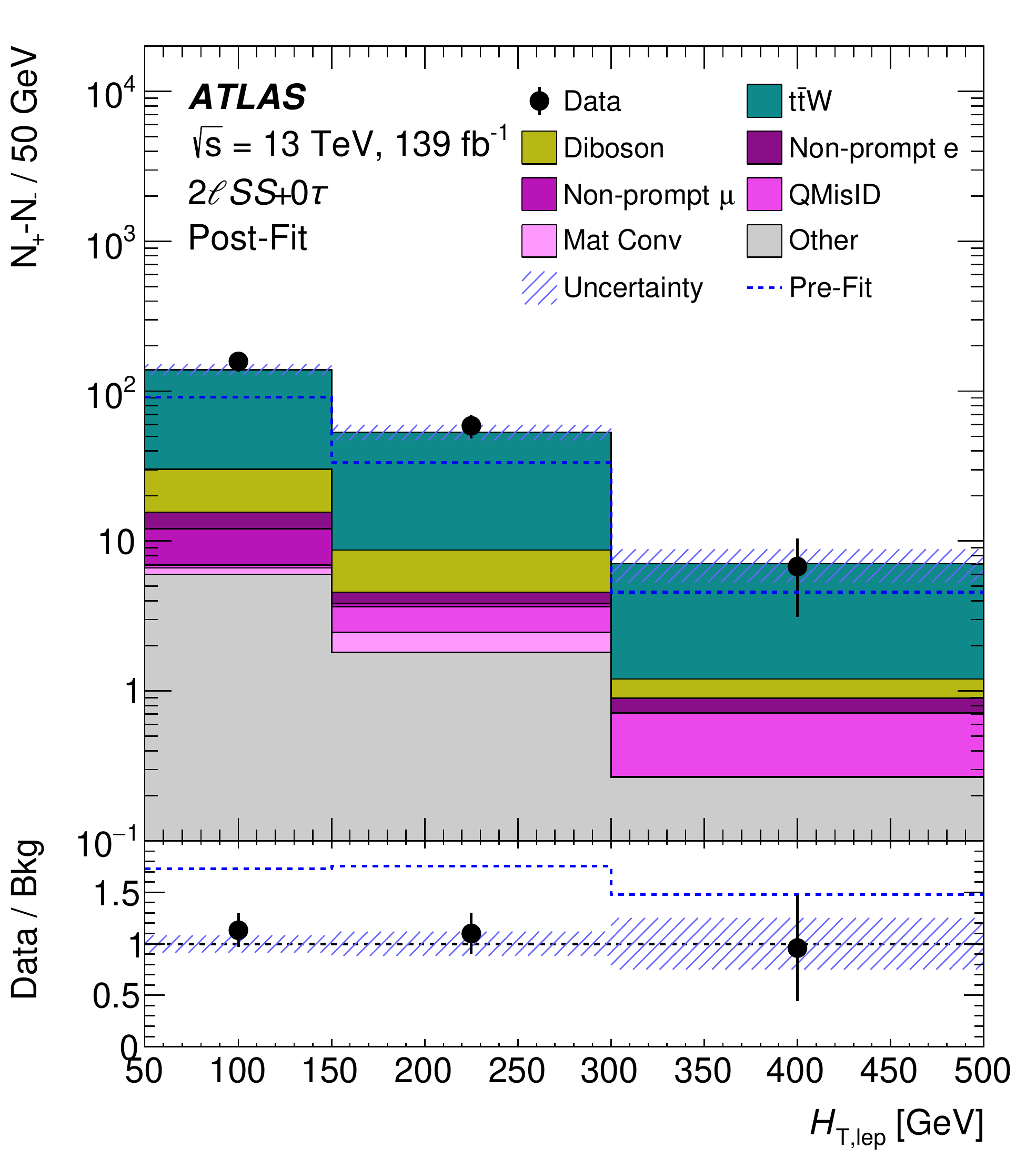}\label{fig:ttW_htlep}}
\subfloat[]{\includegraphics[width=0.40\textwidth]{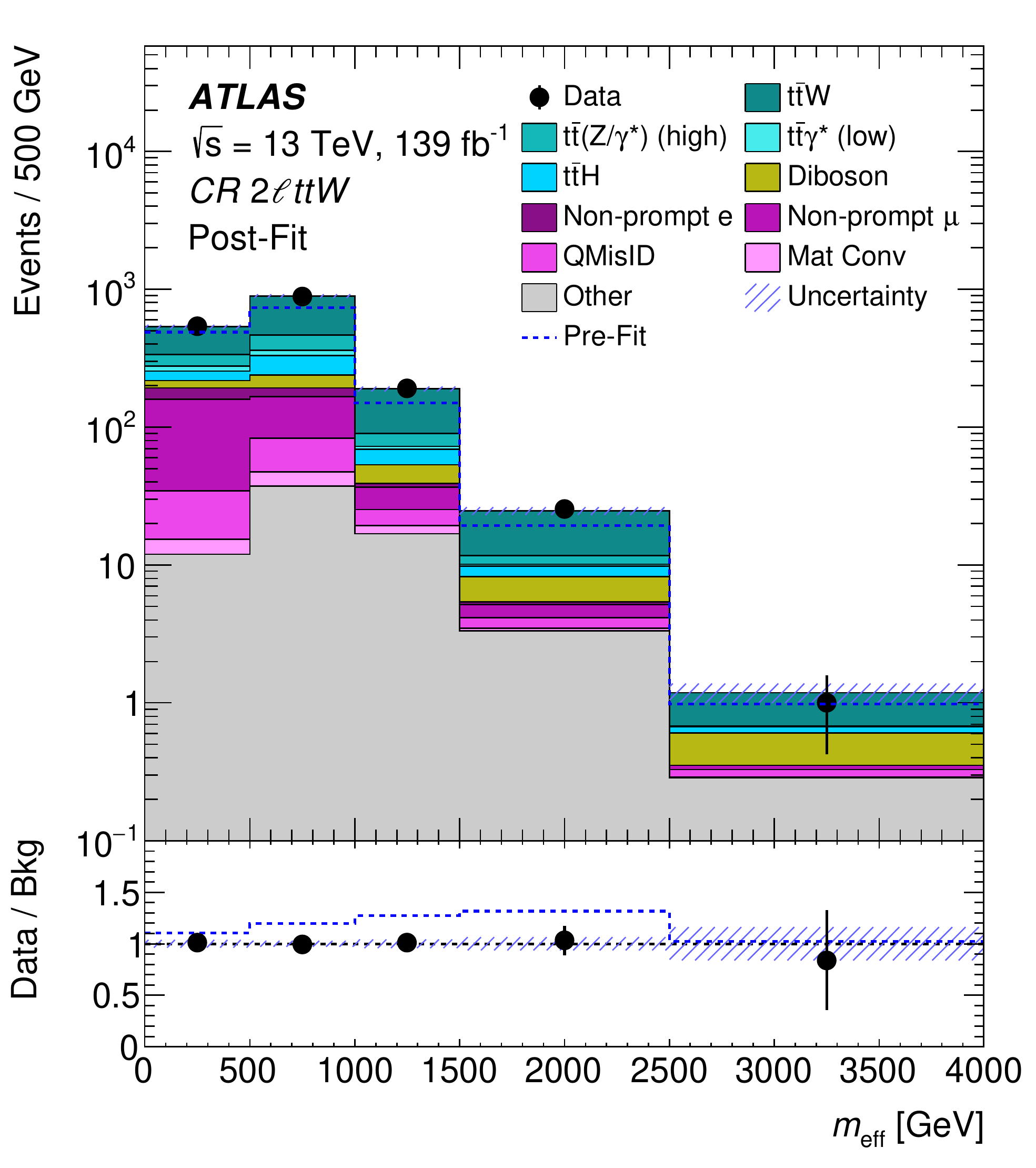}\label{fig:ttW_meff}}
\caption{(a) Comparison between data and the background prediction for the distribution of the scalar sum of the lepton $\pt$ ($H_\textrm{T, lep}$) in the \llSSnotau channel, obtained by subtracting the corresponding distributions for events with positive and negative total charge. In (b) the comparison is performed for the $\Meff$ distribution in the \llttW CR without splitting according to total charge.
The background contributions after the likelihood fit to data (``Post-Fit'') under the background-only hypothesis are shown as filled histograms.
The total background prediction before the likelihood fit to data (``Pre-Fit'') is shown as a dashed blue histogram in the upper panel.
The ratio of the data to the background (``Bkg'') prediction is shown in the lower panel, separately for post-fit background (black points) and pre-fit background (dashed blue line).
The size of the combined statistical and systematic uncertainty in the background prediction is indicated by the blue hatched band.
The last bin in each figure contains the overflow.}
\end{figure}
 
\subsubsection{Other irreducible backgrounds}
\label{sec:other_irreducible_backgrounds}
The total yields in the \lllVV and \lllttZ CRs are used in the likelihood fit to improve the prediction of the background contribution from the $\VV$ and $\ttdy$ processes, respectively.
A comparison of the $\Meff$ distribution between the data and the total prediction in these two CRs exhibits adequate modelling by the simulation even before the likelihood fit to data, as shown in Figure~\ref{fig:VV_ttZ_meff}. The rate of the background from internal conversions with $m(e^+e^-)<1~\gev$ is estimated using the two dedicated CRs (\llintc and \lllintc). The total yield in each category is used in the likelihood fit to determine the following normalisation factor: $\hat{\lambda}^\textrm{IntC}_e = 1.77 \pm 0.32$, where the uncertainty is dominated by the statistical uncertainty.  The normalisation of the internal-conversion background is validated by comparing data and scaled simulation
in a dedicated control sample enhanced in $Z \to \mu^+\mu^- \gamma^* (\to e^+e^-)$ candidate events, defined by requiring two OS Tight muons and one electron satisfying the Very Tight requirements, except for the internal conversion veto.
The level of agreement found between observed and predicted yields is within 25\%, which is assigned as a systematic uncertainty associated with the extrapolation of the estimate from the \llintc and \lllintc CRs to the other event categories.
 
\begin{figure}[t]
\centering
\subfloat[]{\includegraphics[width=0.40\textwidth]{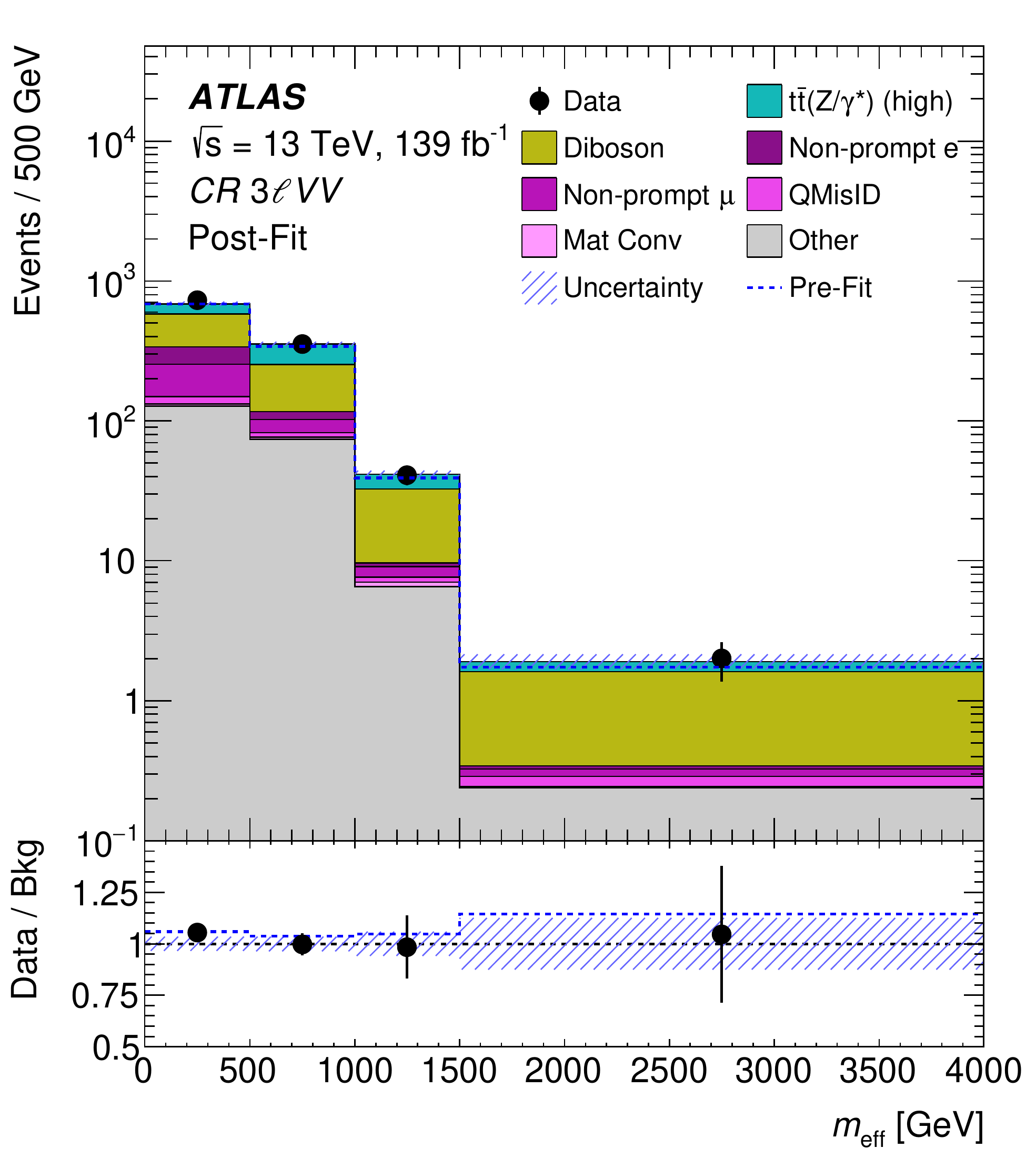}\label{fig:VV_meff}}
\subfloat[]{\includegraphics[width=0.40\textwidth]{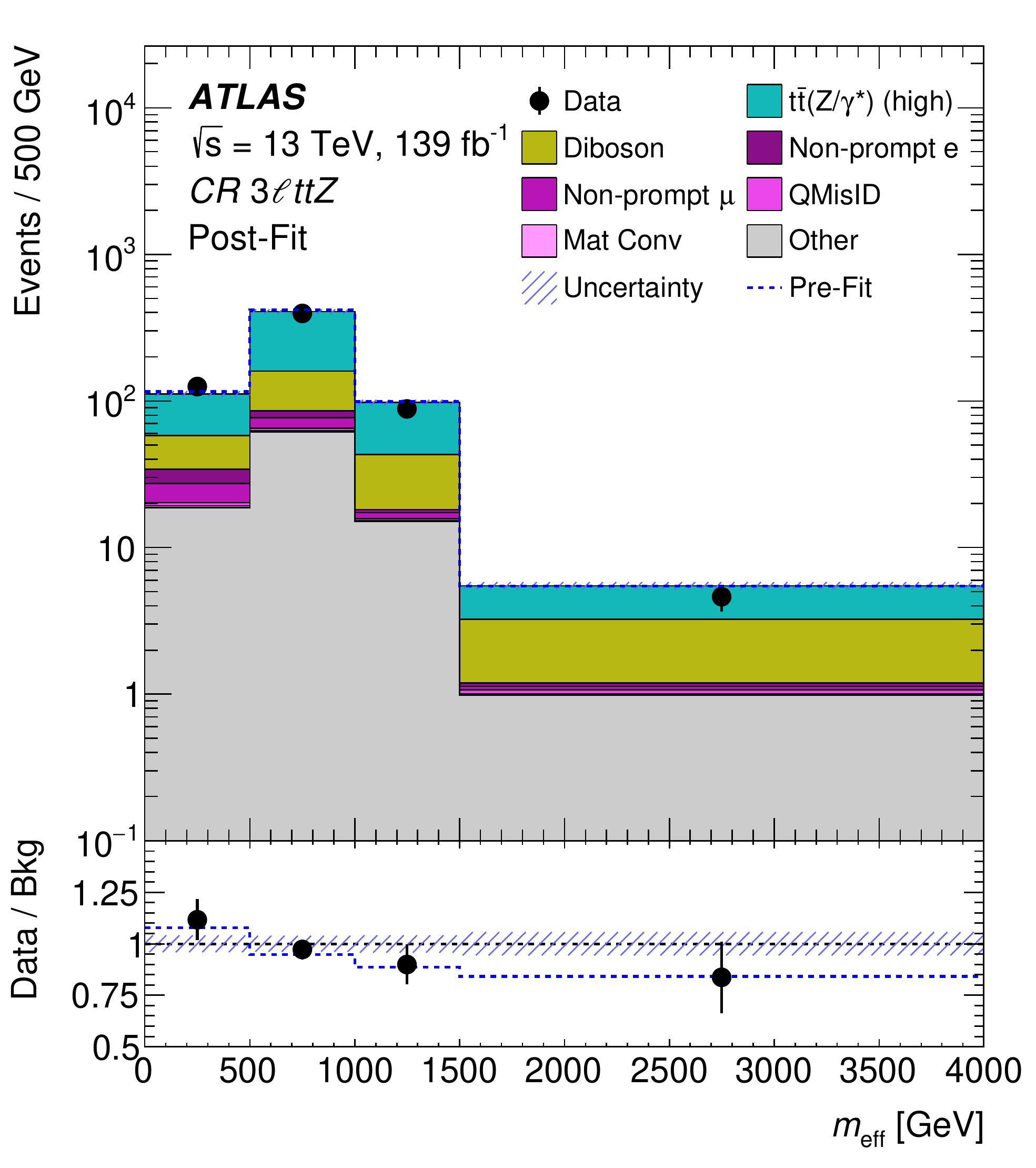}\label{fig:ttZ_meff}}
\caption{Comparison between data and the background prediction for the $\Meff$ distribution in (a) the \lllVV CR and (b) the \lllttZ CR.
The background contributions after the likelihood fit to data (``Post-Fit'') under the background-only hypothesis are shown as filled histograms.
The total background prediction before the likelihood fit to data (``Pre-Fit'') is shown as a dashed blue histogram in the upper panel.
The ratio of the data to the background (``Bkg'') prediction is shown in the lower panel, separately for post-fit background (black points) and pre-fit background (dashed blue line).
The size of the combined statistical and systematic uncertainty in the background prediction is indicated by the blue hatched band.
The last bin in each figure contains the overflow.}
\label{fig:VV_ttZ_meff}
\end{figure}
 
\subsection{Reducible backgrounds}
\label{sec:bkg_reducible}
 
\subsubsection{Fake $\tauh$ candidates}
\label{sec:bkg_reducible_tau}
In most event categories requiring at least one $\tauh$, the dominant background originates from $\ttbar$ production with at least one fake $\tauh$ candidate. Consequently, the estimation of fake-$\tauh$ background relies heavily on the simulation accurately modelling the $\ttbar$ event kinematics and the $\tauh$ misidentification rate from jets. As discussed in Section~\ref{sec:bkg_irreducible_ttbar}, a kinematic reweighting is applied to $\ttbar$ simulated events in order to improve the description of the event kinematics. In order to evaluate such a correction factor, which depends on the jet multiplicity of the events, fake $\tauh$ candidates in $\ttbar$ simulated events are considered as additional jets. After the kinematic reweighting is applied, a suitable correction to the fake-$\tauh$ rate in the simulation is measured. A CR is defined by requiring an OS $e\mu$ pair, at least two jets, at least one $b$-tagged jet, at least one Loose or Medium $\tauh$ candidate, and $\Meff <1~\tev$ (denoted by CR$^{t\bar{t}}$ in Table~\ref{tab:2leptau}). The upper bound on $\Meff$ ensures that any potential $\LQ$ signal contamination would be negligible.
This CR is enriched in dileptonic $\ttbar$ events, such that the selected $\tauh$ candidates primarily originate from jets, and are used to determine a normalisation factor to correct for possible mismodelling of the fake-$\tauh$ rate in the simulation per $\tauh$ candidate. According to the simulation, the flavour composition of the jets giving a fake $\tauh$ candidate in this CR is similar to that in the SRs considered.
This normalisation factor is measured as a function of $p_\mathrm{T}^{\tauh}$, and for one-prong and three-prong $\tauh$ candidates separately.  In the case of one-prong (three-prong) $\tauh$ candidates satisfying the Loose requirement, the normalisation factors range from $1.07 \pm 0.06$ ($1.10 \pm 0.31$) for $p_\mathrm{T}^{\tauh}$ in the range of 25--45~\gev\ (25--50~\gev),
to $0.57 \pm 0.19$ ($0.80 \pm 0.30$) for $p_\mathrm{T}^{\tauh} \geq 100~\gev$ $(75~\gev)$.
The quoted uncertainty includes the statistical uncertainty in the CR, the uncertainty in the contribution from real $\tauh$ candidates that is subtracted in the CR, and the difference between this normalisation factor and one measured in a separate CR enhanced in $Z$+jets events (denoted by CR$^Z$ in Table~\ref{tab:2leptau}), which has a different jet-flavour composition of fake $\tauh$ candidates than CR$^{t\bar{t}}$.
No statistically significant differences are found between the normalisation factors for Loose and Medium $\tauh$ candidates; therefore, the above normalisation factors are applied to all channels requiring at least one $\tauh$ candidate.
All simulated background events with at least one fake $\tauh$ candidate\footnote{This includes $t\bar{t}$ as well as other subleading processes such as single top, $V$+jets, etc.}
are scaled by the product of the corresponding per-candidate normalisation factors calculated according to the multiplicity of fake $\tauh$ and non-prompt light leptons (see Section~\ref{sec:bkg_reducible_light}) before the likelihood fit to data.
After applying the kinematic reweighting and the $\pt$-dependent fake-$\tauh$ normalisation factors discussed above, the simulation is found to provide good modelling of relevant kinematic distributions for the fake-$\tauh$ background before the likelihood fit to data, as shown in Figures~\ref{fig:CR1l1tauSS_taupt} and~\ref{fig:VR2lOS1tau_njets}.
The uncertainties associated with the normalisation factors are accounted for as nuisance parameters in the likelihood fit (see Section~\ref{sec:results}).
To account for the approximation of treating fake $\tauh$ candidates as jets in the $\ttbar$ kinematic reweighting, in the statistical analysis, the uncertainties associated with the PS and hadronisation model, the ME-to-PS matching, and the modelling of QCD radiation, are treated as uncorrelated between $\ttbar$ events with at least one fake $\tauh$ candidate and the rest of the $\ttbar$ events.

\begin{figure}[t]
\centering
\subfloat[]{\includegraphics[width=0.40\textwidth]{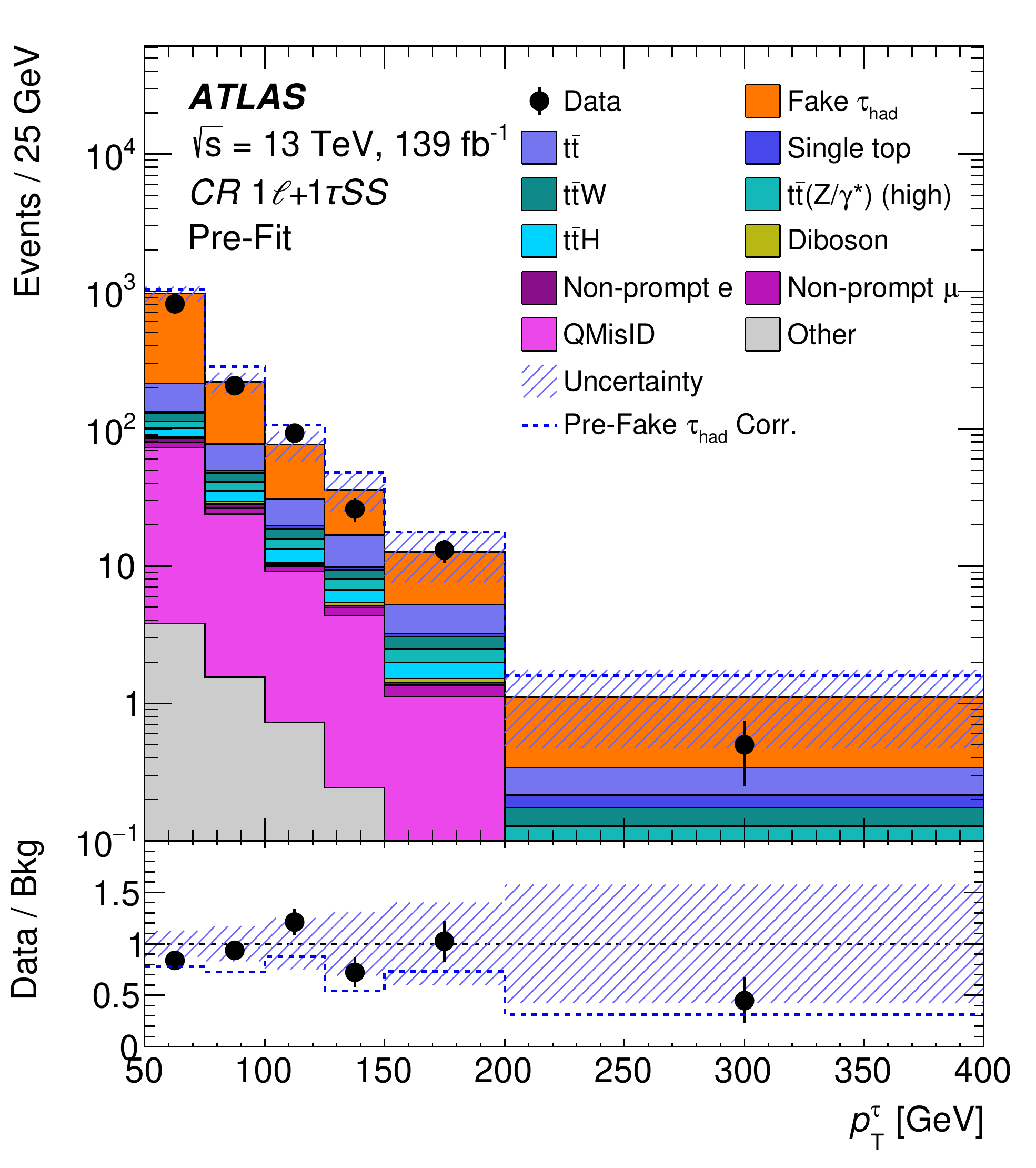}\label{fig:CR1l1tauSS_taupt}}
\subfloat[]{\includegraphics[width=0.40\textwidth]{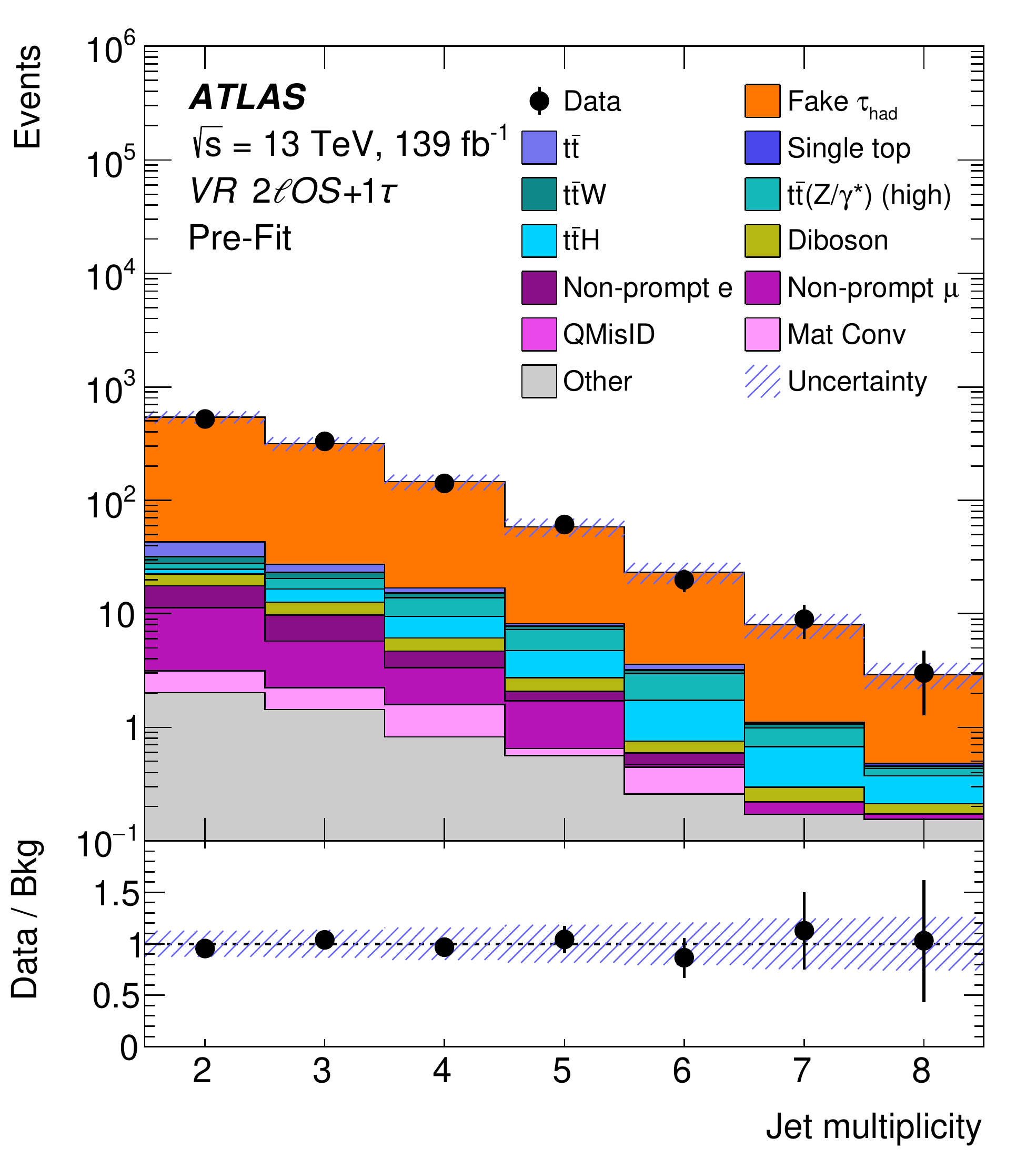}\label{fig:VR2lOS1tau_njets}}
\caption{Comparison between data and the background prediction for (a) the $\tauh$ $\pt$ ($p_{\mathrm T}^{\tau}$) distribution in the \lonetauSS CR, and (b) the jet multiplicity distribution in the
\llOSonetau VR.
The background contributions before the likelihood fit to data (``Pre-Fit'') are shown as filled histograms.
The ratio of the data to the background (``Bkg'') prediction is shown in the lower panel.
In (a), the total background prediction before the correction with the per-candidate normalisation factors (``Pre-Fake $\tauh$ Corr.'') is shown as a dashed blue histogram (line) in the upper (lower) panel.
The size of the combined statistical and systematic uncertainty in the background prediction is indicated by the blue hatched band.
The last bin in each figure contains the overflow.}
\end{figure}
 
\subsubsection{Non-prompt light leptons and charge misassignment}
\label{sec:bkg_reducible_light}
 
Non-prompt leptons originate from material conversions, heavy-flavour hadron decays, or the improper reconstruction of other particles, with an admixture strongly depending on the lepton quality requirements and varying across event categories. These backgrounds are in general very small in all SRs and thus are estimated from simulation, with the normalisation determined by the likelihood fit. The main contribution to the non-prompt-lepton background is from $\ttbar$ production, followed by much smaller contributions from $V$+jets and single-top-quark processes. The non-prompt light leptons in the simulated samples are labelled according to whether they originate from heavy-flavour (HF) or light-flavour (LF) hadron decays, or from a material conversion candidate. The HF category includes leptons from both bottom and charm decays.
QMisID backgrounds arise mainly from $t\bar{t}$ production, with one electron having a hard bremsstrahlung emission followed by an asymmetric conversion ($e^\pm \to e^\pm \gamma^* \to e^\pm e^+ e^-$) or a mismeasured track curvature.
The muon charge misassignment rate is negligible in the $\pt$ range relevant to this analysis.
 
Several of the event categories introduced in Section~\ref{sec:events} were designed to be enriched in specific processes and are used to derive normalisation factors to improve their modelling by the simulation.
The \llmatc and \lllmatc CRs are enriched in MatC and QMisID backgrounds and only the total event yield is used. There are four CRs enriched in contributions from HF non-prompt leptons in \ttbar events, i.e.\ \llttep, \llttem, \llttmup, and \llttmum. In these CRs, the $H_\textrm{T, lep}$ distribution is used to provide separation from the $\ttW$ background and thus optimise the sensitivity to the HF non-prompt electron and muon contributions. The simultaneous fit to these regions, split by total charge, provides additional separation due to the charge asymmetry of the $\ttW$ process. Normalisation factors for three non-prompt-lepton background contributions are estimated from the likelihood fit. The normalisation factor for HF non-prompt leptons is estimated separately for electrons and muons, $\lambda^\textrm{had}_e$ and $\lambda^\textrm{had}_\mu$ respectively. An additional normalisation factor is determined for the sum of MatC and QMisID backgrounds, $\lambda^\textrm{MatC}_e$. The measured normalisation factors are:  $\hat{\lambda}^\textrm{had}_e = 1.06 \pm 0.30$,  $\hat{\lambda}^\textrm{had}_\mu = 0.81 \pm 0.12$, and $\hat{\lambda}^\textrm{MatC}_e = 1.03 \pm 0.24$, where the uncertainties are dominated by the statistical uncertainty.
The systematic uncertainties considered are discussed in the following, although they have a negligible impact on the final result.
The background estimation procedure for non-prompt light leptons relies on the simulation to predict the kinematic distributions of the \ttbar process, and thus is affected by related modelling uncertainties (see Section~\ref{sec:datamc}). Additional uncertainties are estimated by relaxing lepton criteria to enrich the samples in the different types of non-prompt leptons, and comparing the data with the simulation.
A 25\% uncertainty is estimated for material conversions, based on a comparison between data and simulation in
a dedicated control sample enhanced in $Z \to \mu^+\mu^- \gamma (\to e^+e^-)$ candidate events, defined by requiring two OS Tight muons and one Tight electron that fails the material conversion veto requirement.
This uncertainty is applied to all categories except for \llmatc and \lllmatc as thus acts as an extrapolation uncertainty.
Figures~\ref{fig:HFe_derivation} and ~\ref{fig:HFmu_validation} display the $H_\textrm{T, lep}$ distribution in the \llttem and \llttmum CRs after the likelihood fit to data.
As shown in the figures,
the spectra for the HF non-prompt electron and muon contributions are softer than those for the $t\bar{t}W$ and $VV$ backgrounds. For this comparison, the CRs with negative total charge are selected, as this requirement suppresses the $t\bar{t}W$ and $VV$ contributions, due to their charge asymmetry, thus increasing the fraction of non-prompt-lepton background.
 
\begin{figure}[t]
\centering
\subfloat[]{\includegraphics[width=0.40\textwidth]{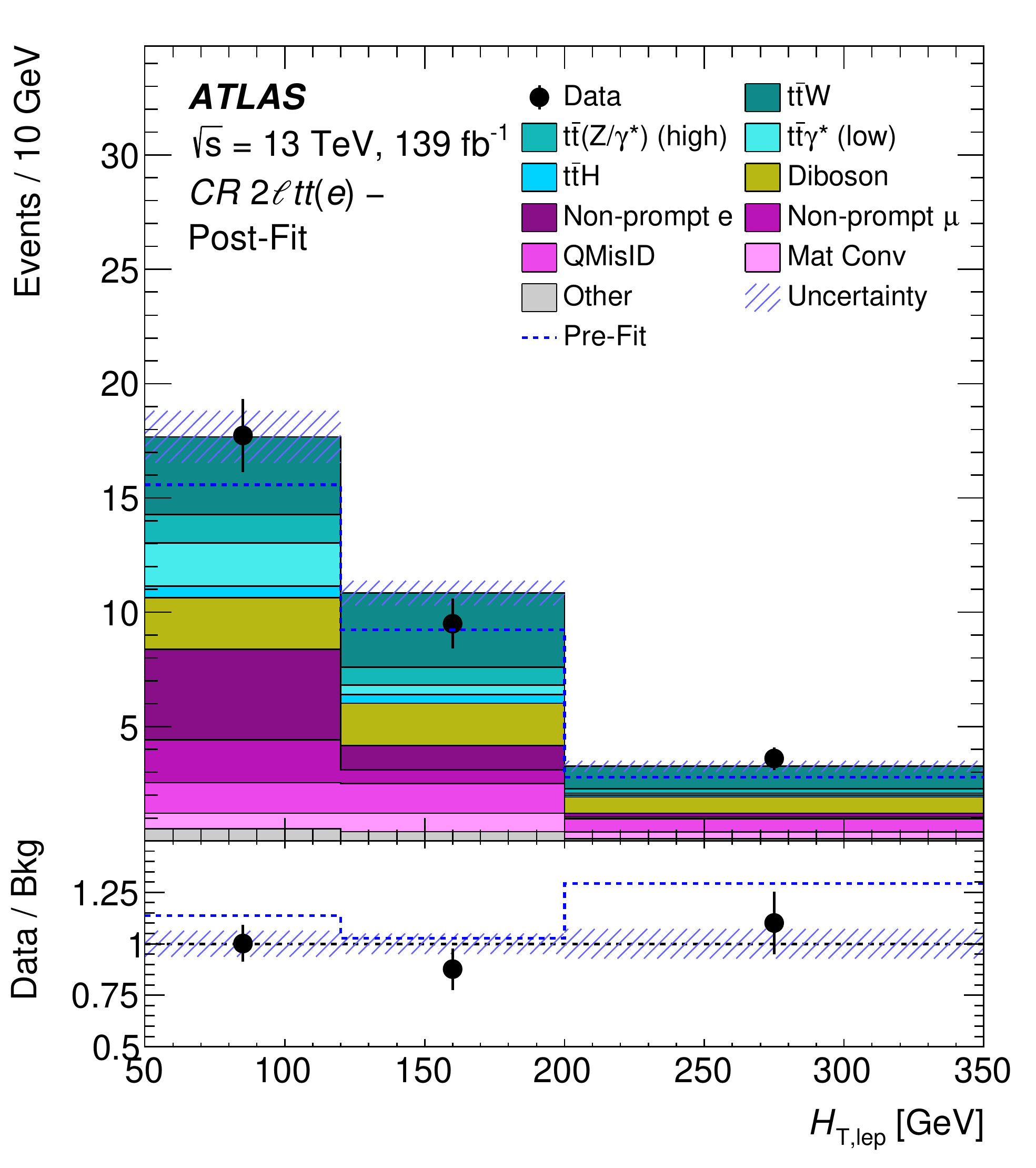}\label{fig:HFe_derivation}}
\subfloat[]{\includegraphics[width=0.40\textwidth]{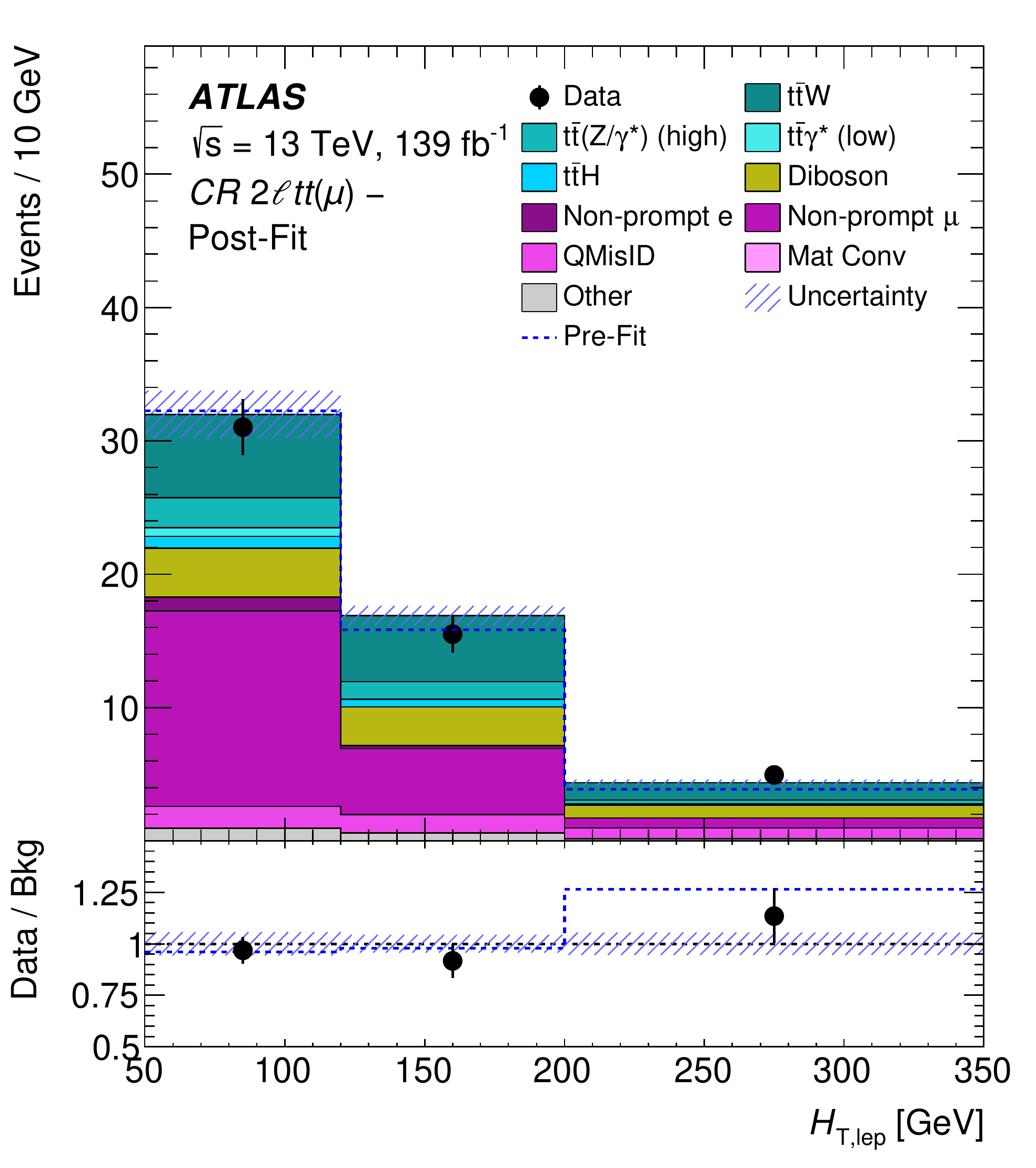}\label{fig:HFmu_validation}}
\caption{
Comparison between data and the background prediction for the distribution of the scalar sum of the lepton $\pt$ ($H_\textrm{T, lep}$) in (a) the \llttem CR and (b) the \llttmum CR.
The background contributions after the likelihood fit to data (``Post-Fit'') under the background-only hypothesis are shown as filled histograms.
The total background prediction before the likelihood fit to data (``Pre-Fit'') is shown as a dashed blue histogram in the upper panel.
The ratio of the data to the background (``Bkg'') prediction is shown in the lower panel, separately for post-fit background (black points) and pre-fit background (dashed blue line).
The size of the combined statistical and systematic uncertainty in the background prediction is indicated by the blue hatched band.
The last bin in each figure contains the overflow.}
\end{figure}

 
\section{Analysis model and results}
\label{sec:results}
\begin{table*}
\centering
\caption{\label{tbl:fitconfiguration}
Summary of the event categories per channel, discriminating variables per event category, and number of bins used in the statistical analysis.
}
\vspace*{0.25cm}
\small
\begin{tabular}{l|ccc|cc}
\hline\hline
&
\ltau &
\llOStau &
\llSSorllltau &
\llSSnotau &
\lllnotau \\ \hline \hline
Number of event categories & 6 & 2  & 2  & 8  & 4   \\ \hline
$\Meff$ spectrum & 3 SRs & 2 SRs & 2 SRs & --- & ---  \\
$H_{\mathrm{T, lep}}$ spectrum & --- & --- & --- & 6 CRs & 2 CRs \\
Event yield & 3 CRs & --- & --- & 2 CRs & 2 CRs  \\ \hline
Total number of bins & 16 & 9 & 6 & 20 & 10 \\
\hline \hline
\end{tabular}
\end{table*}

A maximum-likelihood fit is performed on all bins in the 22 event categories considered, consisting of 15 CRs and 7 SRs (see Table~\ref{tbl:fitconfiguration}), to simultaneously determine the background and $\LQ$ signal yields that are most consistent with the data. The $\Meff$ spectrum is used in the SRs to maximise the sensitivity to the $\LQ$ signal, while the CRs are used to either determine or constrain different backgrounds.
In the eight CRs from the \llSSnotau and \lllnotau channels that require very tight selection criteria for light leptons including the internal and material conversion vetoes, the $H_\textrm{T, lep}$ spectrum is used to discriminate between, and separately normalise, the $\ttbar$ (with non-prompt electrons and muons) and $\ttW$ backgrounds, as well as to constrain the $\ttdy$, and $\VV$ background predictions.
In the remaining seven CRs, the total event yield (i.e. a single bin) is used: three CRs are used to constrain the $\ttbar$ background prediction with either real or fake $\tauh$ candidates, and four CRs are used to normalise the backgrounds with an internal conversion, and with a material conversion or QMisID.

The likelihood function ${\cal L}(\mu,\vec{\lambda}, \vec{\theta})$ is constructed as a product of Poisson probability terms over all bins considered in the search, and depends on the signal-strength parameter, $\mu$, defined as a multiplicative factor applied to the predicted yield for the $\LQ$ signal (depending on the assumed $\LQ$ mass and the $\LQ \rightarrow t\tau$ branching fraction), $\vec{\lambda}$, the normalisation factors for several backgrounds (see Section~\ref{sec:background}), and $\vec{\theta}$, a set of nuisance parameters (NP) encoding systematic uncertainties in the signal and background expectations~\cite{Cranmer:2012sba}. Systematic uncertainties can impact the estimated signal and background rates, the migration of events between categories, and the shape of the fitted distributions; they are summarised in Table~\ref{tab:SystSummary}.
Both $\mu$ and $\vec{\lambda}$ are treated as free parameters in the likelihood fit.  The NPs $\vec{\theta}$ allow variations of the expectations for signal and background according to the systematic uncertainties, subject to Gaussian
constraints in the likelihood fit.
Their fitted values represent the deviations from the nominal expectations that globally provide the best fit to the data.
Statistical uncertainties in each bin due to the limited size of the simulated samples are taken into account by dedicated parameters
using the Beeston--Barlow technique~\cite{Barlow:1993dm}.

\begin{table*}
\centering
\caption{Sources of systematic uncertainty considered in the analysis.
``N'' means that the uncertainty is taken as normalisation-only for all processes and channels affected.
Some of the systematic uncertainties are split into several components, as indicated by the number in the rightmost column.
}
\resizebox{\textwidth}{!}{
\begin{tabular}[t]{lc}
\hline\hline
Systematic uncertainty                       & Components  	      \\
\hline
\hline
Luminosity                                   & 1		      \\
Pile-up reweighting                          & 1		      \\
\textbf {Physics objects}                    &		      \\
\ \ Electron                                 & 6		      \\
\ \ Muon                                     & 16~~	      \\
\ \ $\tau$-leptons                                     & 21~~		      \\
\ \ Jet energy scale and resolution          & 34~~                  \\
\ \ Jet vertex fraction                      & 1		      \\
\ \ Jet flavour tagging                      & 22~~	      \\
\ \ \met                                     & 3		      \\
\hline
Total (Experimental)                         & 105~~~~		     \\
\hline
\hline
\textbf {Data-driven reducible background estimates}   &		     \\
\ \ \ttbar kinematic reweighting				 & 2         \\
\ \ Fake-\tauh estimates                     & 14~~          \\
\ \ Material conversions modelling                         	& 1		     \\
\ \ Internal conversions modelling                         	& 1		      \\
\hline
Total (Data-driven reducible background)     & 18~~		   \\
\hline
\hline
\end{tabular}
\begin{tabular}[t]{lc}
\hline\hline
Systematic uncertainty                       & Components  	      \\
\hline
\hline
\textbf {Signal modelling}                       &                     \\
\ \ Cross section (N)                           & 1                   \\
\ \ QCD scale ($\muF$, $\muR$)              & 1		      \\
\ \ PDFs+$\alpha_\mathrm{S}$         & 1                   \\
\ \ QCD radiation                  & 1                   \\
\textbf {\ttbar modelling}                        	&		      \\
\ \ Cross section (N)                        		& 1		     \\
\ \ Parton shower and hadronisation model           & 2	      \\
\ \ Generator                                		& 2		      \\
\ \ QCD radiation                                    	& 2		     \\
\ \ QED radiative top-quark decay (N) 		& 1		     \\
\textbf {\ttW modelling}                        &		      \\
\ \ QCD scale                                   & 1		      \\
\ \ Generator                                	& 1		      \\
\textbf {$\ttdy$ (high mass) modelling}                        	&		      \\
\ \ Cross section (N)                               & 1		     \\
\ \ Generator                    	& 1		      \\
\textbf {\ttH modelling}                        &		      \\
\ \ Cross section (N)                          	& 1		      \\
\ \ Parton shower and hadronisation model       & 1		      \\
\textbf {$WZ$ modelling}    							&		      \\
\ \ Cross section (N)                               & 1		      \\
\ \ Heavy-flavour composition  (N)       			& 1		      \\
\textbf {Other background modelling}                &		      \\
\ \ Cross section (N)                               & 10~~		      \\
\hline
Total (Signal and background modelling)             & 30~~		     \\
\hline\hline
Total (Overall)                                     & 153~~~~	      \\
\hline\hline
\end{tabular}
}
\label{tab:SystSummary}
\end{table*}

The test statistic $q_\mu$ is defined as the profile likelihood ratio:
$q_\mu = -2\ln({\cal L}(\mu,{\hat{\vec{\lambda}}}_\mu, {\hat{\vec{\theta}}}_\mu)/{\cal L}(\hat{\mu},\hat{\vec{\lambda}}_{\hat{\mu}}, \hat{\vec{\theta}}_{\hat{\mu}}))$,
where $\hat{\mu}$, $\hat{\vec{\lambda}}_{\hat{\mu}}$, and $\hat{\vec{\theta}}_{\hat{\mu}}$ are the values of the parameters that
maximise the likelihood function,
and ${\hat{\vec{\lambda}}}_\mu$ and ${\hat{\vec{\theta}}}_\mu$ are the values of the parameters that maximise the likelihood function  for a given value of $\mu$.
The test statistic $q_\mu$ is evaluated with the {\textsc RooFit} package~\cite{Verkerke:2003ir}.
A related statistic is used to determine the probability that the observed data are compatible with the background-only hypothesis (i.e.~the discovery test)
by setting $\mu=0$ in the profile likelihood ratio ($q_0$).
The $p$-value (referred to as $p_0$) representing the probability of the data being compatible with the
background-only hypothesis is estimated by integrating
the distribution of $q_0$ from background-only pseudo-experiments, approximated using the asymptotic formulae given in Ref.~\cite{Cowan:2010js}, above the observed value of $q_0$. Some model dependence exists
in the estimation of the $p_0$, as a given signal scenario needs to be assumed in the calculation of the denominator of $q_0$, even
if the overall signal normalisation is allowed to float and is fitted to data. The observed $p_0$ is checked for each explored signal scenario.
Upper limits on the signal production cross section for each of the
signal scenarios considered are derived by using $q_\mu$ in the CL$_{\textrm{s}}$ method~\cite{Junk:1999kv,Read:2002hq}.
For a given signal scenario, values of the production cross section (parameterised by $\mu$) yielding
CL$_{\textrm{s}} < 0.05$, where CL$_{\textrm{s}}$ is computed using the asymptotic approximation~\cite{Cowan:2010js}, are excluded at $\geq 95\%$ confidence level (CL).
The upper limits derived with the asymptotic approximation agree very well with those estimated via background-only pseudo-experiments.
 
\begin{figure*}[t!]
\centering
\subfloat[]{\includegraphics[width=0.6\textwidth]{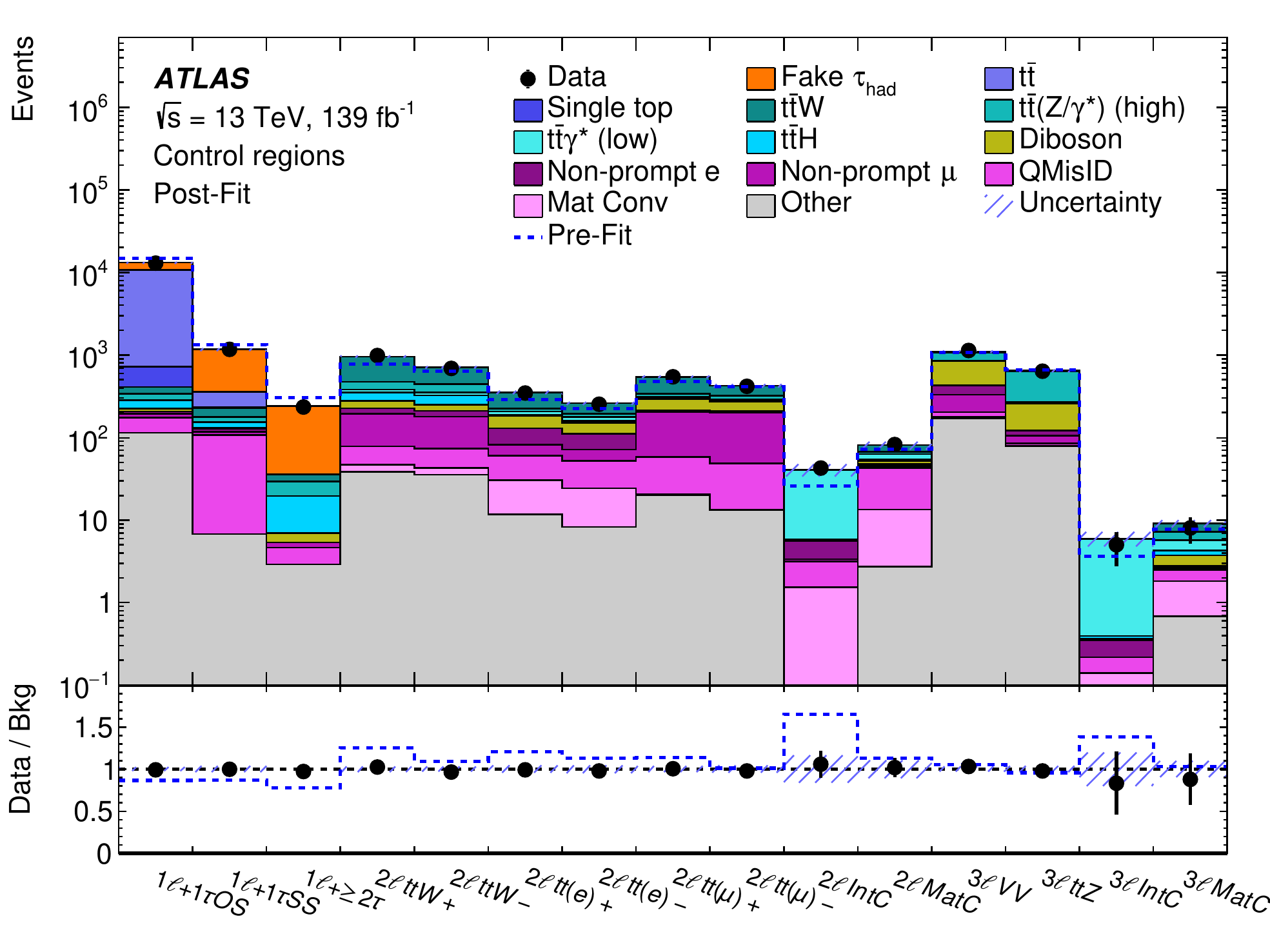}\label{fig:summaryCR}} \\
\subfloat[]{\includegraphics[width=0.6\textwidth]{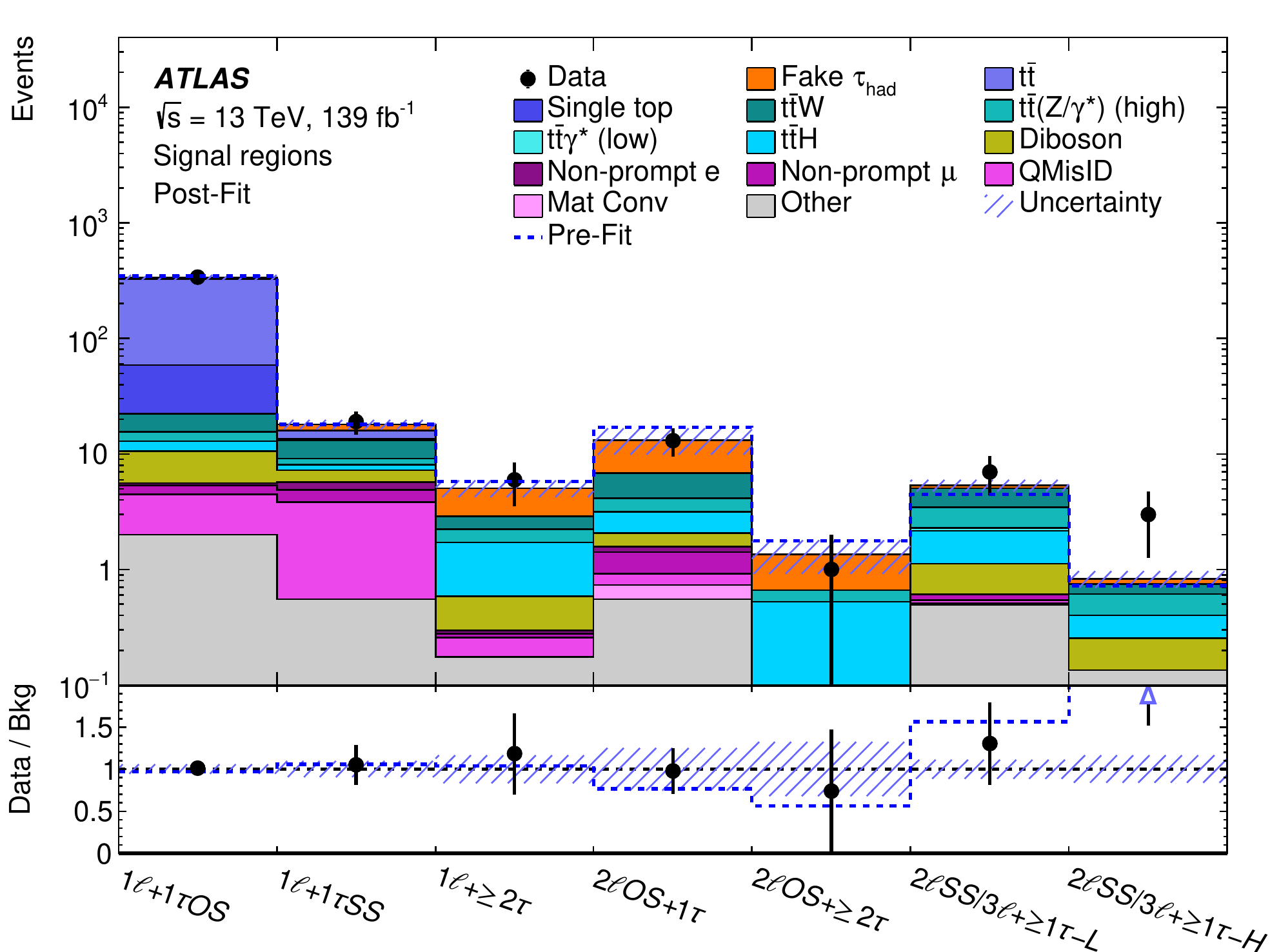}\label{fig:summarySR}}
\caption{Comparison between data and the background prediction for the event yields in (a) the 15 control region categories and (b) the 7 signal region categories.
The background contributions after the likelihood fit to data (``Post-Fit'') under the background-only hypothesis are shown as filled histograms.
The total background prediction before the likelihood fit to data (``Pre-Fit'') is shown as a dashed blue histogram in the upper panel.
The ratio of the data to the background (``Bkg'') prediction is shown in the lower panel, separately for post-fit background (black points) and pre-fit background (dashed blue line).
The size of the combined statistical and systematic uncertainty in the background prediction is indicated by the blue hatched band.
The blue triangles indicate points that are outside the vertical range of the figure.}
\end{figure*}
 
A comparison of the distributions of observed and expected yields in the 15 CRs and the 7 SRs after the combined likelihood fit under the background-only hypothesis
is shown in Figures~\ref{fig:summaryCR} and~\ref{fig:summarySR}, respectively. The corresponding post-fit yields for the SRs can be found in Table~\ref{tab:postfityields}.
In general, good agreement between the data and predicted background yields is found across all event categories. As shown in Figure~\ref{fig:summaryVR},
good agreement is also obtained between the data and post-fit background prediction in the VRs, which were not used in the fit, giving confidence in the overall procedure.
 
The comparison between data and the background prediction for the $\Meff$ distributions used in the different SRs is shown in Figures~\ref{fig:meffspectrum1} and~\ref{fig:meffspectrum2}. The binning used for the $\Meff$ distributions in the different SRs represents a compromise between preserving enough discrimination in the fit between the background and the signal for the different values of LQ mass considered, and keeping the statistical uncertainty of the background prediction per bin well below 30\%. No significant excess is observed in any of the SRs. The observed $p_0$ is found to be consistent with the background-only hypothesis for all values of $m_{\LQ}$ and $\BR$ considered. The observed and expected $p_0$ as a function of $m_{\LQ}$ are shown in Figure~\ref{fig:p0scan}, assuming values of $\BR=1$ and $\BR=0.5$. This illustrates the significant expected sensitivity of the search, which for $\BR=1$ exceeds 5 standard deviations for $m_{\LQ}<1.21~\tev$ and 3 standard deviations for $m_{\LQ}<1.36~\tev$.

\begin{table*}
\centering
\caption{
Summary of observed and predicted yields in the seven signal region categories. The background prediction is shown after the combined likelihood fit to data under the background-only hypothesis across all control region and signal region categories. The expected signal yields that are obtained by using their theoretical cross sections are also shown with their pre-fit uncertainties, assuming $\BR=1$. Dashes refer to components that are negligible or not applicable.
}
\vspace*{0.25cm}
\small
\begin{tabular}{l|ccc}
\hline \hline
& \lonetauOS & \lonetauSS  & \ltwotau \\
\hline\hline
Data &
$339$ &
$19$ &
$6$ \\
\hline
Total background &
$340\pm20$~~ &
$17.1\pm2.1$~~ &
$5.2\pm1.2$ \\
\hline
Fake $\tauh$ &
$10.2\pm6.5$~~ &
$1.8\pm1.7$ &
$1.9\pm1.0$ \\
$\ttbar$ &
$270\pm21$~~ &
$2.4\pm1.0$ &
--- \\
Single top &
$37.4\pm5.4$~~ &
$0.79\pm0.56$ &
--- \\
$\ttW$ &
$6.7\pm1.0$ &
$3.93\pm0.59$ &
$0.66\pm0.13$ \\
$\ttbar Z/\gamma^{*}$ (high mass) &
$2.38\pm0.65$ &
$1.11\pm0.30$ &
$0.57\pm0.15$ \\
$\ttbar \gamma^{*}$ (low mass) &
--- &
$0.03\pm0.01$ &
--- \\
$\ttH$ &
$2.38\pm0.44$ &
$0.87\pm0.21$ &
$1.18\pm0.34$ \\
Diboson &
$5.02\pm0.66$ &
$1.70\pm0.27$ &
$0.31\pm0.07$ \\
QMisID &
$2.59\pm0.68$ &
$3.38\pm0.90$ &
$0.10\pm0.08$ \\
Other &
$3.50\pm0.91$ &
$1.14\pm0.38$ &
$0.45\pm0.13$ \\
\hline\hline
$\LQ$ ($0.9~\tev$) &
$80.3\pm9.2$~~ &
$25.1\pm2.6$~~ &
$51.9\pm9.3$~~ \\
$\LQ$ ($1.1~\tev$) &
$20.9 \pm 2.5$~~ &
$6.92\pm0.74$ &
$11.4\pm2.1$~~ \\
$\LQ$ ($1.3~\tev$) &
$6.02\pm0.75$ &
$1.93\pm0.25$ &
$2.89\pm0.57$ \\
\hline \hline
\end{tabular}
\vspace*{0.5cm}
 
\begin{tabular}{l|cccc}
\hline \hline
& \llOSonetau & \llOStwotau & \llSSorllltau-L & \llSSorllltau-H \\
\hline\hline
Data &
$13$ &
$1$ &
$7$ &
$3$ \\
\hline
Total background &
$14.6\pm3.3$~~ &
$1.66\pm0.58$ &
$5.38\pm0.68$ &
$0.83\pm0.14$ \\
\hline
Fake $\tauh$ &
$8.0\pm3.4$ &
$0.96\pm0.54$ &
$0.33\pm0.07$ &
$0.09\pm0.04$ \\
$\ttW$ &
$2.68\pm0.51$ &
--- &
$1.57\pm0.39$ &
$0.13\pm0.05$ \\
$\ttbar Z/\gamma^{*}$ (high mass)&
$0.86\pm0.26$ &
$0.15\pm0.06$ &
$1.10\pm0.20$ &
$0.20\pm0.06$ \\
$\ttbar \gamma^{*}$ (low mass) &
$0.02\pm0.01$ &
--- &
$0.03\pm0.02$ &
--- \\
$\ttH$ &
$1.07\pm0.19$ &
$0.51\pm0.14$ &
$1.09\pm0.27$ &
$0.14\pm0.03$ \\
Diboson &
$0.51\pm0.08$ &
--- &
$0.53\pm0.10$ &
$0.13\pm0.03$ \\
Non-prompt $e$ &
$0.14\pm0.13$ &
--- &
--- &
--- \\
Non-prompt $\mu$ &
$0.44\pm0.28$ &
--- &
$0.06\pm0.06$ &
--- \\
QMisID &
$0.19\pm0.13$ &
--- &
$0.04\pm0.05$ &
--- \\
Mat Conv&
$0.18\pm0.15$ &
--- &
$0.01\pm0.02$ &
--- \\
Other &
$0.55\pm0.22$ &
$0.04\pm0.03$ &
$0.61\pm0.28$ &
$0.13\pm0.06$ \\
\hline\hline
$\LQ$ ($0.9~\tev$) &
$26.0\pm3.9$~~ &
$6.1\pm1.1$ &
$6.02\pm0.90$ &
$13.6\pm2.1$~~ \\
$\LQ$ ($1.1~\tev$) &
$6.4\pm1.0$ &
$1.35\pm0.26$ &
$1.09\pm0.17$ &
$3.44\pm0.57$ \\
$\LQ$ ($1.3~\tev$) &
$1.69\pm0.31$ &
$0.36\pm0.08$ &
$0.20\pm0.05$ &
$1.07\pm0.19$ \\
\hline \hline
\end{tabular}
\label{tab:postfityields}
\end{table*}

\begin{figure*}[t!]
\centering
\includegraphics[width=0.6\textwidth]{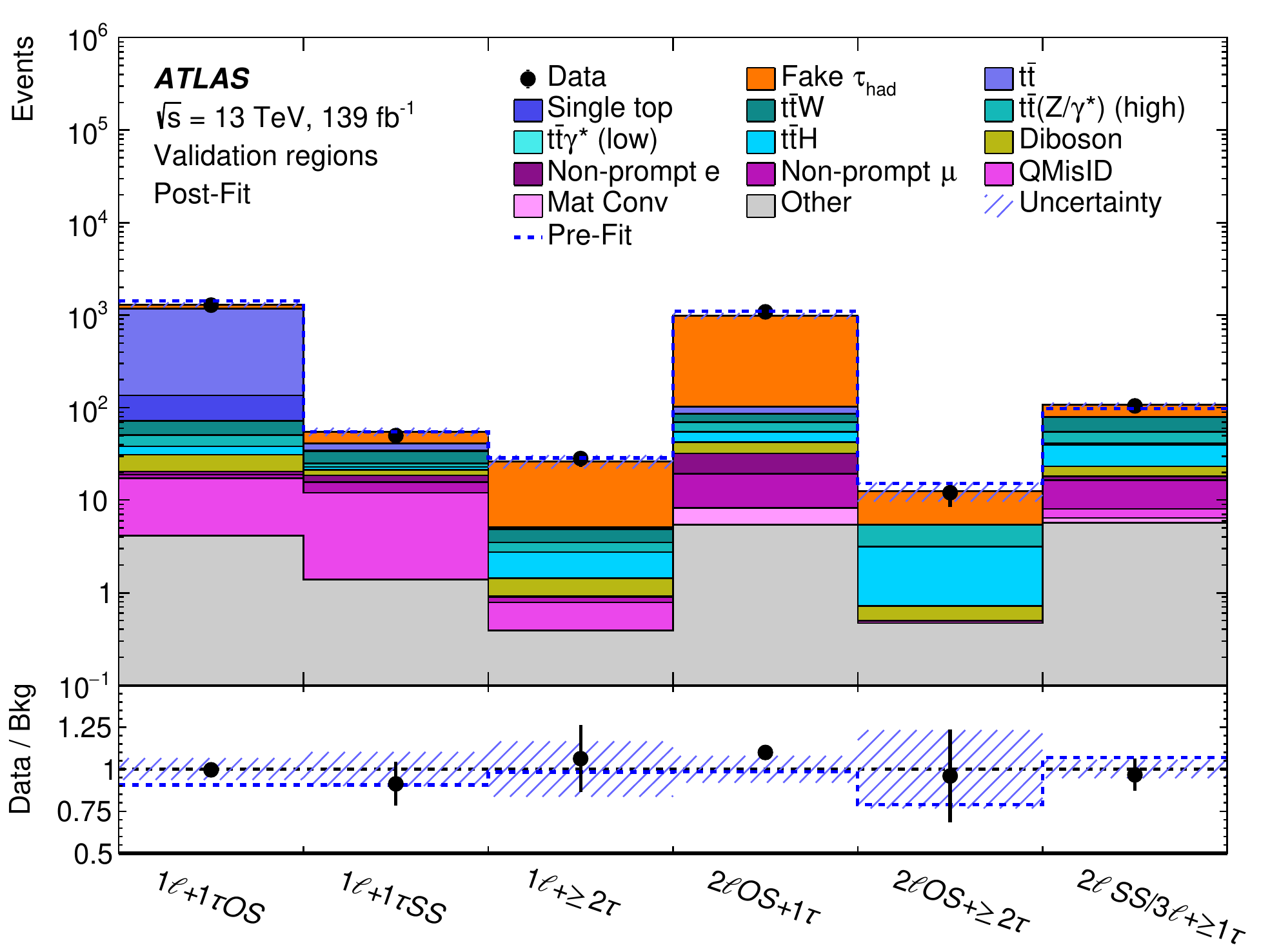}
\caption{Comparison between data and the background prediction for the event yields in the six validation region categories.
The background contributions after the likelihood fit to data (``Post-Fit'') under the background-only hypothesis are shown as filled histograms.
The ratio of the data to the background (``Bkg'') prediction is shown in the lower panel.
The total background prediction before the likelihood fit to data (``Pre-Fit'') is shown as a dashed blue histogram (line) in the upper (lower) panel.
The size of the combined statistical and systematic uncertainty in the background prediction is indicated by the blue hatched band.}
\label{fig:summaryVR}
\end{figure*}

\begin{figure*}[t]
\centering
\subfloat[]{\includegraphics[width=0.40\textwidth]{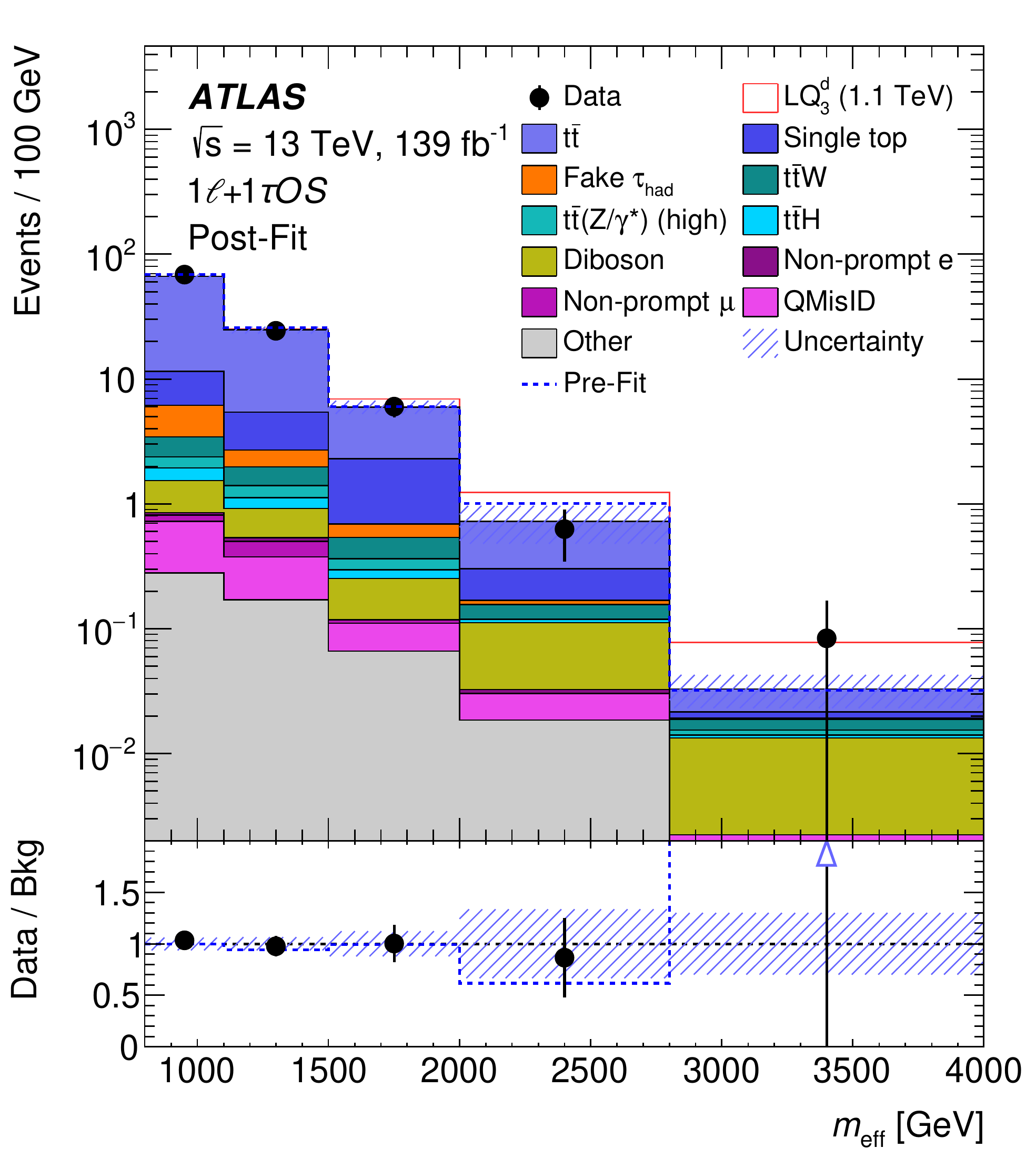}}
\subfloat[]{\includegraphics[width=0.40\textwidth]{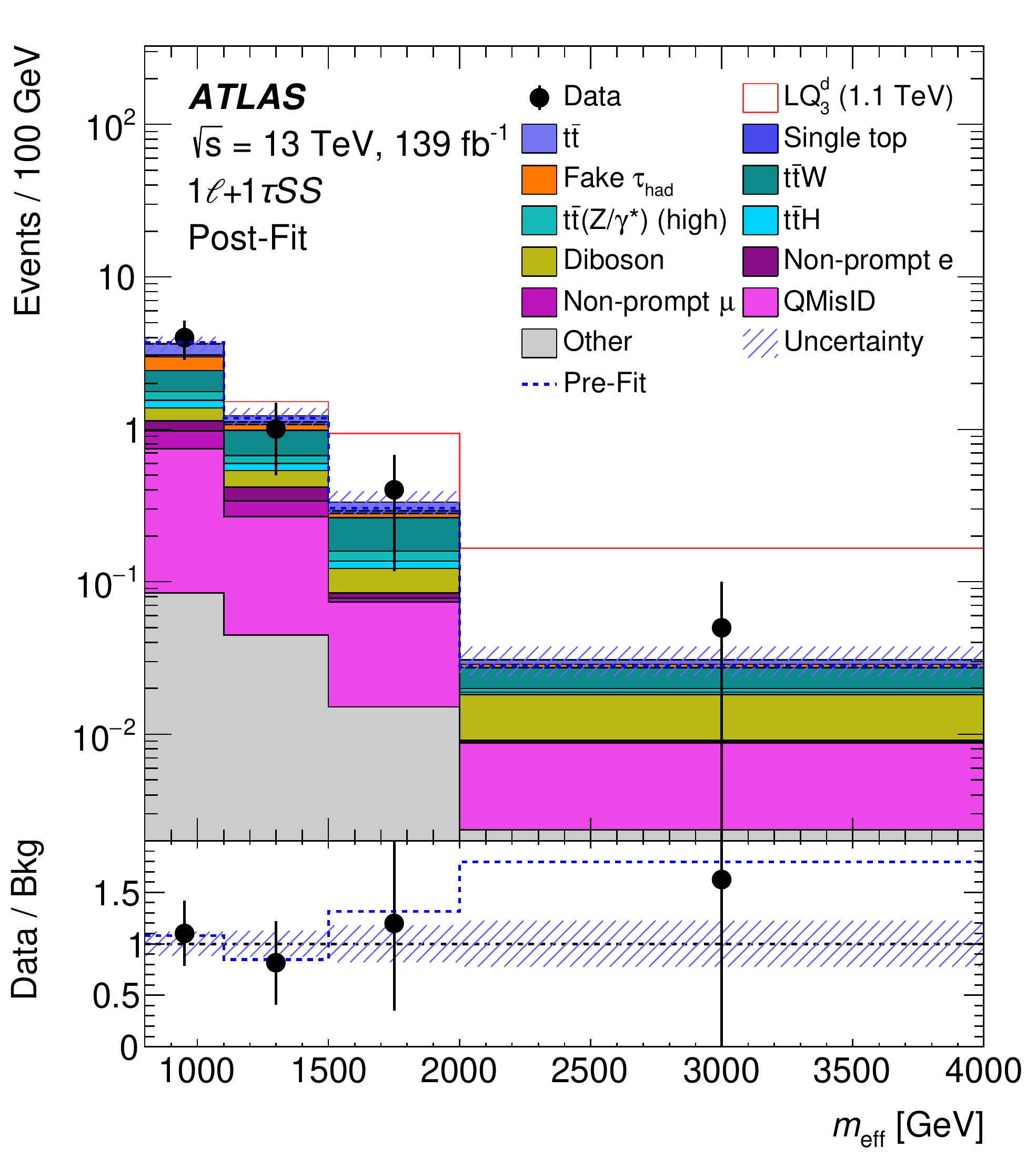}} \\
\subfloat[]{\includegraphics[width=0.40\textwidth]{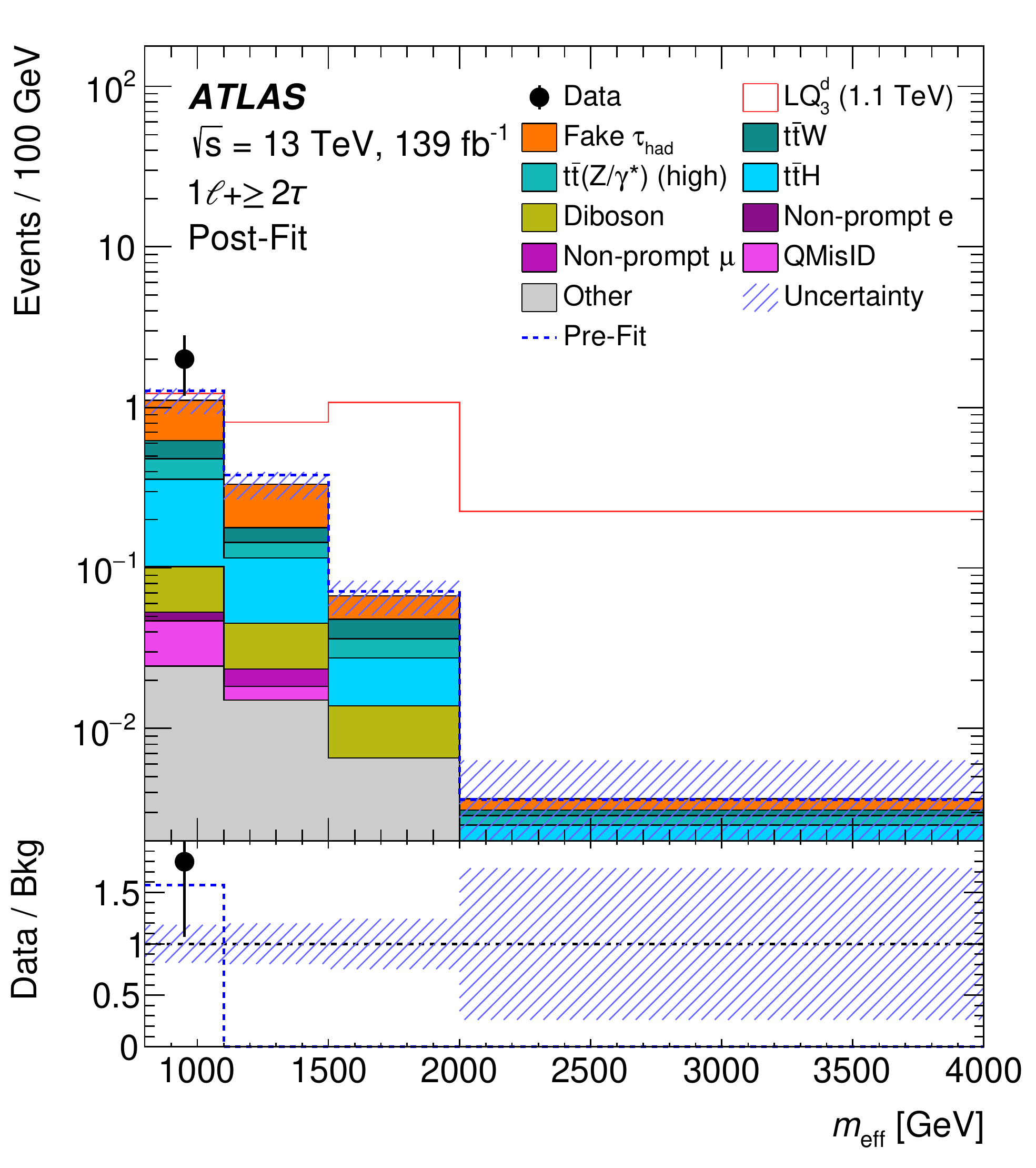}}
\caption{
Comparison between data and prediction for the $\Meff$ distribution used in different signal region categories of the \ltau channel:
(a) \lonetauOS, (b) \lonetauSS, and (c) \ltwotau.
The background contributions after the likelihood fit to data (``Post-Fit'') under the background-only hypothesis are shown as filled histograms.
For illustrative purposes, the expected signal for $m_{\LQ}=1.1~\tev$ and $\BR = 1$ is shown as a unfilled red histogram added to the post-fit background.
The total background prediction before the likelihood fit to data (``Pre-Fit'') is shown as a dashed blue histogram in the upper panel.
The ratio of the data to the background (``Bkg'') prediction is shown in the lower panel, separately for post-fit background (black points) and pre-fit background (dashed blue line).
The size of the combined statistical and systematic uncertainty in the background prediction is indicated by the blue hatched band.
The blue triangles indicate points that are outside the vertical range of the figure.
The last bin in each figure contains the overflow.}
\label{fig:meffspectrum1}
\end{figure*}
 
\begin{figure*}[t]
\centering
\subfloat[]{\includegraphics[width=0.40\textwidth]{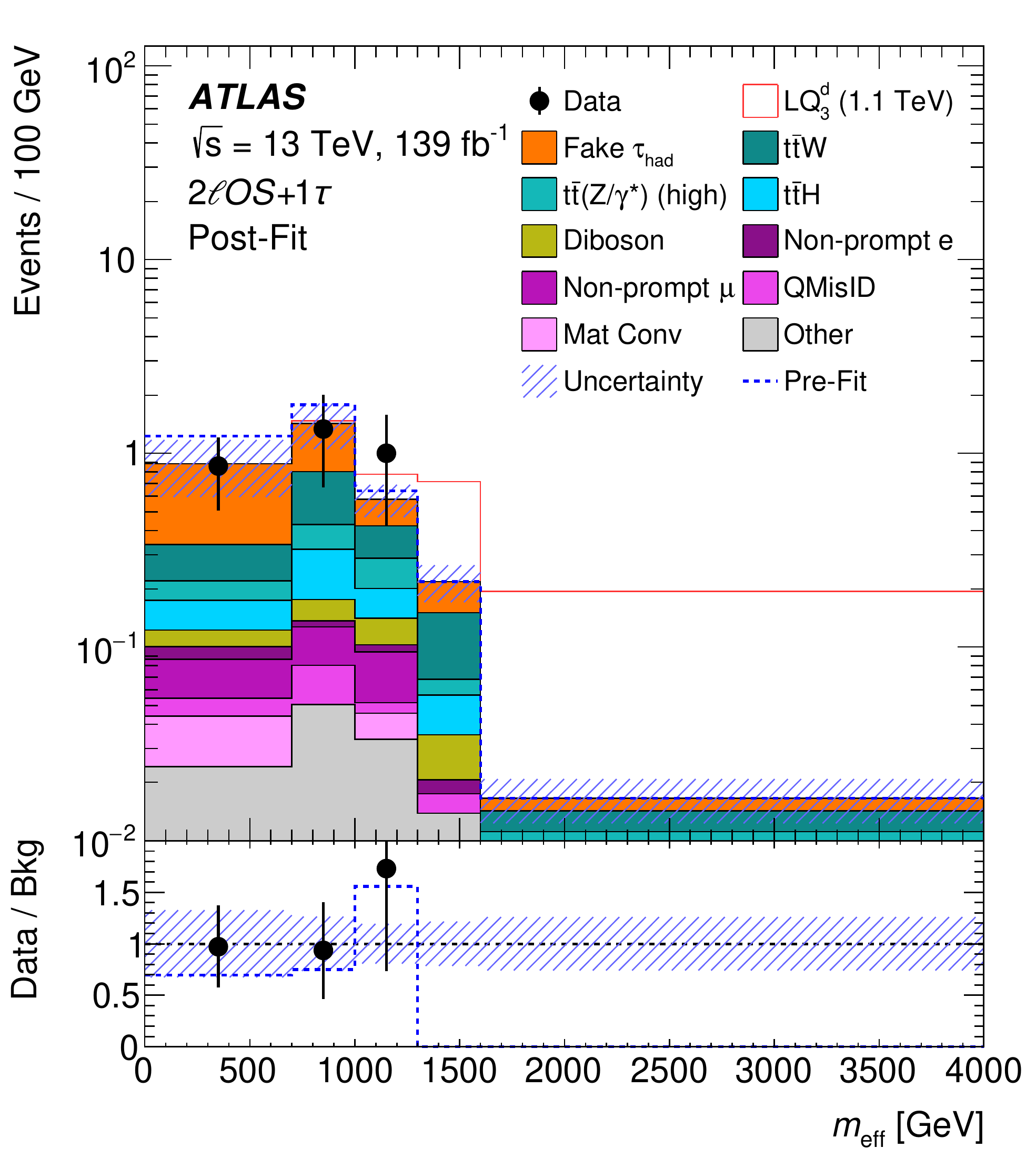}}
\subfloat[]{\includegraphics[width=0.40\textwidth]{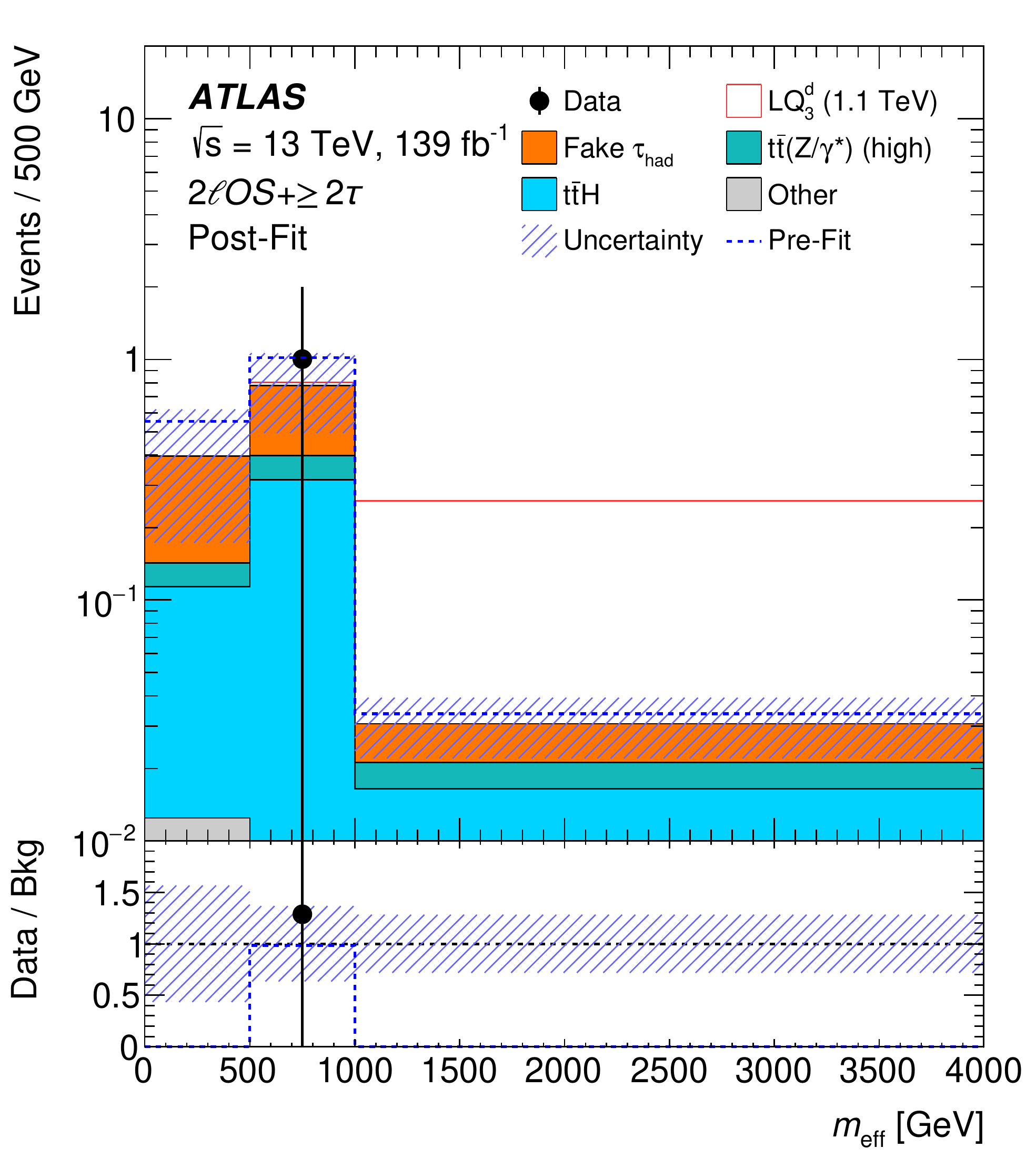}} \\
\subfloat[]{\includegraphics[width=0.40\textwidth]{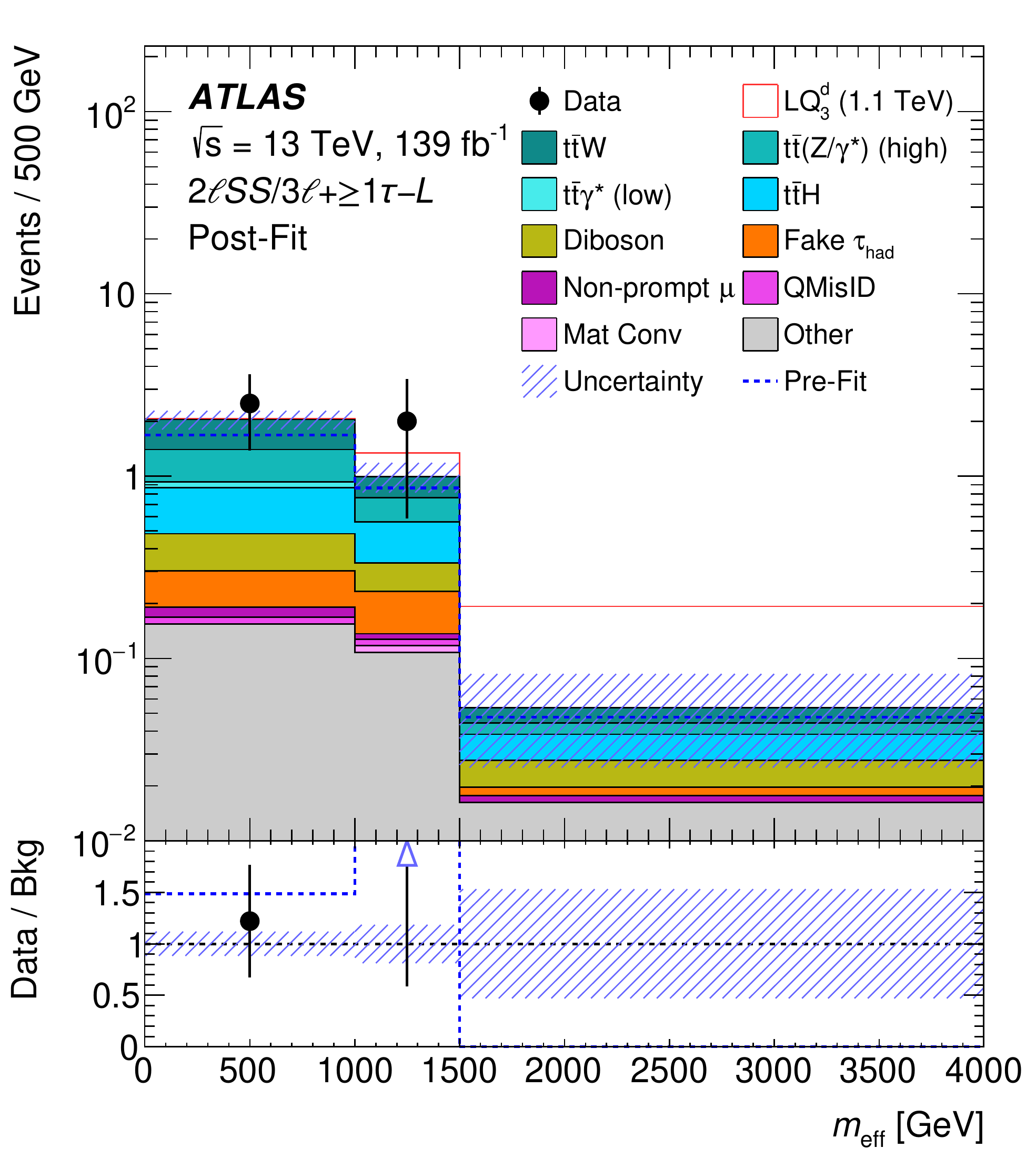}}
\subfloat[]{\includegraphics[width=0.40\textwidth]{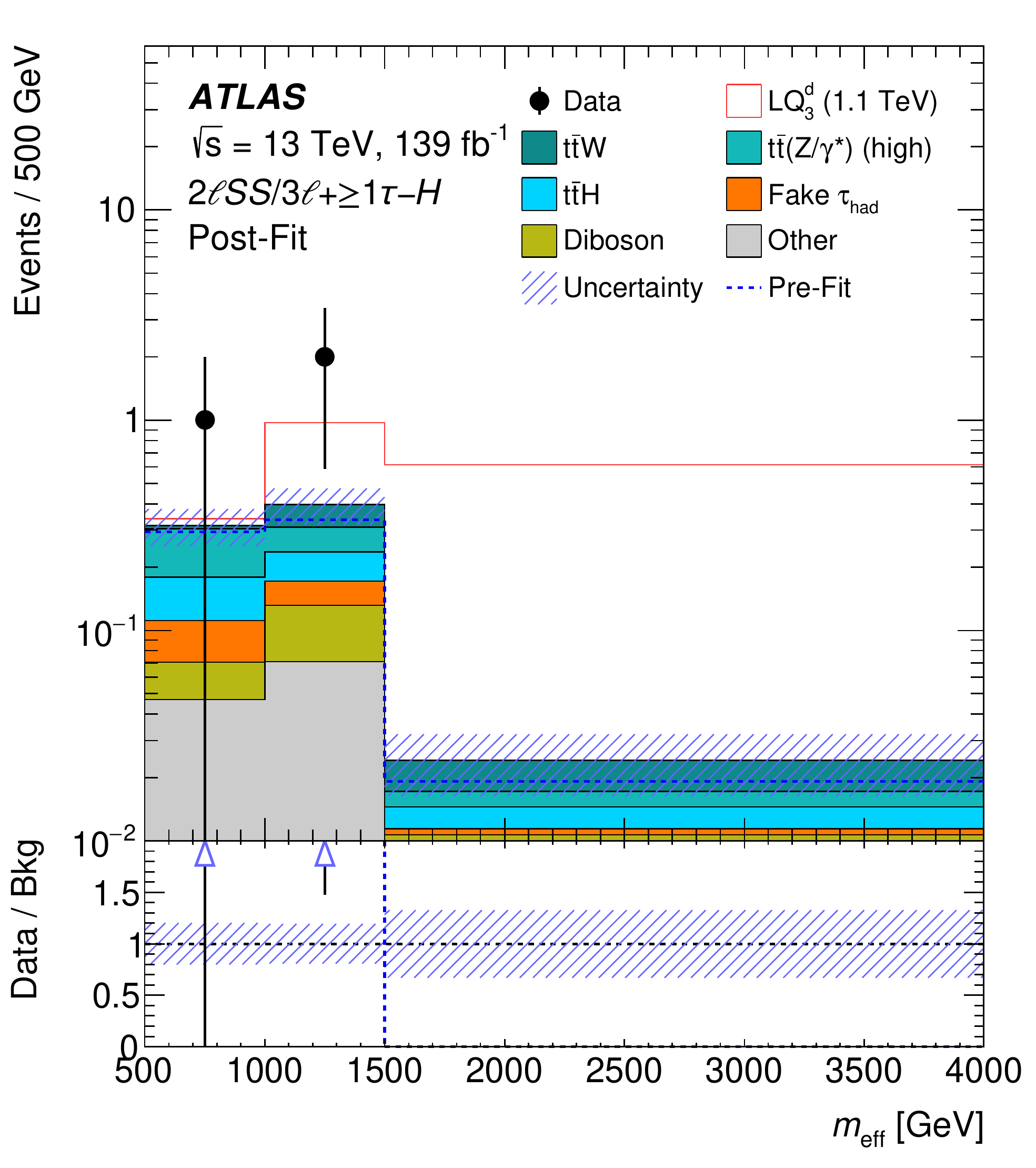}} \\
\caption{
Comparison between data and prediction for the $\Meff$ distribution used in different signal region categories of the \llOStau and \llSSorllltau channels:
(a) \llOSonetau, (b) \llOStwotau, (c) \llSSorllltau-L, and (d) \llSSorllltau-H.
The background contributions after the likelihood fit to data (``Post-Fit'') under the background-only hypothesis are shown as filled histograms.
For illustrative purposes, the expected signal for $m_{\LQ}=1.1~\tev$ and $\BR = 1$ is shown as unfilled red histogram added to the post-fit background.
The total background prediction before the likelihood fit to data (``Pre-Fit'') is shown as a dashed blue histogram in the upper panel.
The ratio of the data to the background (``Bkg'') prediction is shown in the lower panel, separately for post-fit background (black points) and pre-fit background (dashed blue line).
The size of the combined statistical and systematic uncertainty in the background prediction is indicated by the blue hatched band.
The blue triangles indicate points that are outside the vertical range of the figure.
The last bin in each figure contains the overflow.}
\label{fig:meffspectrum2}
\end{figure*}
 
\begin{figure*}[t]
\centering
\includegraphics[width=0.6\textwidth]{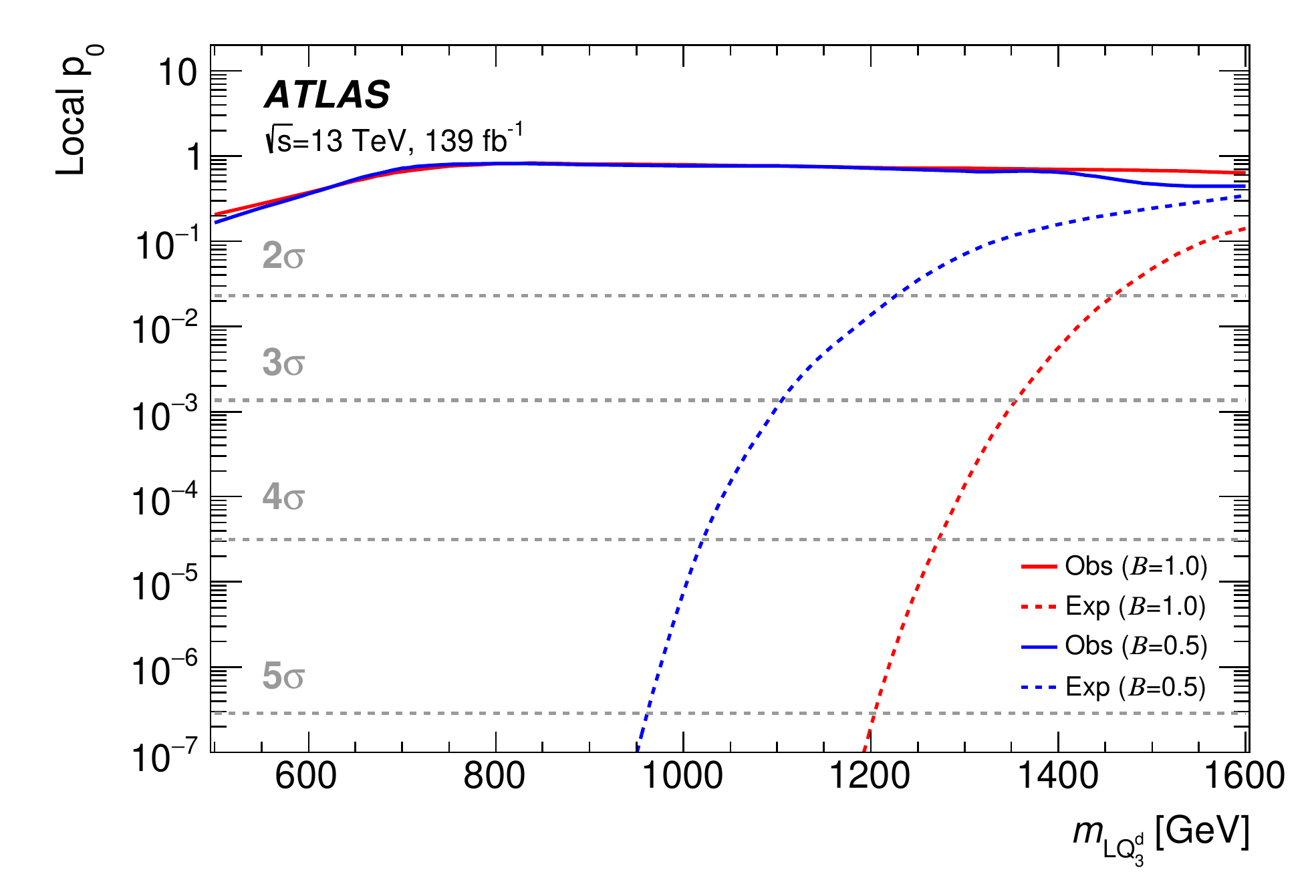}
\caption{The observed (solid) local $p_0$ as a function of $\LQ$ mass ($m_{\LQ}$) assuming $\BR=0.5$ (blue) and $\BR=1$ (red). The dashed curve shows the expected local $p_0$ under the hypothesis of a $\LQ$ signal at that mass. The horizontal dashed lines indicate the $p$-values corresponding to significances of 2 to 5 standard deviations.
}
\label{fig:p0scan}
\end{figure*}
 
In absence of any significant excess above the SM background prediction, 95\% CL upper limits are set on the cross section for the $\LQ$ pair production as a function of the assumed $m_{\LQ}$ and $\BR$. Figure~\ref{fig:limit1d} shows the 95\% CL upper limits on the $\LQ$ pair production cross section as a function of $m_{\LQ}$ resulting from the combination of all analysis channels, assuming $\BR=1$. The sensitivity is dominated by the \ltau channel, although the \llOStau and \llSSorllltau channels bring a significant improvement to the combined limit. The result is completely limited by the statistical uncertainty of the data, with the impact of systematic uncertainties being only to raise the expected cross-section upper limit by 2.3\% at $m_{\LQ}=1~\tev$, and more at lower and higher masses, reaching 4.5\% at $m_{\LQ}=500~\gev$ and $m_{\LQ}=1.6~\tev$.
The leading source of systematic uncertainty arises from $\tauh$ identification and energy scale calibration, followed by $\ttbar$ modelling.
A comparison of the cross-section limits with the theoretical prediction is used to derive 95\% CL limits on $\BR$ as a function of $m_{\LQ}$, as shown in Figure~\ref{fig:limit2d}.
Assuming that $\BR=1$, the observed and expected 95\% CL lower limits on $m_{\LQ}$ are $1.43~\tev$ and $1.41~\tev$, respectively. The corresponding limits for $\BR=0.5$ are $1.22~\tev$ and $1.19~\tev$, respectively.
The above limits assume that the only possible decay modes are $\mathrm{LQ} \to t\tau / b\nu$. In the case of a non-negligible contribution from the $\mathrm{LQ} \to q\tau$ ($q=u,c$) decay mode, more stringent limits could be derived for intermediate values of $\BR$, since $\mathrm{LQ}\mathrm{\overline{LQ}} \to t\tau q\tau$ final states would be probed by the \ltwotau SR, which dominates the sensitivity of this search.
 
\begin{figure*}[t]
\centering
\subfloat[]{\includegraphics[width=0.45\textwidth]{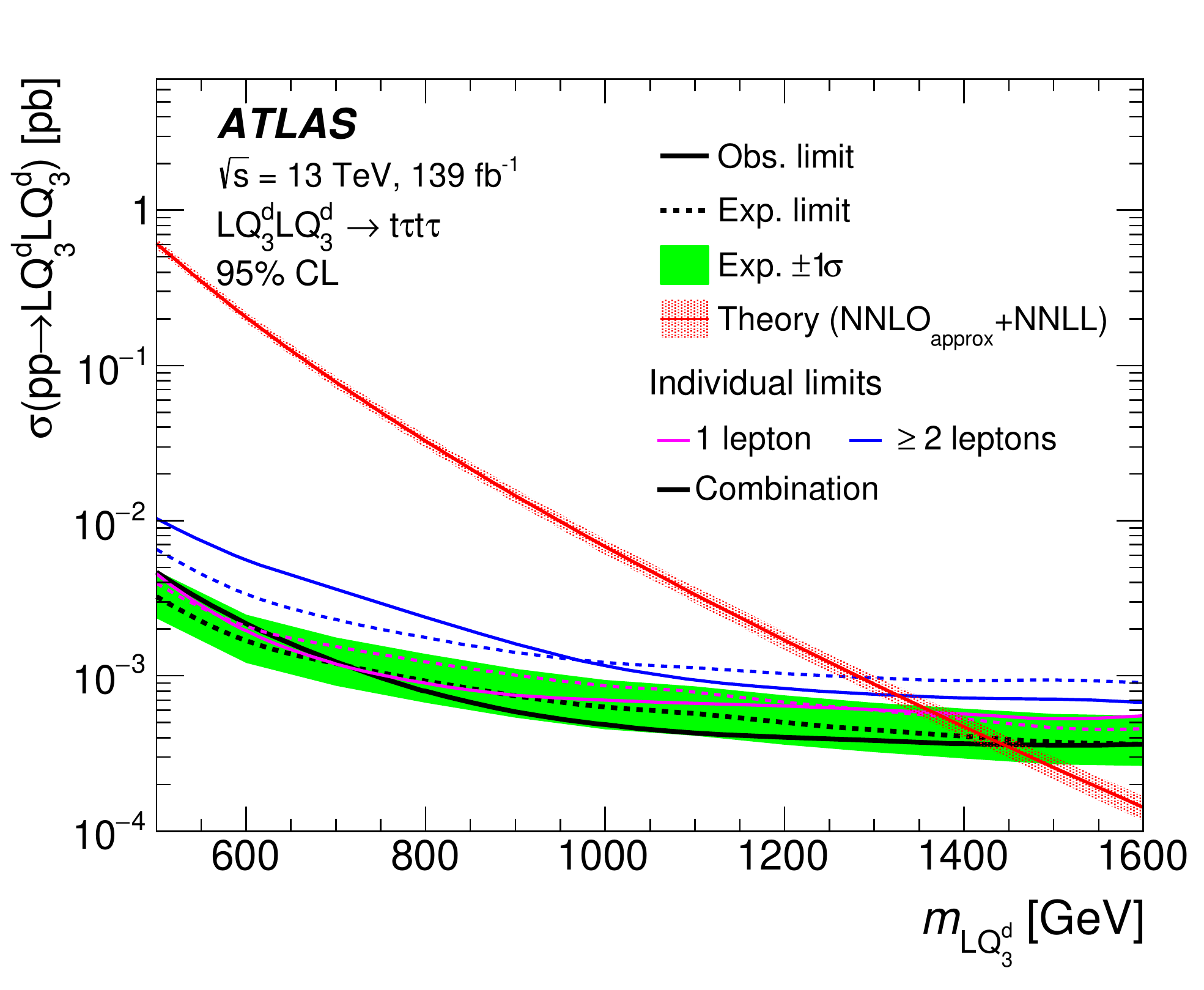}\label{fig:limit1d}}
\subfloat[]{\includegraphics[width=0.45\textwidth]{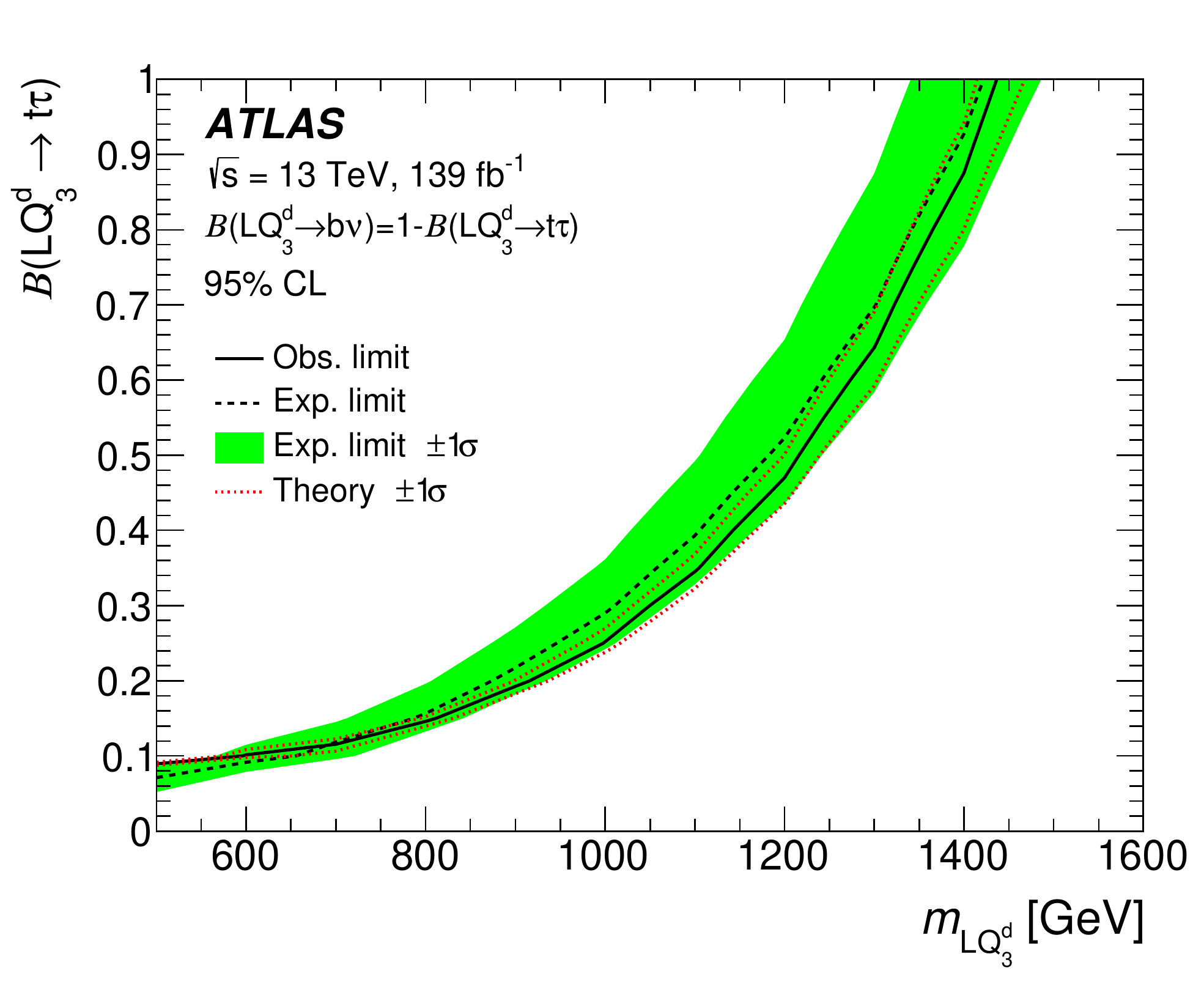}\label{fig:limit2d}}
\caption{(a) Observed (solid line) and expected (dashed line) 95\% CL upper limits on the $\LQ$ pair production cross section as a function of $m_{\LQ}$ resulting from the combination of all analysis channels, assuming $\BR=1$.
The surrounding shaded band corresponds to the $\pm$1 standard deviation ($\pm$1\,$\sigma$) uncertainty around the combined expected limit, as estimated using the asymptotic approximation (see text). This approximation is found to overestimate the +1\,$\sigma$ ($-$1\,$\sigma$) uncertainty of the combined expected limit by about 5\%--15\% (15\%--30\%), depending on $m_{\LQ}$.
The red line and band show the theoretical prediction and its $\pm$1\,$\sigma$ uncertainty. The individual expected limits for the \ltau channel and the combination of the \llOStau and \llSSorllltau channels are shown as the magenta and blue dashed lines, respectively. (b) Observed (solid line) and expected (dashed line) 95\% CL upper limits on $\BR$ as a function of $m_{\LQ}$ resulting from the combination of all analysis channels.
The surrounding shaded band corresponds to the $\pm$1\,$\sigma$ uncertainty around the combined expected limit. The same statement regarding the asymptotic approximation given for (a) applies.
The dotted red line around the observed limit indicates how the observed limit changes when varying the theoretical prediction for the $\LQ$ pair production cross section by its $\pm$1\,$\sigma$ uncertainty.
}
\end{figure*}

\FloatBarrier
 
\section{Conclusion}
\label{sec:conclusion}
A search for pair production of third-generation scalar leptoquarks with a significant branching fraction into a top quark and a $\tau$-lepton has been presented.
The search is based on the full Run~2 dataset recorded with the ATLAS detector at Large Hadron Collider, which corresponds to 139~fb$^{-1}$ of $pp$ collisions at $\sqrt{s}=13~\tev$.
Events are selected if they have one light lepton (electron or muon) and at least one hadronically decaying $\tau$-lepton, or at least two light leptons, and additional jets.
Six final states, defined by the multiplicity and flavour of lepton candidates, are considered in the analysis. Each of them is split into multiple event categories used to search for the signal and improve the modelling of several leading backgrounds.
The signal-rich event categories require at least one hadronically decaying $\tau$-lepton candidate and
employ the total effective mass distribution to discriminate between the signal and the background. The search reaches an expected significance of 5 standard deviations for a
scalar leptoquark decaying exclusively into $t\tau$ and with mass below about $1.2~\tev$, which represents a significant improvement compared to previous searches.
This results from a combination of the higher integrated luminosity used, a significantly improved identification of hadronically decaying $\tau$-leptons, and the sophisticated event selection and categorisation employed, which ensures a high signal acceptance and low background yields.
No significant excess above the Standard Model expectation is observed in any of the considered event categories, and 95\% CL upper limits are set on the production cross section as a function of the leptoquark mass, for different assumptions about the branching fractions into $t\tau$ and $b\nu$. Scalar leptoquarks decaying exclusively into $t\tau$ are excluded up to masses of $1.43~\tev$ while, for a branching fraction of 50\% into $t\tau$, the lower mass limit is $1.22~\tev$. The corresponding expected mass exclusions are  $1.41~\tev$ and $1.19~\tev$, respectively.
 
 
\section*{Acknowledgements}

We thank CERN for the very successful operation of the LHC, as well as the
support staff from our institutions without whom ATLAS could not be
operated efficiently.
 
We acknowledge the support of ANPCyT, Argentina; YerPhI, Armenia; ARC, Australia; BMWFW and FWF, Austria; ANAS, Azerbaijan; SSTC, Belarus; CNPq and FAPESP, Brazil; NSERC, NRC and CFI, Canada; CERN; ANID, Chile; CAS, MOST and NSFC, China; COLCIENCIAS, Colombia; MSMT CR, MPO CR and VSC CR, Czech Republic; DNRF and DNSRC, Denmark; IN2P3-CNRS and CEA-DRF/IRFU, France; SRNSFG, Georgia; BMBF, HGF and MPG, Germany; GSRT, Greece; RGC and Hong Kong SAR, China; ISF and Benoziyo Center, Israel; INFN, Italy; MEXT and JSPS, Japan; CNRST, Morocco; NWO, Netherlands; RCN, Norway; MNiSW and NCN, Poland; FCT, Portugal; MNE/IFA, Romania; JINR; MES of Russia and NRC KI, Russian Federation; MESTD, Serbia; MSSR, Slovakia; ARRS and MIZ\v{S}, Slovenia; DST/NRF, South Africa; MICINN, Spain; SRC and Wallenberg Foundation, Sweden; SERI, SNSF and Cantons of Bern and Geneva, Switzerland; MOST, Taiwan; TAEK, Turkey; STFC, United Kingdom; DOE and NSF, United States of America. In addition, individual groups and members have received support from BCKDF, CANARIE, Compute Canada, CRC and IVADO, Canada; Beijing Municipal Science \& Technology Commission, China; COST, ERC, ERDF, Horizon 2020 and Marie Sk{\l}odowska-Curie Actions, European Union; Investissements d'Avenir Labex, Investissements d'Avenir Idex and ANR, France; DFG and AvH Foundation, Germany; Herakleitos, Thales and Aristeia programmes co-financed by EU-ESF and the Greek NSRF, Greece; BSF-NSF and GIF, Israel; La Caixa Banking Foundation, CERCA Programme Generalitat de Catalunya and PROMETEO and GenT Programmes Generalitat Valenciana, Spain; G\"{o}ran Gustafssons Stiftelse, Sweden; The Royal Society and Leverhulme Trust, United Kingdom.
 
The crucial computing support from all WLCG partners is acknowledged gratefully, in particular from CERN, the ATLAS Tier-1 facilities at TRIUMF (Canada), NDGF (Denmark, Norway, Sweden), CC-IN2P3 (France), KIT/GridKA (Germany), INFN-CNAF (Italy), NL-T1 (Netherlands), PIC (Spain), ASGC (Taiwan), RAL (UK) and BNL (USA), the Tier-2 facilities worldwide and large non-WLCG resource providers. Major contributors of computing resources are listed in Ref.~\cite{ATL-SOFT-PUB-2020-001}.
 

\clearpage
\printbibliography

\clearpage \input{atlas_authlist}

\end{document}

%% file: atlas_authlist.tex
 
\begin{flushleft}
\hypersetup{urlcolor=black}
{\Large The ATLAS Collaboration}

\bigskip

\AtlasOrcid[0000-0002-6665-4934]{G.~Aad}$^\textrm{\scriptsize 102}$,    
\AtlasOrcid[0000-0002-5888-2734]{B.~Abbott}$^\textrm{\scriptsize 128}$,    
\AtlasOrcid[0000-0002-7248-3203]{D.C.~Abbott}$^\textrm{\scriptsize 103}$,    
\AtlasOrcid[0000-0002-2788-3822]{A.~Abed~Abud}$^\textrm{\scriptsize 36}$,    
\AtlasOrcid[0000-0002-1002-1652]{K.~Abeling}$^\textrm{\scriptsize 53}$,    
\AtlasOrcid[0000-0002-2987-4006]{D.K.~Abhayasinghe}$^\textrm{\scriptsize 94}$,    
\AtlasOrcid[0000-0002-8496-9294]{S.H.~Abidi}$^\textrm{\scriptsize 167}$,    
\AtlasOrcid[0000-0002-8279-9324]{O.S.~AbouZeid}$^\textrm{\scriptsize 40}$,    
\AtlasOrcid{N.L.~Abraham}$^\textrm{\scriptsize 156}$,    
\AtlasOrcid[0000-0001-5329-6640]{H.~Abramowicz}$^\textrm{\scriptsize 161}$
% ...
\bigskip
\\

% ...
$^{36}$CERN, Geneva; Switzerland.\\
% ...
$^{40}$Niels Bohr Institute, University of Copenhagen, Copenhagen; Denmark.\\
% ...
$^{53}$II. Physikalisches Institut, Georg-August-Universit\"{a}t G\"ottingen, G\"ottingen; Germany.\\
% ...
$^{94}$Department of Physics, Royal Holloway University of London, Egham; United Kingdom.\\
% ...
$^{102}$CPPM, Aix-Marseille Universit\'e, CNRS/IN2P3, Marseille; France.\\
$^{103}$Department of Physics, University of Massachusetts, Amherst MA; United States of America.\\
% ...
$^{128}$Homer L. Dodge Department of Physics and Astronomy, University of Oklahoma, Norman OK; United States of America.\\
% ...
$^{156}$Department of Physics and Astronomy, University of Sussex, Brighton; United Kingdom.\\
% ...
$^{161}$Raymond and Beverly Sackler School of Physics and Astronomy, Tel Aviv University, Tel Aviv; Israel.\\
% ...
$^{167}$Department of Physics, University of Toronto, Toronto ON; Canada.\\
% ...

\end{flushleft}
